\documentclass[twocolumn,amssymb, nobibnotes, aps, prd, superscriptaddress]{revtex4-2}

\usepackage{graphicx}
\usepackage{amsmath, amsthm, amssymb}
\usepackage{mathdots}
\usepackage{float}
\theoremstyle{definition}

\usepackage{pgfplots} 
\usepackage{tikz-3dplot}
\usepgfplotslibrary{patchplots}
\usepgfplotslibrary{groupplots}
\pgfplotsset{compat=1.6,ylabsh/.style={every axis y label/.style={at={(0,0.5)}, xshift=#1, rotate=90}}}
\usetikzlibrary{positioning}
\usetikzlibrary{shapes}
\usetikzlibrary{shapes.geometric}
\usetikzlibrary{decorations.text}
\usetikzlibrary{decorations.pathreplacing,shapes.misc}
\usetikzlibrary{matrix}
\usepackage{caption,subcaption}
\newcommand*{\addheight}[2][.5ex]{
	\raisebox{0pt}[\dimexpr\height+(#1)\relax]{#2}
}

\begin{document}

	\title{Mean field elastic moduli of a three-dimensional cell-based vertex model}
	\author{Kyungeun Kim}
	\affiliation{Physics Department, Syracuse University, Syracuse, NY 13244 USA}
	\author{Tao Zhang}
	\affiliation{Department of Polymer Science and Engineering, Shanghai Jiao Tong University, Shanghai, 200240 China}
	\author{J. M. Schwarz}
	\affiliation{Physics Department, Syracuse University, Syracuse, NY 13244 USA}
	\affiliation{Indian Creek Farm, Ithaca, NY 14850 USA}
	
	\date{\today}
		\begin{abstract}
  The mechanics of a foam typically depends on the bubble geometry, topology, and the material at hand, be it metallic or polymeric, for example. While the foam energy functional for each bubble is typically minimization of surface area for a given volume, biology provides us with a wealth of additional energy functionals, should one consider biological cells as a foam-like material. Here, we focus on a mean field approach to obtain the elastic moduli, within linear response, for an ordered, three-dimensional vertex model using the space-filling shape of a truncated octahedron and whose energy functional is characterized by a restoring surface area spring and a restoring volume spring. The tuning of the three-dimensional shape index exhibits a rigidity transition via a compatible-incompatible transition. Specifically, for smaller shape indices, both the target surface area and volume cannot be achieved, while beyond some critical value of the three-dimensional shape index, they can be, resulting in a zero-energy state. As the elastic moduli depend on curvatures of the energy when the system, we obtain these as well. In addition to analytically determining the location of the transition in mean field, we find that the rigidity transition and the elastic moduli depend on the parameterization of the cell shape with this effect being more pronounced in three dimensions given the array of shapes that a polyhedron can take on (as compared to a polygon). We also uncover nontrivial dependence on the deformation protocol in which some deformations result in affine motion of the vertices, while others result in nonaffine motion. Such dependencies on the shape parameterization and deformation protocol give rise to a nontrivial shape landscape and, therefore, nontrivial mechanical response even in the absence of topology changes.
		\end{abstract}
		
		\maketitle
		
		\section{Introduction}

    Different types of objects fill three-dimensional space. Such objects include foams, metallic grains, and biological cells. There are indeed constraints on the shapes of these objects to be space-filling. Example shapes include the cube, the dodecahedron, and the truncated octahedron, all of which are convex shapes. See Fig. 1 for a three-dimensional example as well as a two-dimensional one. Efforts to put further constraints on the shapes that, for example, have the least possible surface area for a given volume, otherwise known as Kelvin cells, also exist~\cite{Kelvin1887}. In 1994, Weaire and Phelan discovered an space-filling arrangement with two types of cells whose surface area was 0.3\% smaller than the originally-proposed polyhedron by Kelvin~\cite{Weaire1994,Kelvin1887}. 

    As all objects are deformable to greater or lesser extents, it, therefore, makes sense to explore the shape-changing capability of such objects and ask how does this capability affect the mechanics of the material? More specifically, how does the shear modulus differ between a tessellation of truncated octahedra and a tessellation of cubes? We will address this question for a more recent energy functional, as compared to foams, that is inspired by biological cells. More specifically, this energy functional is quadratic in both surface area and volume, each with a respective target surface area and target volume~\cite{Bi_2015,Merkel_2018,Zhang2023}.  The area "spring", if you will, captures the elasticity of the cell cortex lying just beneath the cell membrane that is, in part, responsible to the structural integrity of a cell. The volume "spring", on the other hand captures the volume compressibility of a cell given that it contains water, proteins, fats, etc. For foams, on the other hand, the energy functional scales linearly with the surface area~\cite{Kelvin1887,Cantat2013}. 

\begin{center}
	\begin{tabular}{cc}
		\addheight{	\includegraphics[width=3.0cm]{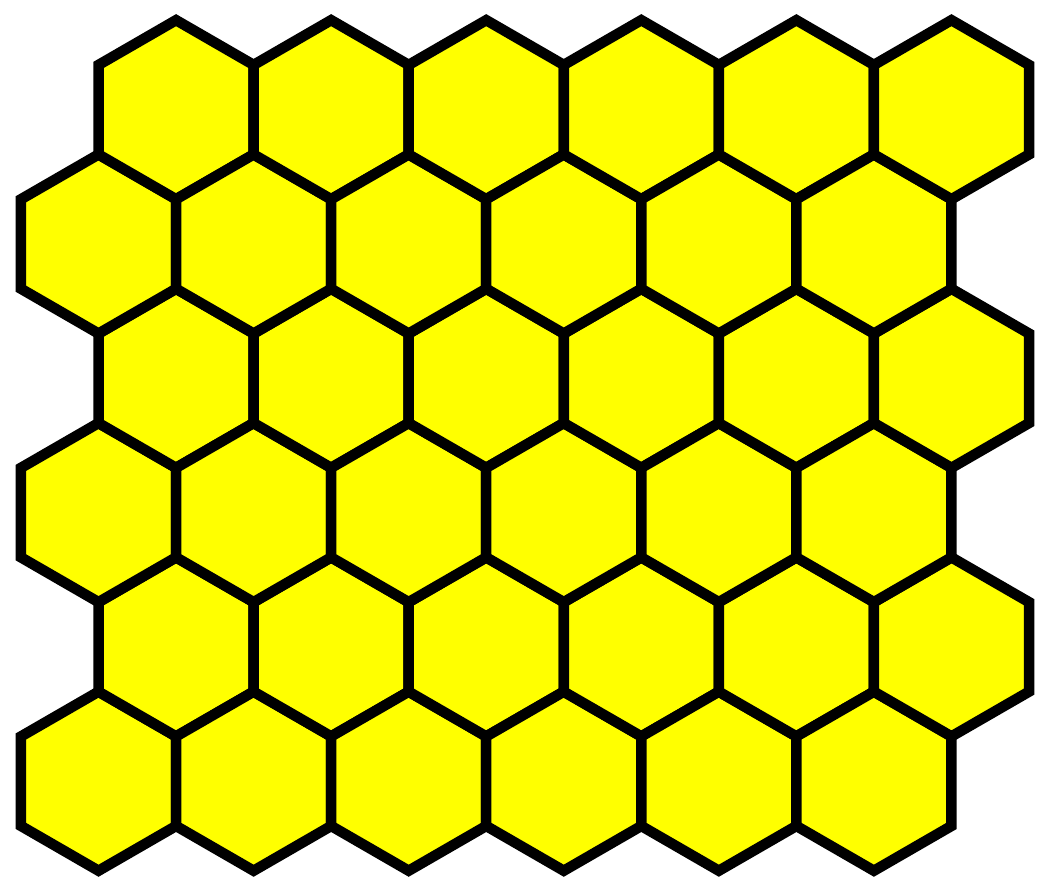}
		} &
		\addheight{\includegraphics[width=3.0cm]{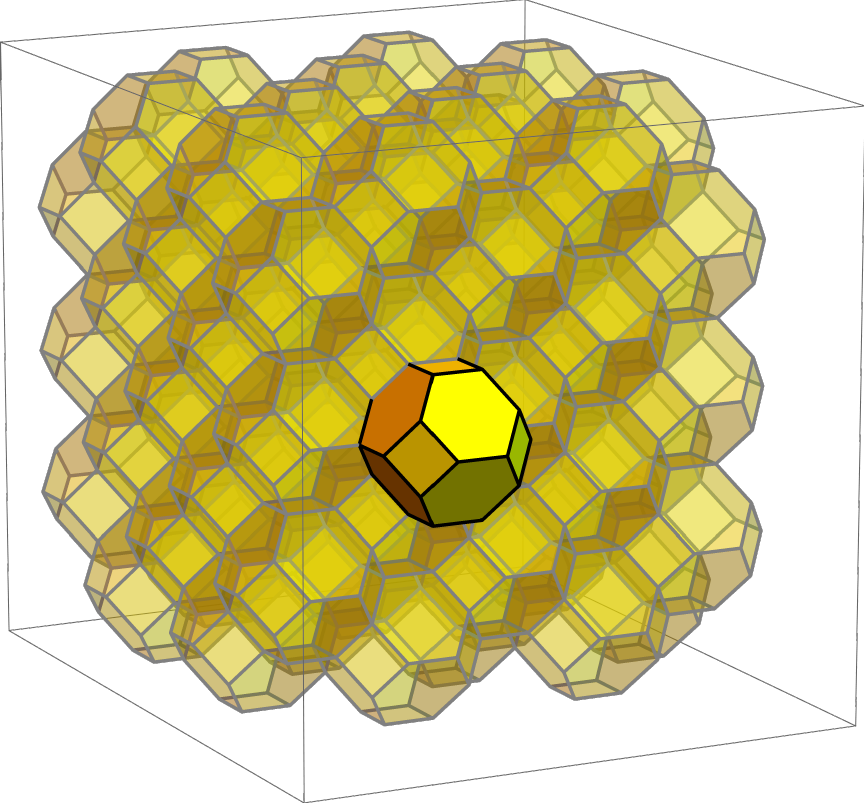}
		} \\
	\end{tabular}
	\captionsetup{type=figure}
	\captionof{figure}{{\it Space-filling packings.} Left: A packing of hexagons in two dimensions. Right: A three-dimensional packing built out of truncated octahedrons.}
	\label{fig:lattice-model}
\end{center}		

    Some of the earliest work considering a space-filling collection of cells as a foam-like material in two dimensions, otherwise known as vertex models, was pioneered by Honda and collaborators to quantify, for example, the cellular response to a wound in the eye of a cat~\cite{Honda1982,Honda2022}. Later work began to consider an energy functional with a quadratic contribution of the surface area~\cite{Farhadifar2007,Staple2010}. Certainly, two-dimensional, open-source code bases has helped accelerate this work~\cite{Mirams2013,Sussman2017}.  Such two-dimensional models have demonstrated rather rich collective behavior, to name a few~\cite{Staple2010,Moshe2018,Sahu2020,Huang2022,Chen2022}. To be more specific, in a ordered system and in the absence of cellular rearrangements, there exists a rigidity transition from a rigid phase to a floppy phase as a dimensionless version of the target surface area, otherwise known as the shape index, is increased~\cite{Staple2010}. Such a transition signifies the breakdown of nonlinear elasticity with an emergent, anomalous coupling between compression and shear~\cite{Moshe2018}. There also exists another rigidity transition from a rigid phase to a fluid phase when cellular rearrangements are allowed~\cite{Bi_2015}.  In addition, such models exhibit the phenomenon of micro-demixing~\cite{Sahu2020}, shear-driven solidification~\cite{Huang2022} and even fracture~\cite{Chen2022}. Several aspects of the theoretical richness have been confirmed experimentally make the vertex model approach a compelling one~\cite{Park2015,Sahu2020}, though other complimentary theoretical approaches, such as Voronoi models~\cite{Bi2016,Barton2017}, do exist. 
    
    There also exists much progress on three-dimensional vertex model approaches. For instance, over ten years ago Okuda and collaborators outlined an algorithm to simulate such models that included reconnection events, or cells moving past each other~\cite{Okuda2012}. They also explored cellular swelling to determine how an initially flat monolayer of cells deformed into a curved monolayer of cells~\cite{Okuda2012}. This exploration was done using a different energy functional from the one that will be implemented here~\cite{Okuda2012}. More recently, Okuda and Sato have demonstrated that polarized interfacial tension can induce motion of clusters of cells within a model tissue~\cite{Okuda2022}. Combining Turing models with three-dimensional vertex models as well as branching in three-dimensional vertex models has also been explored, to name a few~\cite{Okuda2018,Rozman2020}. 
    
    Given these initial results, one wonders how the energy functional affects the physics of such three-dimensional vertex models. Very recent work begins to address that very question using the restoring surface area and volume springs to find that there exists a rigidity transition in a cellular-based vertex model~\cite{Zhang2023}. This work also identified a new boundary-bulk effect in the spatial organization of cells in which boundary cells are much more aligned than bulk cells~\cite{Zhang2023}. Even more recent work on multiscale modeling addresses how such a three-dimensional, cellular-based vertex model can be coupled with a minimal mechanical model for a cell nucleus to address foundational questions in biology~\cite{ZhangSarthak2023}. 

While there are a multitude of directions that can be explored in this three-dimensional, cellular-based vertex model, inspired by recent work focusing on elastic moduli in an ordered version in two dimensions~\cite{Staddon2023}, here, we focus on a mean field approach to determine the elastic moduli in three dimensions.  To do so, we will revisit the two-dimensional calculations for hexagonal cells and expand the results to include octogonal cells. This expansion will help us better understand our three-dimensional results for truncated octahedra. We will determine the shear modulus, bulk modulus, and Young's modulus as a function of a dimensionless shape index in both two and three dimensions for the three different models. In doing so, we will be able to formulate some closed form estimates for the location of the rigidity transition. The manuscript, thus, continues with describing the model and the parameterization of the cell shape in both two and three dimensions, followed by a description of deformation protocol in both dimensions. We then present our results in both dimensions and conclude with a discussion of the implications of our results.

\section{A cellular-based Vertex Model} 	
\subsection{The two-dimensional version}\label{sec:vm2d}
		Consider a cross-section of a monolayer of cells. The $i$th cellular cross-section is represented as as a polygon with $A_i$ area and $P_i$ perimeter. Each cell is assumed to have a target area of $A_0$ and a target perimeter of $P_0$. Assuming all edges of a polygon are equivalent, the elasticity of the cortical tension can, in its simplest form, be expressed as a perimeter spring (with a target perimeter), while an area spring encodes the bulk compressibility of a cell. One can, therefore, express the energy of a collection of $N$ cells, or the tissue energy, $E_T$ as $E_T=\frac{1}{2}K_A\sum_{i=1}^{N} (A_i-A_0)^2+\frac{1}{2}K_P\sum_{i=1}^{N}(P_i-P_0)^2$ where $K_P$ denotes cell perimeter stiffness/contractility and $K_A$, cell area stiffness~\cite{Bi_2015}. Note that the edges between the cells are shared, as indicated in Fig. 1, representing the adhesiveness of the cell. 
  
  To build more intuition about this model, one can readily implement a mean-field theory approach for an ordered cellular packing in which each cell is equivalent and, therefore, contributes equally to the packing. In other words, $E_T=N\,E_S$, where $E_S$ represents the energy of a single cell. Moreover, we use a dimensionless form of cell energy $e_s=\frac{E_s}{K_A A_0^2}$ to arrive at 
  \begin{equation}
   e_S=\frac{1}{2}(a-1)^2+\frac{1}{2}k_r(p-p_0)^2,   
  \end{equation}
  with $a=\frac{A}{A_0},p=\frac{P}{\sqrt{A_0}}$, and $p_0=\frac{P_0}{\sqrt{A_0}}$. In addition, $k_r$ stands for the rigidity stiffness and is expressed as $\frac{K_P}{K_A A_0}$. Furthermore, the single cell's target dimensionless area is unity. We will assign $p$ to be the dimensionless shape index in two dimensions. Now that the energy functional for a single cell has been defined, let move to further characterize its shape. 
		\subsubsection{Hexagonal Model}\label{HM}
		We begin with a single cell as a hexagon, which is parameterized by two edge lengths as illustrated in Fig. 2. The cell is embedded in a box of area $L_x\times L_y$, where $L_x$ and $L_y$ are the lengths of the box in the $x$ and $y$ directions,  respectively. Note that the box is indicated as a shaded gray in Fig. 2. 
		\begin{figure}[h!]
			\begin{tikzpicture}[scale=1.8]
				\fill[gray,opacity=0.5](1,0.866025)--(1,-0.866025)--(-1,-0.866025)--(-1,0.866025)--cycle;
				\draw[-,thick,fill=white](1.0, 0)--(0.5, 0.866025)--(-0.5,  0.866025)--(-1.0, 0)--(-0.5,-0.866025)--(0.5,-0.866025)--cycle;
				\draw[<->](-0.5, -1.0)--(0.5, -1.0);
				\draw[<->](-0.5, 1.0)--(0.5, 1.0);
				\draw[<->](-1.1, 0.1)--(-0.65, 0.9);
				\draw[<->](-1.15, -0.1)--(-0.6, -1.0);
				\draw[<->](1.1, 0.1)--(0.65, 0.9);
				\draw[<->](1.15, -0.1)--(0.6, -1.0);
				\draw[-,dashed](-1,0)--(1,0);
				\node[label={0:{\large$l_1$}},inner sep=2pt] at (-0.2,-1.2) {};
				\node[label={0:{\large$l_1$}},inner sep=2pt] at (-0.2,1.2) {};
				\node[label={90:{\large$l_2$}},inner sep=2pt] at (-1.1,0.4) {};
				\node[label={-90:{\large$l_2$}},inner sep=2pt] at (-1.0,-0.5) {};
				\node[label={90:{\large$l_2$}},inner sep=2pt] at (1.1,0.4) {};
				\node[label={-90:{\large$l_2$}},inner sep=2pt] at (1.0,-0.5) {};
				\node[label={0:{\large$\theta$}},inner sep=2pt] at (-0.5,0.45) {};
				\draw[-](-0.30, 0.85) arc[start angle=9, end angle=-128,radius=0.2cm];
				\node[label={0:{\large$v_1$}},inner sep=2pt] at (1.1,0) {};
				\node[label={45:{\large$v_2$}},inner sep=2pt] at (0.5,0.866025) {};
				\node[label={100:{\large$v_3$}},inner sep=2pt] at (-0.5,0.866025) {};
				\node[label={180:{\large$v_4$}},inner sep=2pt] at (-1.1,0) {};
				\node[label={265:{\large$v_5$}},inner sep=2pt] at (-0.5,-1) {};
				\node[label={-85:{\large$v_6$}},inner sep=2pt] at (0.5,-1) {};
				\node[label={-90:{\color{blue}\large$O$}},inner sep=2pt] at (0,0) {};
				\draw[fill=black](1.0, 0) circle[radius=0.03];
 		        \draw[fill=black](0.5, 0.866025) circle[radius=0.03];
 		        \draw[fill=black](-0.5,  0.866025) circle[radius=0.03];
 		        \draw[fill=black](-0.5,-0.866025) circle[radius=0.03];
 		        \draw[fill=black](0.5,-0.866025) circle[radius=0.03];
 		        \draw[fill=black](-1.0, 0) circle[radius=0.03];
				\draw[fill=blue](0, 0) circle[radius=0.05];
			\end{tikzpicture}
			\captionsetup{type=figure}
			\captionof{figure}{Parameterization of the hexagonal cell model embedded in an $L_x$ by $L_y$ box.}
		\end{figure}
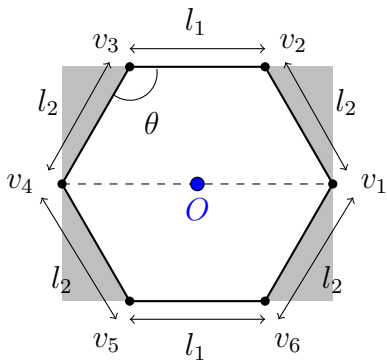
		Moreover, the angle between the two lines $l_1$ and $l_2$ only is denoted by $\theta$.  The vector of each vertex in relation to the origin $O=\{0,0\}$ can be stated as follows (noting that $\cos\theta\leq0$ as $\theta$ is an obtuse angle):
		\begin{align*}
			v_1&=\bigg\{\frac{l_1}{2} - l_2\cos\theta,0\bigg\},v_2=\bigg\{\frac{l_1}{2}, l_2\sin\theta\bigg\}\\
			v_3&=\bigg\{-\frac{l_1}{2}, l_2\sin\theta\bigg\},v_4=\bigg\{-\frac{l_1}{2} + l_2\cos\theta,0\bigg\}\\
			v_5&=\bigg\{-\frac{l_1}{2},-l_2\sin\theta\bigg\},v_6=\bigg\{\frac{l_1}{2},-l_2\sin\theta\bigg\}.
		\end{align*}
The perimeter of the shape is $p=2l_1+4l_2$, while the area is equal to $-2\frac{1}{2}(2l_2 \sin\theta)(l_2 \cos\theta)+2{l_1}l_2 \sin\theta=2l_2\sin\theta(l_1-l_2\cos\theta)$. Accordingly, the dimensionless energy of a single cell (with a unit area) is
		\begin{align}
			e_S&=\frac{1}{2}(a-1)^2+\frac{1}{2}k_r(p-p_0)^2\notag\\
			=&\frac{1}{2}\big(2l_2\sin\theta(l_1-l_2\cos\theta)-1\big)^2+\frac{1}{2}k_r(2l_1+4l_2-p_0)^2.
		\end{align}
		Note that for a regular hexagon with unit area, $p=2l_1+4l_2=\sqrt{8\sqrt{3}}$ and $2l_2\sin(\frac{2\pi}{3})(l_1-l_2\cos(\frac{2\pi}{3}))=1$, such that $l_1=l_2=\frac{\sqrt{2}}{3^{\frac{3}{4}}}$.

  The above energy depends on three parameters, $l_1$, $l_2$, and $\theta$. This parameterization makes it clear that area and perimeter do not alone uniquely define cell shape, as pointed out in Ref.~\cite{Staddon2023}. To look for a compatible-incompatible transition, one can minimize the energy with respect to these three parameters to find that for $p_0<p_0^*(6)$, the energy is nonzero with the cell unable to achieve both the target area and perimeter, where $p_0^*(6)$ is the shape index for a regular hexagon with unit area. While for $p_0\ge p_0^*(6)$, the cell is able to achieve its target perimeter and area, hence, the cell's shape is compatible with its energy.  Please see Appendix A for details. Similar analysis was done in Ref.~\cite{Staddon2023} for a slightly different shape parameterization. 
		\subsubsection{Octagonal Model}\label{OM}
		In looking towards the three-dimensional model, it will be useful to study an octogonal cell model in two dimensions, despite the fact that the octogonal cell alone cannot tile two-dimensional space. Interestingly, however, octogonal cells can achieve rectangular shapes. 
  
  We parameterize the two-dimensional octagonal model in the following way (see Fig. 3).
		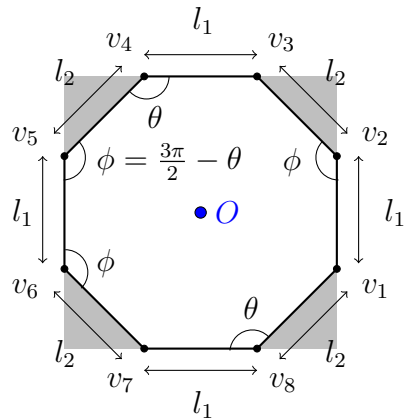
\begin{figure}[h!]
			\begin{tikzpicture}[scale=1.5]
				\fill[gray,opacity=0.5](-1.20711, -1.20711)--(-1.20711, 1.20711)--(1.20711, 1.20711)--(1.20711, -1.20711)--cycle;
				\draw[-,thick,fill=white](-0.5, -1.20711)--(-1.20711, -0.5)--(-1.20711, 0.5)--(-0.5, 1.20711)--(0.5, 1.20711)--(1.20711, 
				0.5)--(1.20711, -0.5)--(0.5, -1.20711)--cycle;
				\draw[<->](-0.5, -1.4)--(0.5, -1.4);
				\draw[<->](-0.5, 1.4)--(0.5, 1.4);
				\draw[<->](-1.4, -0.45)--(-1.4, 0.5);
				\draw[<->](1.4, -0.45)--(1.4, 0.5);
				\draw[<->](-1.3, 0.7)--(-0.7, 1.3);
				\draw[<->](-1.3, -0.7)--(-0.7, -1.3);
				\draw[<->](1.3, 0.7)--(0.7, 1.3);
				\draw[<->](1.3, -0.7)--(0.7, -1.3);
				\draw[fill=blue](0, 0) circle[radius=0.05];
				\node[label={0:{\large$l_1$}},inner sep=2pt] at (-1.8,0) {};
				\node[label={0:{\large$l_1$}},inner sep=2pt] at (1.5,0) {};
				\node[label={0:{\large$l_1$}},inner sep=2pt] at (-0.2,-1.7) {};
				\node[label={0:{\large$l_1$}},inner sep=2pt] at (-0.2,1.7) {};
				\node[label={90:{\large$l_2$}},inner sep=2pt] at (-1.2,1) {};
				\node[label={90:{\large$l_2$}},inner sep=2pt] at (1.2,1) {};
				\node[label={-90:{\large$l_2$}},inner sep=2pt] at (1.2,-1) {};			\node[label={-90:{\large$l_2$}},inner sep=2pt] at (-1.2,-1) {};
				\draw[fill=black](-0.5, -1.20711) circle[radius=0.03];
 		        \draw[fill=black](-1.20711, -0.5) circle[radius=0.03];
 		        \draw[fill=black](-1.20711, 0.5) circle[radius=0.03];
 		        \draw[fill=black](-0.5, 1.20711) circle[radius=0.03];
 		        \draw[fill=black](0.5, 1.20711) circle[radius=0.03];
 		        \draw[fill=black](1.20711, 0.5) circle[radius=0.03];
 		        \draw[fill=black](1.20711, -0.5) circle[radius=0.03];
 		        \draw[fill=black](0.5, -1.20711) circle[radius=0.03];
                \node[label={-60:{\large$v_8$}},inner sep=2pt] at (0.5, -1.3) {};           
                \node[label={-100:{\large$v_7$}},inner sep=2pt] at (-0.5, -1.3) {};           
                \node[label={-160:{\large$v_6$}},inner sep=2pt] at (-1.3, -0.5) {};           
                \node[label={-200:{\large$v_5$}},inner sep=2pt] at (-1.3, 0.5) {};           
                \node[label={110:{\large$v_4$}},inner sep=2pt] at (-0.5, 1.3) {};           
                \node[label={70:{\large$v_3$}},inner sep=2pt] at (0.5, 1.3) {};           
                \node[label={20:{\large$v_2$}},inner sep=2pt] at (1.3, 0.5) {};           
                \node[label={-20:{\large$v_1$}},inner sep=2pt] at (1.3, -0.5) {};           
                \node[label={0:{\large$\theta$}},inner sep=2pt] at (-0.6,0.85) {};
				\node[label={0:{\large$\theta$}},inner sep=2pt] at (0.25,-0.85) {};			\node[label={0:{\large$\phi=\frac{3\pi}{2}-\theta$}},inner sep=2pt] at (-1.05,0.45) {};
				\node[label={0:{\large$\phi$}},inner sep=2pt] at (-1.05,-0.45) {};
				\node[label={0:{\large$\phi$}},inner sep=2pt] at (0.6,0.45) {};
				\node[label={0:{\color{blue}\large$O$}},inner sep=2pt] at (0,0) {};
				\draw[-](-0.28, 1.20711) arc[start angle=0, end angle=-136,radius=0.2cm] ;
				\draw[-](0.6, -1.1) arc[start angle=45, end angle=175,radius=0.2cm] ;
				\draw[-](-1.08, 0.63) arc[start angle=45, end angle=-88,radius=0.2cm] ;
				\draw[-](-1.2, -0.33) arc[start angle=90, end angle=-45,radius=0.2cm] ;
				\draw[-](1.08, 0.63) arc[start angle=135, end angle=270,radius=0.2cm] ;
			\end{tikzpicture}
			\captionsetup{type=figure}
			\captionof{figure}{Parameterization of octagonal cell model.}
		\end{figure}
	Assuming the octagon is symmetric along $x,y$ direction about the origin $O=\{0,0\}$ and setting $\theta$ as the angle between the horizontal $l_1$ and $l_2$, as previously, the remaining angles will be automatically defined from the following coordinate vectors of each vertex: 
		\begin{align*}
			v_1&=\bigg\{\frac{l_1}{2}-l_2\cos\theta,-\frac{l_1}{2}\bigg\},v_2=\bigg\{\frac{l_1}{2}-l_2\cos\theta,\frac{l_1}{2}\bigg\},\\
			v_3&=\bigg\{\frac{l_1}{2},\frac{l_1}{2}+l_2\sin\theta\bigg\},v_4=\bigg\{-\frac{l_1}{2},\frac{l_1}{2} + l_2\sin\theta\bigg\},\\
			v_5&=\bigg\{-\frac{l_1}{2}+ l_2\cos\theta,\frac{l_1}{2}\bigg\},v_6=\bigg\{-\frac{l_1}{2}+l_2\cos\theta,-\frac{l_1}{2}\bigg\},\\
            v_7&=\bigg\{-\frac{l_1}{2}, -\frac{l_1}{2}-l_2\sin\theta\bigg\},v_8=\bigg\{\frac{l_1}{2},-\frac{l_1}{2}-l_2\sin\theta\bigg\}.
		\end{align*}
With the above parametrization, the dimensionless single cell energy function (of unit target area) will be
		\begin{align}
			e_S&=\frac{1}{2}(a-1)^2+\frac{1}{2}k_r(p-p_0)^2\notag\\
			=&\frac{1}{2}\big[l_1(l_1-2l_2\cos\theta)+2l_2\sin\theta(l_1-l_2\cos\theta)-1\big]^2\notag\\
			&+\frac{1}{2}k_r(4l_1+4l_2-p_0)^2.
		\end{align}
		When $\theta=\frac{3\pi}{4}$ and $a=1$, then $l_1=l_2=\frac{1}{\sqrt{2 + 2 \sqrt{2}}}$. We note that there is a similar shape transition from incompatible to compatible shapes as the target shape index is increased with $p<p_0^*(8)$.
  
		\subsection{The three-dimensional version}\label{sec:vm3d}
		To generalize the two-dimensional vertex model discussed in the prior subsections, we will simply transform areas $A$ to volumes $V$ and perimeters $P$ to areas $A$ to arrive at~\cite{Zhang2023} 
  \begin{equation}
   E_T=\frac{1}{2}K_V\sum_i (V_i-V_0)^2+\frac{1}{2}K_S\sum_i(A_i-A_0)^2.   
  \end{equation} 
 Here $K_V$ represents the cell volume stiffness and $K_S$ the cell surface area stiffness. Since $L \sim V^{\frac{1}{3}}$ and $L \sim A^{\frac{1}{2}}$ with $L$ as the length unit in the three-dimensional scenario, we can express a dimensionless version of tissue energy with $e_T=\frac{E_T}{K_V V_0^2}$ as 
 \begin{equation}
    e_T=\frac{1}{2} (v-1)^2+\frac{1}{2}k_\rho(s-s_0)^2, 
\end{equation}
where $v,s,s_0$ are given by $v=\frac{V}{V_0},s=\frac{A}{(V_0)^{\frac{2}{3}}}$, and $s_0=\frac{a_0}{(v_0)^{\frac{2}{3}}}$. Finally, $k_\rho=\frac{K_S}{K_{V}{V_0}^{\frac{3}{2}}}$. Note that we are now using $s$ to denote the dimensionless surface area (as opposed to $a$) and it is also referred to as the shape index. 

To further parameterize the model, we implement the following criteria: 
		\begin{enumerate}			
			\item A single polyhedron whose copies fill three-dimensional space.			
			\item The polyhedron must have flat, non-curved surfaces.			
			\item To ultimately compare with earlier work, each vertex in the tiling has four edges attached to it.
               \item Consider an ordered packing. 
		\end{enumerate}
		The initial requirement fulfills the confluency requirement implemented in standard vertex models. The second criterion is also typically implemented in standard vertex models and helps make for easier computations. The third requirement originates from the model with reconnections that we would like to compare~\cite{Okuda2012,Zhang2023}. As for the last requirement, measuring mechanical properties of disordered cellular packings is more complex than for ordered cellular packings and readily allows us to implement a mean field theory approach, though an effective medium theory has been implemented for disordered packings in two dimensions~\cite{Damavandi2022}.  
  
A truncated octahedron shape (see Fig. 1) meets the three criteria above and so we form an ordered packing of them (fourth criterion). Furthermore, the planar view of the truncated octahedron matches the 4-8 Lattice outlined in Ref.~\cite{Staple2010}.
		\begin{figure}
			\captionsetup{singlelinecheck = false, justification=raggedright}
			\begin{center}
				\includegraphics[width=6cm]{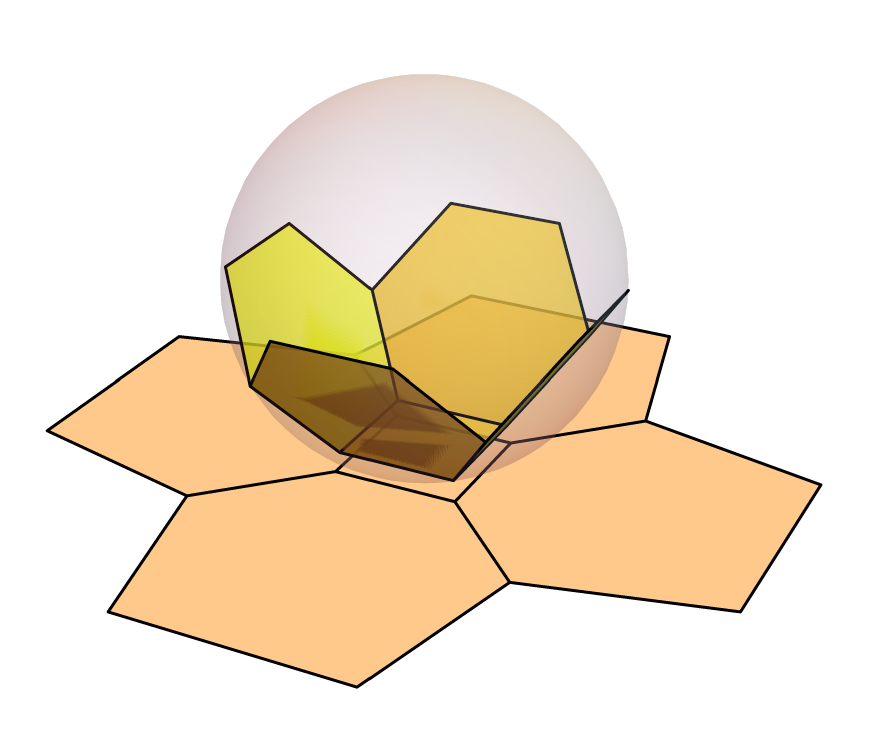}
				\captionsetup{type=figure}
				\captionof{figure}{A graphical representation of the bottom half of a truncated octahedron through stereographic projection. Instead of curvy lines, it is depicted with straight lines between projected points.}
				\label{fig:stereographic-projection}
			\end{center}
		\end{figure}
		A stereographic projection of a bottom half of a truncated octahedron shows a shape with one square and four hexagons, as illustrated in Fig.~\ref{fig:stereographic-projection}. The projection is conformal, meaning it maintains angles and shapes of the  curve. By stretching the four vertices diagonally, we observe a lower hemisphere of a truncated octahedron embedded in an octagon. Fig. \ref{fig:para-truncoct-from-octagon} visualizes this process.
		\begin{figure}[h]
			\captionsetup{singlelinecheck = false, justification=raggedright}
			\centering{
				\resizebox{75mm}{!}{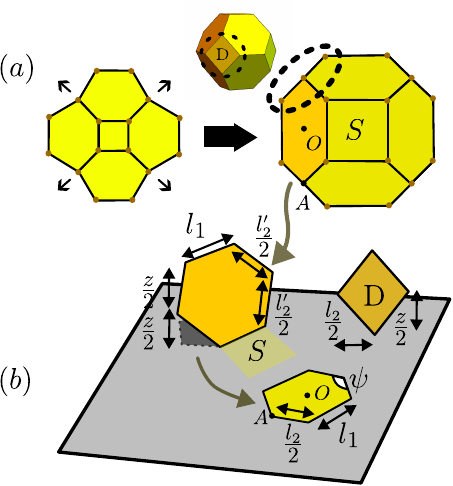}
				\caption{(a) Through stretching the four vertices diagonally, the bottom half of truncated can be expressed as four hexagons and a square $S$ within an octagon. $D$ is used to symbolize a diamond shape on the side connecting the top and bottom halves. (b) Parameters to identify a hexagon tilted in space from a planar view.}
		\label{fig:para-truncoct-from-octagon}
			}
		\end{figure}
		We observe that the right side of Fig. 5a is analogous to a truncated octahedron projected onto the $xy$-plane. Observe that the hexagon is smaller than the octagon due to the fact that it is tilted in space. As in Sec. \ref{OM}, the octagon produced from 
 a projection for a regular truncated octahedron has two separate edge lengths ($l_1,l_2$). Additionally, the side edge length $\frac{l_2'}{2}$ of the hexagon on a plane can be calculated using vertical coordinate $z$ and parameters $l_1,l_2,\psi$ if points $A,O$ are known. Therefore, we can use simple parametrization from the octagon shown below.
		\begin{figure}[h!]
			\captionsetup{singlelinecheck = false, justification=raggedright}
			\centering
			\begin{tikzpicture}[scale=1.5]
				\fill[gray,opacity=0.5](-1.20711, -1.20711)--(-1.20711, 1.20711)--(1.20711, 1.20711)--(1.20711, -1.20711)--cycle;
				\draw[-,thick,fill=white](0.5, 0.5)--(-0.5, 0.5)--(-0.5, -0.5)
				--(0.5, -0.5)--cycle;
				\draw[-,thick,fill=white](0.853553, -0.853553)--(0.5, -1.20711)--(-0.5, -1.20711)--(-0.853553, -0.853553)--(-0.5, -0.5)--(0.5, -0.5)--cycle;
				\draw[-,thick,fill=white](0.5, 0.5)--(-0.5, 0.5)--(-0.853553, 0.853553)--(-0.5, 1.20711)--(0.5, 1.20711)--(0.853553, 0.853553)--cycle;
				\draw[-,thick,fill=white](0.853553, -0.853553)--(1.20711, -0.5)--(1.20711, 0.5)--(0.853553, 0.853553)--(0.5, 0.5)--(0.5, -0.5)--cycle;
				\draw[-,thick,fill=white](-0.853553, -0.853553)--(-1.20711, -0.5)--(-1.20711, 0.5)--(-0.853553, 0.853553)--(-0.5, 0.5)--(-0.5, -0.5)--cycle;
				\draw[<->](-0.5, -1.4)--(0.5, -1.4);
				\draw[<->](-1.4, -0.45)--(-1.4, 0.5);
				\draw[<->](-1.3, 0.7)--(-0.7, 1.3);
				\draw[<->](-1.3, -0.7)--(-0.7, -1.3);
				\node[label={0:{\large$l_1$}},inner sep=2pt] at (-1.8,0) {};
				\node[label={0:{\large$l_1$}},inner sep=2pt] at (-0.2,-1.6) {};
				\node[label={90:{\large$l_2$}},inner sep=2pt] at (-1.2,1) {};
				\node[label={-90:{\large$l_2$}},inner sep=2pt] at (-1.2,-1) {};
				\node[label={0:{\large$\theta$}},inner sep=2pt] at (-0.6,0.85) {};
				\node[label={0:{\large$\phi$}},inner sep=2pt] at (-1.0,0.45) {};
				\draw[-](-0.28, 1.20711) arc[start angle=0, end angle=-136,radius=0.2cm] ;
				\draw[-](-1.08, 0.63) arc[start angle=45, end angle=-88,radius=0.2cm] ;
			\end{tikzpicture}
			\captionsetup{type=figure}
			\captionof{figure}{Octagonal cell model for a truncated octahedron.}
		\end{figure}
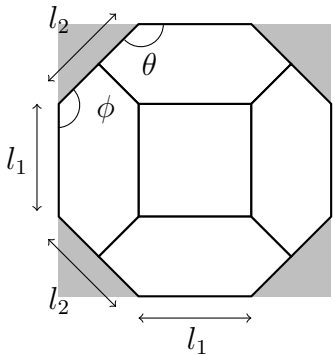

  As defined in Section \ref{HM}, it is possible to define a hexagon using two edge lengths and one angle. The octagonal cell model from Fig. 5b, however, is characterized by $l_1,\frac{l_2}{2}$, and $\theta$ (or $\phi=\frac{3\pi}{2}-\theta$) which coincides with the octagonal model from Section \ref{OM}. The octagonal cell model from Fig. 5b  is managed by $l_1,\frac{l_2}{2}$, and $\theta$ (or $\phi=\frac{3\pi}{2}-\theta$), which coincides with the octagonal model from Section \ref{OM}. The $z$ directional coordinate is chosen to satisfy the regular octahedron at $l_1=1, l_2=\sqrt{2},\theta=\frac{3\pi}{4}$, and $z=\sqrt{2}$. 
  Within mean field, we are now ready to specify the dimensionless single cell energy as  \begin{equation}
  e_S=\frac{1}{2}(v-1)^2+\frac{1}{2}k_\rho(s-s_0)^2,
  \end{equation}
		where 
  \begingroup
  \allowdisplaybreaks
		\begin{align*}
			&v=2 z (l_1 - l_2 \cos\theta) (l_1 + l_2 \sin\theta)\\
			&s=2\bigg[\frac{l_2 z^2 \sin\theta}{\sqrt{z^2 + l_2^2 \cos\theta^2}}-l_2 \cos\theta \sqrt{z^2 + l_2^2 \sin\theta^2}+l_2 \cos\theta^2 \big(\\
			&\frac{2 l_1 l_2}{\sqrt{z^2 + l_2^2 \cos\theta^2}}+ \frac{l_2^2 \sin\theta}{\sqrt{z^2 + l_2^2 \cos\theta^2}} + \frac{z}{\sqrt{\cos\theta^2 + \sin\theta^2}}\big) \\
			&+l_2 \sin\theta^2\big (\frac{z}{\sqrt{\cos\theta^2 + \sin\theta^2}} + \frac{2 l_1 l_2}{\sqrt{z^2 + l_2^2 \sin\theta^2}}\big) \\
			&+l_1 \bigg(l_1 +2 z^2 \big(\frac{1}{\sqrt{z^2 + l_2^2 \cos\theta^2}} 
			+ \frac{1}{\sqrt{z^2 + l_2^2 \sin\theta^2}}\big)\bigg)\bigg].	
		\end{align*}
  \endgroup
	Both $v$ and $s$ above have been computed using surface triangulation from Mathematica~\cite{Mathematica}. As before, one can also find $l_1,l_2$ values for unit volume with $l_1=\frac{1}{2\cdot2^\frac{1}{6}},l_2=\frac{\sqrt{2}}{2 \cdot2^\frac{1}{6}}$. The dimensionless shape index $s$ is $s\simeq 5.31474$. Just in the case of the two-dimensional hexagon, there is a compatible-incompatible transition at the target shape index of the regular truncated octahedron.  See Appendix A for the analysis.

	\section{Methods: Deformation Protocol and Determining Elastic Moduli}
		To determine the elastic moduli in mean field, we have implemented a similar deformation protocol in Ref.~\cite{Staddon2023}, which is a basic linear deformation for an affine transformation. In two dimensions, this can be expressed by $F\mathbf{x_i}+g$ with $g=0$ and $i$ is the index of a vertex of a polygon. To express it simply, $F\mathbf{x}$ is used to refer to all coordinate vectors of the vertices for $n-$gons transformed by the matrix $F=\begin{bmatrix} C_x & 0 \\ 0& C_y\end{bmatrix}$  $\mathbf{x_i}=\begin{bmatrix} x_i\\ y_i\end{bmatrix}$. 
		\begin{figure}[h]
			\centering{
				\resizebox{75mm}{!}{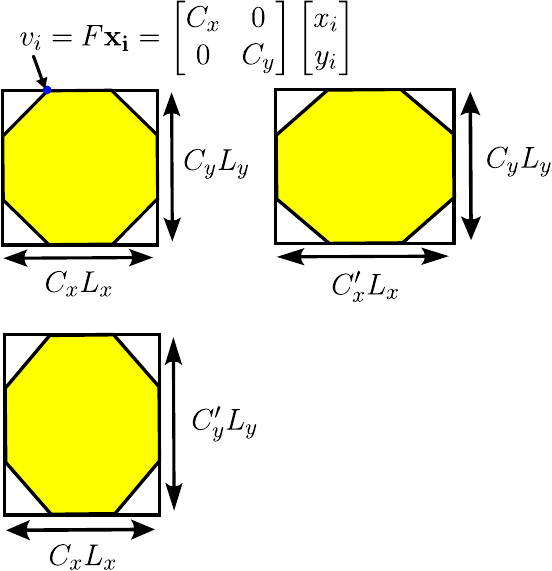}
				\caption{Affine deformation of the box.}
				\label{fig:affine-deformation}
			}
		\end{figure}
	To measure the elastic moduli, for a given set of parameters, each cell initially has a unit area or volume, depending on the dimension, with $C_x=C_y=1$. Then, a small change of $\delta=0.0001$ is made to either one or both sides, depending on the elastic modulus under exploration. Please see Table I for the different types of deformations. The size of box $L_x,L_y$ is derived from the coordinates' maximum and minimum. 
 
 With this deformation, the vertices move in response. Given where the vertices are now positioned, we compute the perimeter/area and area/volume to determine the change in the dimensionless single cell energy $e_S$ without any energy minimization. As the configuration may not necessarily be an optimal one with force-balance no longer achieved, this configuration, and its associated energy, is dubbed the constrained state, following Ref.~\cite{Staddon2023}. We also study the energy minimized state given the hidden $\theta$ degree of freedom, for example, which is the relaxed state. We then apply additional deformation by increasing $\delta$. The single cell energy $e_S$ is then re-calculated for both the constrained state and the relaxed state for each strain step.  
 
 We then compute the curvature of the energy function $e_S$ as a function of $\delta$, which is proportional to the elastic moduli.  Since the dimensionless shear modulus $G$ is defined as $G=\frac{1}{a(0)}\frac{\partial^2 e_s}{\partial \delta^2}$ in two dimensions, with $a(0)=1$, the curvature equates to the dimensionless shear modulus. Notating $K(e)$ as a curvature of the energy function, here is how the two states are characterized:
		\begin{align*}
			\text{Constrained}\colon \lim_{C_x\to 1}&K(e_S(\text{fixed vertex coordinates}))\\
			\text{Relaxed}\colon \lim_{C_x\to1}&K(\min{[e_S (\text{vertex coordinates }}\\
            &\text{with a given parameterization})]).
		\end{align*}

		In the constrained case, we simply evaluate the curvature as a function of strain to compute the dimensionless elastic moduli for each target $p_0/s_0$, the parameter that allows the system to toggle between compatible and incompatible states, should the boundary exist. We also compute the elastic moduli for the relaxed case using energy minimization. As Table \ref{table:2} illustrates, in the relaxed case, some non-affine deformations may result given the hidden degree of freedom. We chose unbounded relaxation of energy, in contrast to Ref.~\cite{Staddon2023} who measured relaxation with boundary conditions. The number of variables will drop if we fix the boundary for a single cell, and we can only use non-affine relaxation. When the box size is determined, our only option is to discover a shape that meets the target parameters($a,p_0/v,s_0$) within the box. As for the additional details on the energy minimization procedure with Mathematica~\cite{Mathematica}, we evaluated the solutions based on these four criteria:  
		\begin{itemize}
			\item We assume small shifts $\delta$ will result in a continuous deformation, or relaxation, of the structure.
			\item Given the possibility of a transition between compatible and incompatible states, we assume continuous energy values for the target $p_0$ such that the configurations $l_1,l_2,\theta,\dots$ of $p_0$ and $p_0+\epsilon$ ($\epsilon\simeq 0.1$) are similar. 
            \item Minimization is allowed within the convex region of the shape ($\frac{\pi}{2}\leq\theta\leq\pi$) as long as it stays within the physically reasonable range ($l_1,l_2\geq0$).
			\item In order to find solutions, we lowered the accuracy of minimization function when the minimum energy value was not zero.
		\end{itemize}

Since the moduli depend on curvatures of the energy, we compare two methods for computing the curvature using a discrete method and a continuous method. Following the parameters discussed in Sec. \ref{HM}, we simulate the mechanical responses of a hexagonal cell by measuring the curvature of energy function according to the deformation protocol shown in Table \ref{table:1}. We examine two distinct methods for the constrained case to verify the shape before examining the relaxed case. 

\begin{figure}[h]
			\captionsetup{singlelinecheck = false, justification=raggedright}
			\centering
			\includegraphics[width=0.5\textwidth]{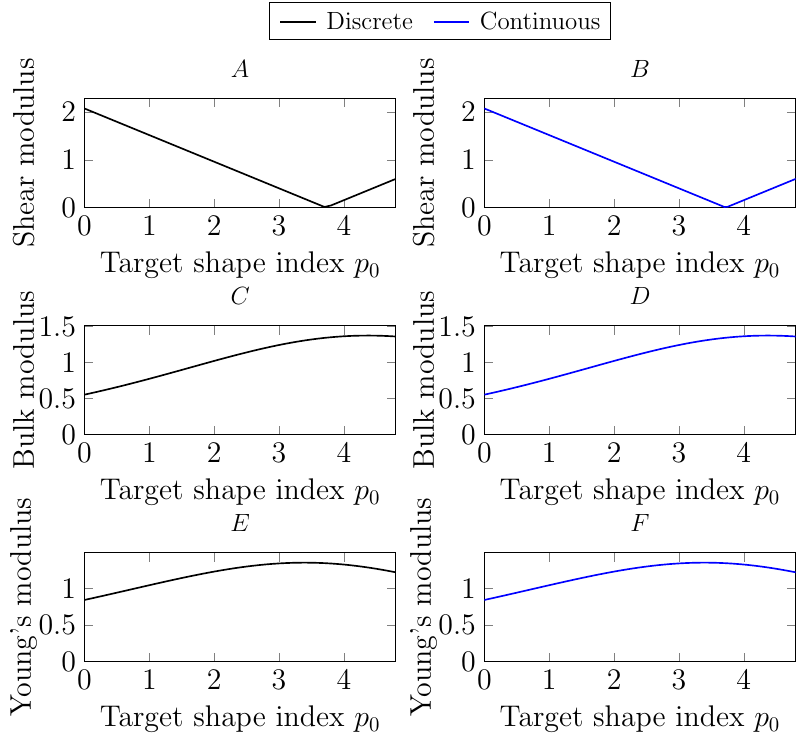}
				\caption{Curvature for the constrained case using two different methods: Curves A,C,E are generated using discrete curvature, while curves B,D,F are formed using continuous curvature.}
			\label{Fig:discrete-continuous}
		\end{figure}

We start by computing the discrete curvature of the energy function. To calculate Young’s modulus, we measure $e_s$ by varying $C_x$ over $(1-0.0005,1+0.0005)$. We then calculate the discrete curvature $K$ for every data point. We acquire the value of $K$ for $C_x=1$ through linear fitting. The second case is fully analytical. Referring to Section \ref{HM}, where we have the function $e_S$, defined by parameters $l_1,l_2,\theta$, the continuous arc curvature is calculated and displayed in Fig. 8.
		\begin{align*}
			\text{Young's modulus}\colon &K(e_s(l_{1},l_{2},C_x,1,\theta,k_r,p_0))\\
			\text{Shear modulus}\colon &K(e_s(l_{1},l_{2},C_x,1/C_x,\theta,k_r,p_0))\\
			\text{Bulk modulus}\colon &K(e_s(l_{1},l_{2},\sqrt{C_x},\sqrt{C_x},\theta,k_r,p_0)).
		\end{align*} 
We varied $C_x$ and $p_0$ and held the other parameters fixed ($l_{1}=l_{2}=\frac{\sqrt{2}}{3^{3/4}},\theta=\frac{2\pi}{3},k_r=0.1$). Figure \ref{Fig:discrete-continuous} demonstrates that there is effectively no difference between the two approaches, so we move forward using the discrete method. 

  Note that when performing the computations, we typically acquire approximately 10-20 points close to $C_x=1$ (for instance $C_x=1\pm 0.0005$). In some instances, there may be distinct solutions for $C_x=1+\delta$ and $C_x=1$. For some cases, it is related to extreme values (such as $\theta=\pi,\pi/2$). The higher number of samples is selected when there are two groups of solutions from the minimizer.
  
  We have picked $k_r=0.1$ for the two-dimensional model and $K_S=1,K_V=10$ for the three-dimensional one. We then measure discrete curvature and obtain a value at $C_x=1$ via linear fitting. A simple algorithm has been put in place for the discrete curvature from a triangle. The formula for $K(\Delta)$ is $\frac{4\text{ triangle area}}{\text{product of $3$ edges}}$, where$\Delta$ is a triangle with neighboring points. 
  
  In addition to measuring the dimensionless shear modulus, bulk modulus, and Young's modulus for a given deformation protocol, Poisson's ratio can be calculated from $\frac{1}{2}\frac{(C_y'-C_y) L_y/(C_y L_y)}{ (C_x'-C_x)L_x/(C_x L_x) }=\frac{1}{2}\frac{\Delta C_y/C_y}{\Delta C_x/C_x}$. In order to measure the deformation of one side, we divided $\Delta C_y$ by 2, since $C_y$ applied to both sides of the box. 
  
  Table \ref{table:1} and \ref{table:2} summarize the deformation protocol and control parameters.
		\begin{table}[ht]\caption{Deformation Protocol}
			\centering 
			\begin{tabular}{c c c c c }
				\hline\hline                        
				& Young's & Shear & Bulk & Poisson's ratio \\ [0.5ex]
				\hline 
				$C_x$ & $1\pm\delta$ & $1\pm\delta$ & $\sqrt{1\pm\delta}$ & $1\pm\delta$ \\
				$C_y$ & 1 & $\frac{1}{1\pm\delta}$ & $\sqrt{1\pm\delta}$ & -\\
				[1ex]   
				$C_z$ (for 3d) & - & -& $\sqrt{1\pm\delta}$ & -\\
				[1ex]
				\hline
			\end{tabular}\label{table:1}
		\end{table}
		\begin{table}[ht]\caption{Control parameters}
			\centering % used for centering table
			\begin{tabular}{c c c}
				\hline\hline                        
				Parameters & Purpose & Usage\\
				[0.5ex]\hline
				$C_y$ 		 & Relaxed (affine) & Young's\\
				$C_y,\theta$ & Relaxed (non-affine) & Young's\\
				$l_2,\theta$ & Relaxed (non-affine) & Shear/Bulk\\
				$\theta$ &  Relaxed (angle) & Transition \\
				[1ex]
				\hline
			\end{tabular}\label{table:2}
		\end{table}

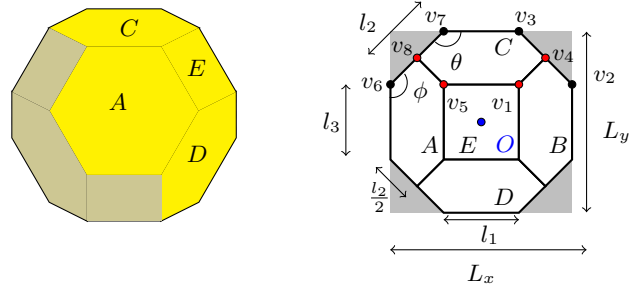
\begin{figure}[h!]
			\centering
			\vspace{10pt}
			\begin{minipage}[t]{0.2\textwidth}
				\vspace{0pt}
				\centering
				\begin{tikzpicture}[tdplot_main_coords,scale=0.7]
					\coordinate (O) at (0,0,0);
					\coordinate (A) at (1, 2, 3, 4); 
					\coordinate (B) at (2, 1, 3, 4); 
					\coordinate (C) at (3, 1, 2, 4); 
					\coordinate (D) at (4, 1, 2, 3);
					\coordinate (E) at (4, 2, 1, 3);
					\coordinate (F) at (3, 2, 1, 4);
					\coordinate (G) at (2, 3, 1, 4);
					\coordinate (H) at (1, 3, 2, 4);
					\coordinate (I) at (1, 4, 2, 3);
					\coordinate (J) at (2, 4, 1, 3);
					\coordinate (K) at (3, 4, 1, 2);
					\coordinate (L) at (4, 3, 1, 2);
					\coordinate (M) at (4, 3, 2, 1);
					\coordinate (N) at (3, 4, 2, 1);
					\coordinate (O) at (2, 4, 3, 1);
					\coordinate (P) at (1, 4, 3, 2);
					\coordinate (Q) at (1, 3, 4, 2);
					\coordinate (R) at (2, 3, 4, 1);
					\coordinate (S) at (3, 2, 4, 1);
					\coordinate (T) at (4, 2, 3, 1);
					\coordinate (U) at (4, 1, 3, 2);
					\coordinate (V) at (3, 1, 4, 2);
					\coordinate (W) at (2, 1, 4, 3);
					\coordinate (X) at (1, 2, 4, 3);
					
					\node[above=2pt, right=2pt] at (A) {};
					\node[above=2pt, left=2pt] at (B) {};
					\node[above=2pt, right=2pt] at (C) {};
					\node[above=2pt, right=2pt] at (D) {};
					\node[above=3pt, left=2pt] at (E) {};
					\node[above=5pt, left=2pt] at (F) {};
					\node[above=2pt, right=2pt] at (G) {};
					\node[above=2pt, right=2pt] at (H) {};
					\node[above=2pt, right=2pt] at (I) {};
					\node[above=2pt, left=2pt] at (J) {};
					\node[above=3pt, left=2pt] at (K) {};
					\node[above=2pt, left=2pt] at (L) {};
					\node[above=2pt, left=2pt] at (M) {};
					\node[above=2pt, left=2pt] at (N) {};
					\node[below=2pt, right=2pt] at (O) {};
					\node[above=2pt, right=2pt] at (P) {};
					\node[above=6pt, left=0pt] at (Q) {};
					\node[below=2pt, left=2pt] at (R) {};
					\node[above=2pt, left=2pt] at (S) {};
					\node[above=2pt, left=2pt] at (T) {};
					\node[above=2pt, left=2pt] at (U) {};
					\node[above=2pt, left=2pt] at (V) {};
					\node[above=6pt, left=0pt] at (W) {};
					\node[above=2pt, left=2pt] at (X) {};
					
					\draw[thick] (M) -- (N) -- (O) -- (R) -- (S) -- (T) -- cycle;
					\draw[thick] (S) -- (R) -- (Q) -- (X) -- (W) -- (V) -- cycle;
					\draw[thick] (O) -- (N) -- (K) -- (J) -- (I) -- (P) -- cycle;
					\draw[thick] (M) -- (L) -- (E) -- (D) -- (U) -- (T) -- cycle;
					\draw[dashed, opacity=0.3] (A) -- (B) -- (C) -- (F) -- (G) -- (H) -- cycle;
					\draw[dashed, opacity=0.3] (E) -- (F) -- (G) -- (J) -- (K) -- (L) -- cycle;
					\draw[dashed, opacity=0.3] (W) -- (B) -- (C) -- (D) -- (U) -- (V) -- cycle;
					\draw[dashed, opacity=0.3] (Q) -- (X) -- (A) -- (H) -- (I) -- (P) -- cycle;
					\draw[thick] (L) -- (K);
					\draw[thick] (Q) -- (P);
					\draw[thick] (V) -- (U);
					
					\fill[yellow, opacity=1] (M) -- (N) -- (O) -- (R) -- (S) -- (T) -- cycle;
					\fill[yellow, opacity=1] (S) -- (R) -- (Q) -- (X) -- (W) -- (V) -- cycle;
					\fill[yellow, opacity=1] (O) -- (N) -- (K) -- (J) -- (I) -- (P) -- cycle;
					\fill[black!40!yellow!60, opacity=1] (M) -- (L) -- (E) -- (D) -- (U) -- (T) -- cycle;
					\fill[black!40!yellow!60, opacity=1] (U) -- (V) -- (S) -- (T) -- cycle;
					\fill[yellow, opacity=1] (R) -- (Q) -- (P) -- (O) -- cycle;
					\fill[black!40!yellow!60, opacity=1] (M) -- (N) -- (K) -- (L) -- cycle;
					\node[label={0:{$A$}},inner sep=2pt] at (2.8,2.0,2.7,0) {};
					\node[label={0:{$C$}},inner sep=2pt] at (2.5,2.0,4.2,0) {};
					\node[label={0:{$D$}},inner sep=2pt] at (0.7,2.0,0.7,0) {};
					\node[label={0:{$E$}},inner sep=2pt] at (1.2,2.5,3.0,0) {};
				\end{tikzpicture}
				\vspace{4mm}
				\subcaption{Truncated octahedron.}
			\end{minipage}
			\hfill
			\begin{minipage}[t]{0.25\textwidth}
				\vspace{0pt}
				\centering
				\begin{tikzpicture}[scale=1.0]
					\fill[gray,opacity=0.5](-1.20711, -1.20711)--(-1.20711, 1.20711)--(1.20711, 1.20711)--(1.20711, -1.20711)--cycle;
					\draw[-,thick,fill=white](0.5, 0.5)--(-0.5, 0.5)--(-0.5, -0.5)
					--(0.5, -0.5)--cycle;
					\draw[-,thick,fill=white](0.853553, -0.853553)--(0.5, -1.20711)--(-0.5, -1.20711)--(-0.853553, -0.853553)--(-0.5, -0.5)--(0.5, -0.5)--cycle;
					\draw[-,thick,fill=white](0.5, 0.5)--(-0.5, 0.5)--(-0.853553, 0.853553)--(-0.5, 1.20711)--(0.5, 1.20711)--(0.853553, 0.853553)--cycle;
					\draw[-,thick,fill=white](0.853553, -0.853553)--(1.20711, -0.5)--(1.20711, 0.5)--(0.853553, 0.853553)--(0.5, 0.5)--(0.5, -0.5)--cycle;
					\draw[-,thick,fill=white](-0.853553, -0.853553)--(-1.20711, -0.5)--(-1.20711, 0.5)--(-0.853553, 0.853553)--(-0.5, 0.5)--(-0.5, -0.5)--cycle;
					\draw[<->](1.4, -1.2)--(1.4, 1.2);
					\draw[<->](-0.5, -1.3)--(0.5, -1.3);
					\draw[<->](-1.21, -1.7)--(1.4, -1.7);
					\draw[<->](-1.8, -0.45)--(-1.8, 0.5);
					\draw[<->](-1.4, -0.6)--(-1.0, -1.0);
					\draw[<->](-1.5, 0.9)--(-0.8, 1.6);
					\node[label={0:{$\frac{l_2}{2}$}},inner sep=2pt] at (-1.7,-1) {};
					\node[label={0:{$l_3$}},inner sep=2pt] at (-2.3,0) {};
					\node[label={0:{$l_1$}},inner sep=2pt] at (-0.2,-1.5) {};
					\node[label={0:{$\theta$}},inner sep=2pt] at (-0.6,0.8) {};
					\node[label={0:{$\phi$}},inner sep=2pt] at (-1.1,0.4) {};
					\node[label={90:{$l_2$}},inner sep=2pt] at (-1.5,1) {};
					\node[label={-90:{$L_x$}},inner sep=2pt] at (0,-1.7) {};
					\node[label={-90:{$L_y$}},inner sep=2pt] at (1.8,0.2) {};
				\node[label={0:{$A$}},inner sep=2pt] at (-1.0,-0.3) {};
				\node[label={0:{$B$}},inner sep=2pt] at (0.7,-0.3) {};
				\node[label={90:{$C$}},inner sep=2pt] at (0.3,0.7) {};
				\node[label={-90:{$D$}},inner sep=2pt] at (0.3,-0.7) {};
				\node[label={0:{$E$}},inner sep=2pt] at (-0.5,-0.3) {};
				\node[label={-40:{\color{blue}$O$}},inner sep=2pt] at (0,0) {};
					\draw[-](-0.27, 1.20711) arc[start angle=0, end angle=-135,radius=0.2cm] ;
					\draw[-](-1.02, 0.64) arc[start angle=35, end angle=-100,radius=0.2cm] ;			
				\node[label={-120:{$v_1$}},inner sep=2pt] at (0.6, 0.5) {};
				\node[label={0:{$v_2$}},inner sep=2pt] at (1.3, 0.6) {};
				\node[label={90:{$v_3$}},inner sep=2pt] at (0.6, 1.1) {};
				\node[label={20:{$v_4$}},inner sep=2pt] at (0.75, 0.65) {};
				\node[label={-70:{$v_5$}},inner sep=2pt] at (-0.6, 0.5) {};
				\node[label={-180:{$v_6$}},inner sep=2pt] at (-1.1, 0.5) {};
				\node[label={90:{$v_7$}},inner sep=2pt] at (-0.6, 1.1) {};
				\node[label={110:{$v_8$}},inner sep=2pt] at (-0.75, 0.7) {};
				\draw[fill=black](1.20711, 0.5) circle[radius=0.05];
				\draw[fill=black](0.5, 1.20711) circle[radius=0.05];
				\draw[fill=red](0.853553, 0.853553) circle[radius=0.05];
				\draw[fill=red](0.5,0.5) circle[radius=0.05];
				\draw[fill=black](-1.20711, 0.5) circle[radius=0.05];
				\draw[fill=black](-0.5, 1.20711) circle[radius=0.05];
				\draw[fill=blue](0, 0) circle[radius=0.05];
				\draw[fill=red](-0.853553, 0.853553) circle[radius=0.05];
				\draw[fill=red](-0.5,0.5) circle[radius=0.05];
				\end{tikzpicture}
				\subcaption{Schematic of 3d model.}
			\end{minipage}
			\captionsetup{singlelinecheck = false, justification=raggedright}
			\caption{A 3d representation of a vertex model with $l_1, l_2, l_3$ (a) and (planar) schematic (b). The box unit indicated by ($L_x \times L_y$) is represented by the shaded area (b).}
			\label{fig:3d-model-schematic}
		\end{figure}

 For our three-dimensional protocol, taking the truncated octahedron shape as the reference, we set up minimal parameters to control regular polyhedron under an affine transformation with a simple rescaling of the now three-dimensional box. We only analyze the alteration of coordinates $\mathbf{x}$ within a box $(L_x,L_y,L_z)$ defined by $F\mathbf{x}$ for $F=\begin{bmatrix} C_x & 0 & 0\\ 0& C_y& 0\\ 0&0& C_z\end{bmatrix}$  $\mathbf{x}=\begin{bmatrix} x\\ y\\z\end{bmatrix}$. The regular truncated octahedron is linear in the $z$ direction, meaning there are just five values ($-z,-\frac{z}{2},0,\frac{z}{2},z$). The other parameters will be established inside the octagon as indicated in Fig. \ref{fig:3d-model-schematic} (b). Taking a look at Fig. \ref{fig:3d-model-schematic}, square $E$ must be a rectangle with a right angle to prevent a non-convex polyhedron when scaling in $x$ and $y$ with $C_z=1$. Furthermore, this polyhedron will be symmetric along the $xy$-axis, which means two hexagons $A$ and $B$ will be equivalent (symmetric along the $y$-axis) with $C$ and $D$ being equivalent (symmetric along the $x$-axis). As a result, Fig. \ref{fig:3d-model-schematic} (b) can be uniquely defined by $l_1,l_2,l_3,\theta$ (recall $\phi=\frac{3\pi}{2}-\theta$) up to a translation. 

To better understand this simple extension to three dimensions and demonstrate that additional parameterization is not required, consider the following. The truncated octahedron is made up of $24$ vertices. Taking into account affine transformation that are rescalings, all coordinate vectors can be determined by the manipulation of four vertices ($v_1\sim v_4$). As illustrated in Fig. \ref{fig:3d-model-schematic} (b), $8$ vertices are combined to create an octagon (black nodes). The upper half contains eight points placed diagonally and a rectangle. Thus, $8+8\times2$ results in $24$. Hexagons $A$ and $B$ are the same, and we can denote the three-dimensional coordinate of $C$ as $v_1=(r_1,r_2,r_3)$, $v_4=(r_4,r_5,r_6)$, $v_5=(-r_1,r_2,r_3)$, and $v_8=(-r_4,r_5,r_6)$. Using the imposec symmetries for $x,y,z$, all other points can be described in the same way. Additionally, since the octagon's $8$ points are on the $xy$-plane ($z=0$), we can set $v_1=(r_1,r_2,r_3)$, $v_2=(r_7,r_8,0)$, $v_3=(r_9,r_{10},0)$, and $v_4=(r_4,r_5,r_6)$. It is evident that $r_1=r_9$ and $r_2=r_{8}$ due to the fact that they are determined by $l_1,l_3$. In addition, each hexagon is symmetrical in the $x,y$ plane, thus, we get $v_4=(r_4,r_5,r_6)=(\frac{r_1+r_7}{2},\frac{r_1+r_{10}}{2},\frac{r_3}{2})$.  It is possible to visualize that each hexagon on the upper half has one edge (with $v_1$) at $z=r_3$ and its opposite side (with $v_2$) at $z=0$. Rewrite the four vertex points as $v_1=(r_1,r_2,r_3)$, $v_2=(r_7,r_2,0)$, $v_3=(r_1,r_{10},0)$, and $v_4=(\frac{r_1+r_7}{2},\frac{r_1+r_{10}}{2},\frac{r_3}{2})$, and to control all vertices, only $r_1,r_2,r_3,r_7,r_{10}$ must be known, which can be calculated from the parameters $l_1,l_2,l_3,\theta,z$ relatively easily.

		\tdplotsetmaincoords{60}{135}

  Because working out analytical solutions in three dimensions is cumbersome, we will not compare the discrete and analytical data for the constrained case. The shape can easily be predicted from the two-dimensional one. We make use of simple linear deformation, as described in the preceding section. The only alteration is adjusting three-dimensional expansion, or  $C_x=C_y=C_z=\sqrt{1\pm\delta}$, to measure the bulk modulus.

  \section{Results}
		 
		\subsection{The two-dimensional model}
		The moduli and Poisson's ratio are measured for both the constrained and relaxed case as $p_0$ is varied from $0$ to $4.8$. 
		\subsubsection{Hexagonal Model}

\begin{figure}[h!]
	\centering
	\includegraphics[width=0.5\textwidth]{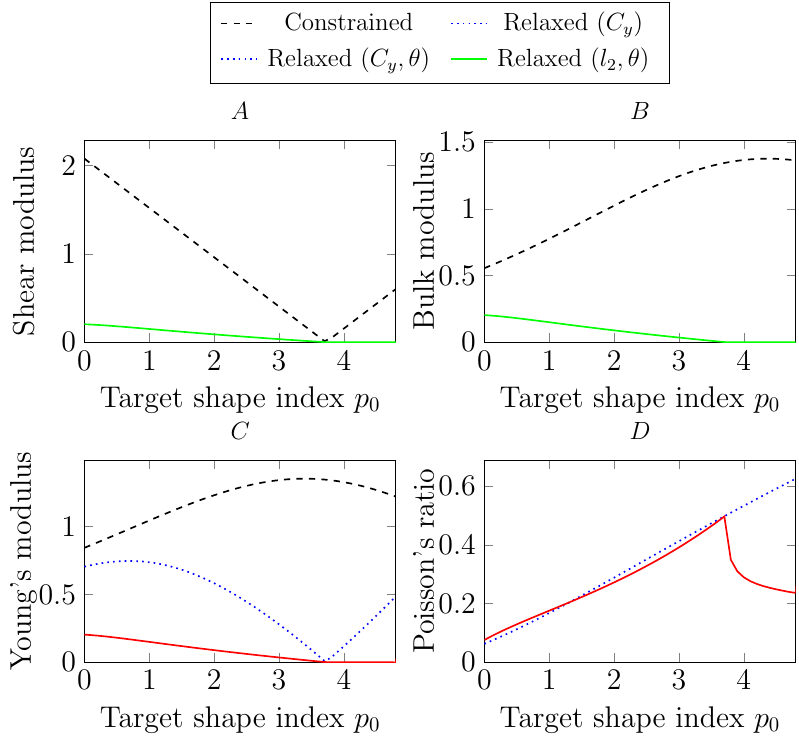}
	\caption{Mechanical response for hexagon.}
	\label{fig:hex-mechanical}
	\end{figure}

  Figure \ref{fig:hex-mechanical} demonstrates the mechanical behavior for both the constrained and relaxed cases. The following plots are generated using discrete curvatures, as specified previously. 
  
  Let us understand the shape of these curves by focusing on the constrained case first. To measure mechanical response, we apply an affine deformation. If we express energy function in the presence of the deformation as $e_S(\delta)$, we write $e_S(\delta)=\frac{1}{2}(a(\delta)-1)^2+\frac{1}{2}k_r(p(\delta)-p_0)^2$ where $\delta$ implies $C_x\to 1+\delta,C_y\to 1/(1+\delta)$ for the shear modulus, $C_x\to \sqrt{1+\delta},C_y\to \sqrt{1+\delta}$ for the bulk modulus, and $C_x\to 1+\delta$ for Young's modulus. Since $C_xC_y=1$ for the shear modulus, $a(\delta)=1$ if $a=1$ (the initial condition). On the other hand, $C_xC_y=1+\delta$ for the bulk modulus and the Young's modulus, so $a(\delta)\neq1$. We can calculate the curvature functions for each case as demonstrated below. Specifically, 
\begin{align}
	K(e_S(\delta))_{\text{B,Y}}&=|(a(\delta)-1)a''(\delta)+(p(\delta)-p_0)p''(\delta)+{a'(\delta)}^2\notag\\
 +{p'(\delta)}^2|
 /(1+&((a(\delta)-1)a'(\delta)+(p(\delta)-p_0)p'(\delta))^2)^{3/2},\\
	K(e_S(\delta))_{\text{G}}&=\frac{|(p(\delta)-p_0)p''(\delta)+{p'(\delta)}^2|}{(1+((p(\delta)-p_0)p'(\delta))^2)^{3/2}}, 
\end{align}
where the primes denote derivatives with respect to $\delta. $ For the hexagon, because $a(\delta)$ is same for both bulk and Young's moduli, we arrive at $(a(\delta)-1)a''(\delta)=0$ and $a'(\delta)=2 l_2 (l_1 - l_2 \cos\theta) \sin\theta$. Moreover, 
\begin{align}
\lim_{\delta\to 0}(p(\delta)-p_0)p'(\delta)_{\text{G}}=& 2k_r (2 l_1 + 4 l_2 - p_0) (l_1 +2l_2 \cos2 \theta)),\\
\lim_{\delta\to 0}(p(\delta)-p_0)p'(\delta)_{\text{B}}=& k_r (2 l_1 + 4 l_2 - p_0) (l_1 + 2l_2 ),\\
\lim_{\delta\to 0}(p(\delta)-p_0)p'(\delta)_{\text{Y}}=&2 k_r (2 l_1 + 4 l_2 - p_0)\notag\\ &\cdot(l_1 + l_2 + l_2 \cos2 \theta).
\end{align}
Finally, we obtain 
\begin{align}
&\lim_{\delta\to 0}K(e_S(\delta))_{\text{G}}=|(2 k_r (2 l_1^2 + 6 l_1 l_2 + 16 l_2^2 - 3 l_2 p_0 \notag\\&+ 
2 l_2 (2 l_1 - 4 l_2 + p_0) \cos2\theta + l_2 (-2 l_1 + p_0) \cos4 \theta))|\notag\\&/(1 + 
4 k_r^2 (-2 l_1 - 4 l_2 + p_0)^2 (l_1 + 2 l_2 \cos2\theta)^2)^{\frac{3}{2}},
\end{align}
\begin{align}
&\lim_{\delta\to 0}K(E(\delta))_{\text{B}}=|(4 (4 l_1^2 l_2^2 + l_2^4 + k_r (l_1 + 2 l_2) p0 \notag\\&- 
4 l_1 l_2^3 \cos\theta - 4 l_1^2 l_2^2 \cos2\theta + 4 l_1 l_2^3 \cos3\theta - l_2^4 \cos4 \theta))|\notag\\&
/(4 + (2 k_r (l_1 + 2 l_2) (2 l_1 + 4 l_2 - p_0)\notag\\
& + 
4 l_2 (-l_1 + l_2 \cos\theta) \sin\theta (1 - 
2 l_1 l_2 \sin\theta + l_2^2 \sin2 \theta))^2)^{\frac{3}{2}},
\end{align}
\begin{align}
&\lim_{\delta\to 0}K(E(\delta))_{\text{Y}}=|(4 (4 k_r l_1 l_2 \cos\theta^4 - 
2 l_1 l_2^3 \cos\theta \sin\theta^2 \notag\\&+ 
l_1^2 (k_r + l_2^2 \sin\theta^2) \notag\\
& + 
l_2 \cos\theta^2 (4 k_r l_2 + (6 k_r l_1 + l_2^3 - 
k_r p_0) \sin\theta^2)))| \notag\\
&/(1 + 
1/4 (4 k_r (2 l_1 + 4 l_2 - p_0) (l_1 + l_2 + l_2 \cos2\theta)  \notag\\&+ 
4 l_2 (-l_1 + l_2 \cos\theta) \sin\theta (1 - 
2 l_1 l_2 \sin\theta + l_2^2 \sin2\theta))^2)^{\frac{3}{2}}.
\end{align}
Notice that effects of $C_x\to 1+\delta$ and $C_x\to \sqrt{1+\delta},C_y\to \sqrt{1+\delta}$ act differently for the cell perimeter. These differences will lead to different curves for the bulk and Young's moduli. For the shear modulus curvature function, if $l_1=l_2=\frac{\sqrt{2}}{3^{3/4}}$, $\theta=\frac{2\pi}{3}$, and $k_r=0.1$, then denominator becomes 1. Thus, we have 
\begin{align}
\lim_{\delta\to 0}K(e_S(\delta))_{\text{G}}\simeq0.2 |2.79181(3.72242 - p_0) |.
\end{align} 
This function becomes linear and has a minimum at $p_0=3.72242$, as indicated in Fig. 9A. However, $\lim_{\delta\to 0}K(E(\delta))_{\text{B}}$ and $\lim_{\delta\to 0}K(E(\delta))_{\text{Y}}$ (Figs. (B and C)) both have curves related to $\frac{|a+bx|}{(c+d(e-x)^2)^{3/2}}$, i.e., they are not linear. More precisely, 
\begin{align}
\lim_{\delta\to 0}K(E(\delta))_{\text{B}}&\simeq	\frac{4|(2+ 0.186121 p_0)|}{(4 + 0.138564 (3.72242 -  p_0)^2)^{3/2}},\\
\lim_{\delta\to 0}K(E(\delta))_{\text{Y}}&\simeq 4|(0.270335 + 
	0.155101 (0.248161  \notag\\
 &+ 0.75 (0.611035 - 0.1 p_0)))| \notag\\&/(1 + 
	0.034641 (3.72242 - p_0)^2)^{3/2}.
\end{align}
The constrained case acts an upper bound on the curves. Note that we did not compute Poisson's ratio as both $C_x$ and $C_y$ are constrained.  

Let us now comment on the less constrained cases. Three separate parameter sets are used for the relaxed case. Changing only the box size in the $y$ direction ($C_y$) softens the Young's modulus, for instance, as compared to the constrained case. However, using two parameters ($C_y,\theta$) allows the system to soften even more so. In addition, relaxing both $l_2$ and $\theta$ softens the shear modulus as compared to the constrained case. The same trend observed in the bulk modulus. It is very reasonable to expect this given the addition degree of freedom. Note that depending in the definition of the moduli, not all parameters can be minimized. For instance, we do not compute the shear modulus by minimizing $C_y$.  

For all three moduli, with a two parameter minimization, a compatible-incompatible transition is observed with the moduli vanishing for $p \ge p_0^*(6)$.  Recall that $e_S$ reaches a minimum at this target shape index. The transition also leads to a drop in Poisson's ratio.  Furthermore, relaxation also takes place with non-affine deformations as $l_2,\theta$ are non-affine parameters. Finally, our curves differ slightly from those in Ref.~\cite{Staddon2023} as our shape parameterization differs slightly from theirs and are only investigating a single cell (as opposed to four cells with relevant boundary conditions). By utilizing an unbounded single cell, our strategy allows for non-affine/affine variables during relaxation and less restrictive geometry constraints. With a fixed boundary, the box size ($L_x\times L_y$) is determined by $C_x,C_y$ values, allowing a cell to relax its energy using non-affine deformation (ratio of $l_1,l_2$ and a relaxation using $\theta$). Therefore, there will be a similar mechanical response in an incompatible state as in a constrained case. The relaxation of energy is more achievable for a cell in a compatible state.

		\subsubsection{Octagonal Model}
		The next step is to compute the mechanical response of the octagonal cell described in Sec. \ref{OM}.
		\begin{figure}[h!]
			\centering
			\includegraphics[width=0.5\textwidth]{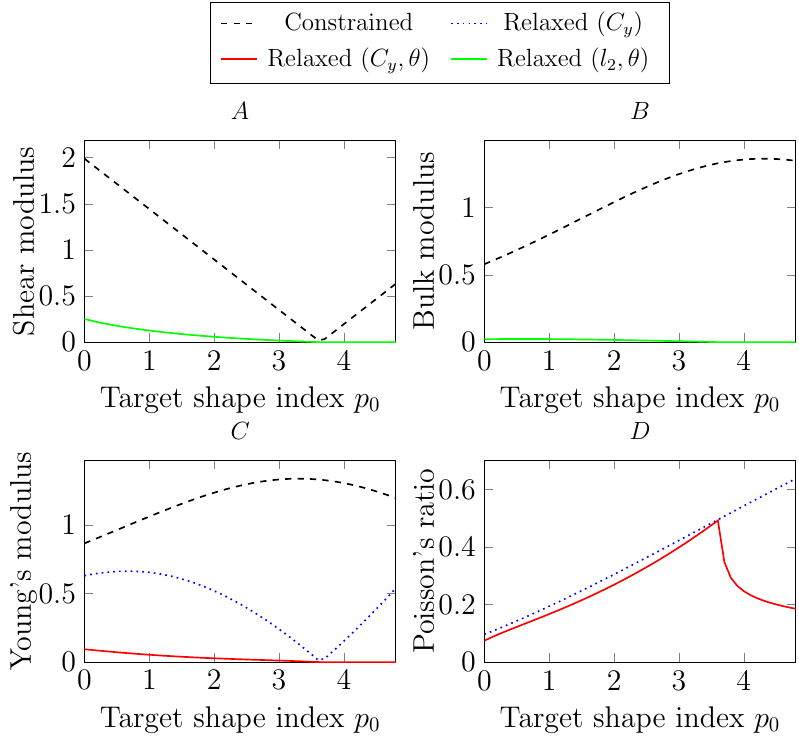}
			\caption{Mechanical response for the octagon cell.}
		\end{figure}
		We observe that the mechanical behavior of the octagonal model is, not suprisingly, rather similar to that of the hexagonal case. See Fig. 10. The modulus of ($C_y,\theta$) or ($l_2,\theta$) will be lower than the hexagonal case when relaxed, as a result of the greater impact of altering $C_y$ or $l_2$ on the perimeter. To directly compare to the two shapes, Fig. \ref{fig:two-models} plots the Young’s modulus for two different cell geometries. The octagonal model has a transition point near $p_0^*(8)\approx 3.64072$, which is denoted by the red dotted line, while the transition point for the hexagonal model is near the shape index for a regular hexagon (with unit area). 
  
  \begin{figure}[h!]
			\captionsetup{singlelinecheck = false, justification=raggedright}
			\begin{center}
				\begin{tikzpicture}
					\begin{axis}[		
						width=0.4\textwidth,
						height=0.25\textwidth,
						xlabel = {Target shape index $p_0$},
						ylabel = {Young's modulus},
						ylabel style={anchor=base, inner ysep=5pt,rotate=0},
						xmin=0, xmax=4.8 ,
						ymin = 0, ymax = 0.3
						]
						\addplot[thick,black,dashed] table {Hex_p0table01.txt};\addlegendentry{Hexagon}
						\addplot[thick,blue] table {Oct_p0table01.txt};\addlegendentry{Octagon}
						\addplot[mark=none, red,dotted,thick] coordinates {(3.64072,0) (3.64072,0.3)};
						\addplot[mark=none, red,thick] coordinates {(sqrt(8*sqrt(3)),0) (sqrt(8*sqrt(3)),0.3)};
					\end{axis}
				\end{tikzpicture}
				\captionsetup{type=figure}
				\captionof{figure}{Young’s modulus for the two cellular models for the relaxed case with ($C_y,\theta$) as the minimized parameters. The solid red line denotes $p_0^*(6)$, while the dashed red line denotes $p_0^*(8)$.}
				\label{fig:two-models}
			\end{center}
		\end{figure}
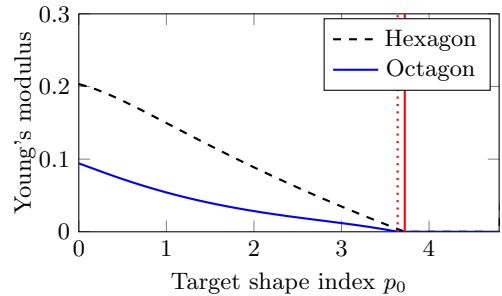
		\subsubsection{Octagonal model with a different parametrization}\label{nOM}
		To better understand how the mechanics depends on the shape parameterization, we include an extra edge length $l_3$ in addition to $l_1$ and $l_2$ to mimic edge length alteration from an affine deformation for a standard octagon. See Fig. 12. Two unique edge lengths will come from different scaling in the $x$ and $y$ directions. Moreover, we alter the parameters of the octagonal cell using the following criteria. 

  \begin{itemize}
			\item We are now employing three edge lengths ($l_1,l_2,l_3$)\\
            instead of two ($l_1,l_2$).
			\item These lengths can be converted to the box-related parameters $w,h_1,h_2$ that were inspired by Ref.~\cite{Staddon2023}. This can be done so via ($L_x,L_y\colon$ total length in x,y of octagon)
		\begin{align*}
			L_x &+l_1=2l_1 - 2 l_2\cos\theta=w,\\
			L_y &=h_1+h_2,\\
			\text{where }&h_1=2 l_2\sin\theta,h_2=l_3.
		\end{align*}
      \item Revise energy function by employing new variables from $e_S(l_1,l_2,l_3,\theta,k_r,p_0)$ to $e_S(w,h_1,h_2,\theta,k_r,p_0)$.
		\end{itemize}
		
  We select $w=L_x+l_1$ instead of $w=L_x$ to form an unbalanced parametrization; the main purpose for this is that the area function, when reparametrized using $w,h$ specified by $w=l_1 - 2 l_2\cos\theta,h=l_1+2 l_2\sin\theta$, has a singularity at $\theta=3\pi/4$ ($a=\frac{(h+w)^2\cot\theta+2hw\csc\theta^2}{2(1+\cot\theta)^2}$ and $\cot(\frac{3\pi}{4})=-1$). With this new parametrization, the cell have the capacity to shift between three distinct shapes. In other words, we have modified the shape landscape and will now determine how the new shape landscape potentially affects the moduli. 
		\begin{itemize}
			\item Octagon
			\item Rectangle ($h_1=0$ or $\theta=\pi/2,\pi$)
			\item Hexagon ($h_2=0$ or $\theta=\pi-\delta$ where $\delta$ is related to $w,h$ ratio).
		\end{itemize}
		The corresponding angle $\phi$ is equal to $\phi=3\pi/2-\theta$. Fig. \ref{fig:oct-mechanical-re} illustrates the mechanical response of a reparametrized octagon. 
		\begin{figure}[h]
			\captionsetup{singlelinecheck = false, justification=raggedright}
			\begin{center}
				\begin{tikzpicture}[scale=1.2]
					\fill[gray,opacity=0.5](-1.20711, -1.20711)--(-1.20711, 1.20711)--(1.20711, 1.20711)--(1.20711, -1.20711)--cycle;
					\draw[-,thick,fill=white](-0.5, -1.20711)--(-1.20711, -0.5)--(-1.20711, 0.5)--(-0.5, 1.20711)--(0.5, 1.20711)--(1.20711, 
					0.5)--(1.20711, -0.5)--(0.5, -1.20711)--cycle;
					\draw[<->](1.4, -1.2)--(2.21, -1.2);
					\draw[<->](-0.5, 1.4)--(0.5, 1.4);
					\draw[<->](-0.5, -1.0)--(0.5, -1.0);
					\draw[<->](-1.21, -1.5)--(2.21, -1.5);
					\draw[<->](-1.5, -0.45)--(-1.5, 0.5);
					\draw[<->](1.5, -0.45)--(1.5, 0.5);
					\draw[<->](-1.5, -0.55)--(-1.5, -1.21);
					\draw[<->](-1.5, 0.6)--(-0.7, 1.4);
					\node[label={0:{$\frac{h_1}{2}$}},inner sep=2pt] at (-2.2,-1) {};
					\node[label={0:{$h_2=l_3$}},inner sep=2pt] at (-2.7,0) {};
					\node[label={0:{$l_3$}},inner sep=2pt] at (1.5,0) {};
					\node[label={0:{$l_1$}},inner sep=2pt] at (1.5,-1.0) {};
					\node[label={0:{$l_1$}},inner sep=2pt] at (-0.2,-0.7) {};
					\node[label={0:{$l_1$}},inner sep=2pt] at (-0.2,1.6) {};
					\node[label={0:{$\theta$}},inner sep=2pt] at (-0.6,0.8) {};
					\node[label={0:{$\phi$}},inner sep=2pt] at (-1.1,0.4) {};
					\node[label={90:{$l_2$}},inner sep=2pt] at (-1.4,1) {};
					\node[label={0:{$w$}},inner sep=2pt] at (0,-1.8) {};
					\draw[-](-0.27, 1.20711) arc[start angle=0, end angle=-135,radius=0.2cm] ;
					\draw[-](-1.02, 0.64) arc[start angle=35, end angle=-100,radius=0.2cm] ;			
				\end{tikzpicture}
			\end{center}
			\caption{Schematic of the reparameterized octogonal model using $h_1,h_2,$ and $w$. Box ($L_x\times L_y$) is represented by the shaded area.}
    \label{fig:reparam-octagon}
		\end{figure}
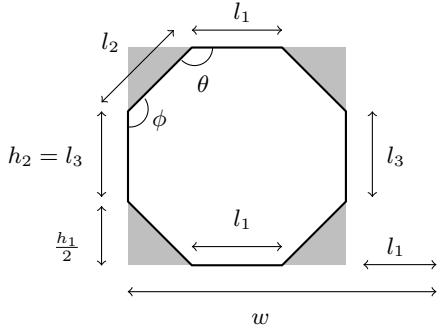
		\begin{figure}[h!]
			\centering
			\includegraphics[width=0.5\textwidth]{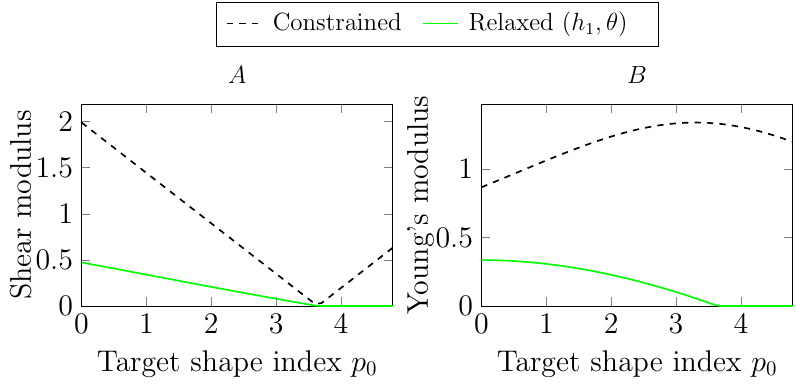}
			\caption{Shear modulus and Young's modulus for the reparametrized octagon model.}
			\label{fig:oct-mechanical-re}
		\end{figure}

  If we take $\theta$ as the sole variable for optimizing energy $e_S$, two peaks show up as presented in Fig. \ref{fig:two-peaks}. The double peak pattern occurs when $(C_y,\theta)$ is minimized in the relaxed case. What is the mechanism behind this two-peaked structure? If we look at Fig. \ref{fig:octagon-transition}, the first case $(C_y,\theta)$ displays the cell shape  transitioning to a rectangular shape, which occurs near $p_0=4.1$. The emergence of the rectangular shape constrains the angle $\theta$, thereby leads to an increase in the modulus and so the minimization is effectively reduced to a one-parameter minimization. Observe that the Young’s modulus continues to increase for Fig. \ref{fig:octagon-transition} (b) as $p_0$ is increased beyond approximately 4.1. 
		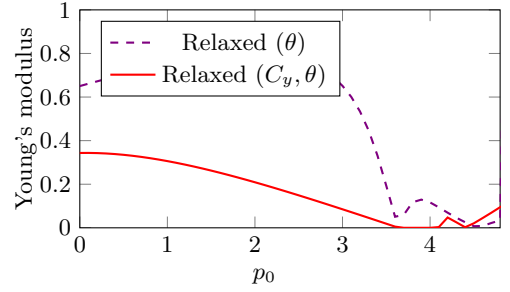
\begin{figure}[h!]
            \captionsetup{singlelinecheck = false, justification=raggedright}
			\begin{center}
				\begin{tikzpicture}
					\begin{axis}[		
						width=0.4\textwidth,
						height=0.25\textwidth,
						xlabel = {$p_0$},
						ylabel = {Young's modulus},
						xlabel style={anchor=north east, inner xsep=5pt},
						ylabel style={anchor=base, inner ysep=5pt,rotate=0},
						xmin=0, xmax=4.8 ,
						ymin = 0, ymax = 1, legend style={at={(0.05,0.95)},anchor=north west}
						]
						\addplot[thick,violet,dashed] table {Octhw_p0table02.txt};\addlegendentry{Relaxed ($\theta$)}
						\addplot[thick,red] table {Octhw_p0table01.txt};\addlegendentry{Relaxed ($C_y,\theta$)}	
					\end{axis}
				\end{tikzpicture}
				\captionsetup{type=figure}
				\captionof{figure}{Young's modulus as a function of $p_0$ for the reparameterized octagonal model and for two different relaxed cases.}
				\label{fig:two-peaks}
			\end{center}
		\end{figure}
		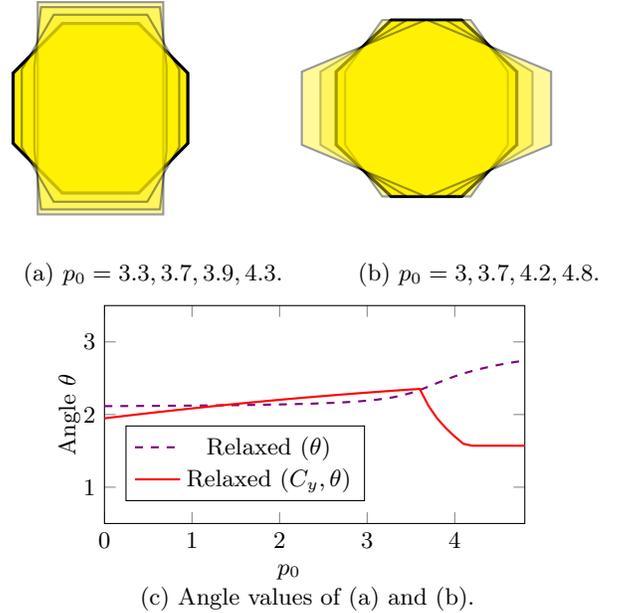
\begin{figure}[!hpb]
			\centering
			\begin{tikzpicture}
				\begin{groupplot}[group style={group name=ch1plots, group size=2 by 1, horizontal sep=0.05\textwidth,vertical     sep=0.\textwidth},xmin=-0.8,xmax=0.8,ymin=-0.8,ymax=0.8,height=5cm,width=0.28\textwidth,no markers,axis lines=none, xtick=\empty, ytick=\empty]
					\nextgroupplot[title={\(\)}]
					\draw[fill=yellow,-,very thick](axis cs:-0.236332, -0.527779)--(axis cs:-0.540555, -0.218613)--(axis cs:-0.540555, 0.218613)--(axis cs:-0.236332, 0.527779)--(axis cs:0.236332, 0.527779)--(axis cs:0.540555, 0.218613)--(axis cs:0.540555, -0.218613)--(axis cs:0.236332, -0.527779)--cycle;
					\draw[fill=yellow,opacity=0.4,-,thick](axis cs:-0.388443, -0.660718)--(axis cs:-0.388443, -0.273678)--(axis cs:-0.388443, 0.273678)--(axis cs:-0.388443, 0.660718)--(axis cs:0.388443, 0.660718)--(axis cs:0.388443, 0.273678)--(axis cs:0.388443, -0.273678)--(axis cs:0.388443, -0.660718)--cycle;
					\draw[fill=yellow,opacity=0.4,-,thick](axis cs:-0.367915, -0.629808)--(axis cs:-0.408972, -0.260875)--(axis cs:-0.408972, 0.260875)--(axis cs:-0.367915, 0.629808)--(axis cs:0.367915, 0.629808)--(axis cs:0.408972, 0.260875)--(axis cs:0.408972, -0.260875)--(axis cs:0.367915, -0.629808)--cycle;
					\draw[fill=yellow,opacity=0.4,-,thick](axis cs:-0.289644, -0.582251)--(axis cs:-0.487243, -0.241176)--(axis cs:-0.487243, 0.241176)--(axis cs:-0.289644, 0.582251)--(axis cs:0.289644, 0.582251)--(axis cs:0.487243, 0.241176)--(axis cs:0.487243, -0.241176)--(axis cs:0.289644, -0.582251)--cycle;
					\nextgroupplot[title={\(\)}]
					\draw[fill=yellow,-,very thick](axis cs:-0.218137, -0.549342)--(axis cs:-0.55875, -0.227545)--(axis cs:-0.55875, 0.227545)--(axis cs:-0.218137, 0.549342)--(axis cs:0.218137, 0.549342)--(axis cs:0.55875, 0.227545)--(axis cs:0.55875, -0.227545)--(axis cs:0.218137,-0.549342)--cycle;
					\draw[fill=yellow,opacity=0.4,-,thick](axis cs:-0.273518, -0.549342)--(axis cs:-0.503369, -0.227545)--(axis cs:-0.503369, 0.227545)--(axis cs:-0.273518, 0.549342)--(axis cs:0.273518, 0.549342)--(axis cs:0.503369, 0.227545)--(axis cs:0.503369, -0.227545)--(axis cs:0.273518, -0.549342)--cycle;
					\draw[fill=yellow,opacity=0.4,-,thick](axis cs:-0.121047, -0.549342)--(axis cs:-0.65584, -0.227545)--(axis cs:-0.65584, 0.227545)--(axis cs:-0.121047, 0.549342)--(axis cs:0.121047, 0.549342)--(axis cs:0.65584, 0.227545)--(axis cs:0.65584, -0.227545)--(axis cs:0.121047, -0.549342)--cycle;
					\draw[fill=yellow,opacity=0.4,-,thick](axis cs:-0.00552226, -0.549342)--(axis cs:-0.771365, -0.227545)--(axis cs:-0.771365, 0.227545)--(axis cs:-0.00552226, 0.549342)--(axis cs:0.00552226, 0.549342)--(axis cs:0.771365, 0.227545)--(axis cs:0.771365, -0.227545)--(axis cs:0.00552226, -0.549342)--cycle;	
				\end{groupplot}
				\tikzset{SubCaption/.style={
						text width=0.28\textwidth,xshift=7mm,yshift=0mm, align=center,anchor=north}}
				\node[SubCaption] at (ch1plots c1r1.south) {\subcaption{$p_0=3.3,3.7,3.9,4.3$.}};
				\node[SubCaption] at (ch1plots c2r1.south) {\subcaption{$p_0=3,3.7,4.2,4.8$.}};
			\end{tikzpicture}
			\hfill
				\vspace{5pt}
				\centering
                \begin{tikzpicture}
					\begin{axis}[		
						width=0.4\textwidth,
						height=0.25\textwidth,
						xlabel = {$p_0$},
						ylabel = {Angle $\theta$},
						xlabel style={anchor=north east, inner xsep=5pt},
						ylabel style={anchor=base, inner ysep=5pt,rotate=0},
						xmin=0, xmax=4.8 ,
						ymin = 0.5, ymax = 3.5, legend style={at={(0.05,0.45)},anchor=north west}
						]
						\addplot[thick,violet,dashed] table {Octhw_p0angtable02.txt};\addlegendentry{Relaxed ($\theta$)}
						\addplot[thick,red] table {Octhw_p0angtable01.txt};\addlegendentry{Relaxed ($C_y,\theta$)}
					\end{axis}
				\tikzset{SubCaption/.style={
						text width=0.28\textwidth,xshift=7mm,yshift=0mm, align=center,anchor=north}}
				\node[SubCaption] at (2,-0.5) 
                {\subcaption{Angle values of (a) and (b).}};
				\end{tikzpicture}
			\caption{(a,b) Optimized shapes for relaxed conditions for target $p_0$ $3-4.8$. (a) For the relaxed case  ($C_y,\theta$), the cell is an octagonal shape when $p_0=3.7$ and is a rectangular shape when $p_0=4.3$ (dark to light) (b) In the relaxed state ($\theta$), when $p_0$ is 3.7, the cell has an octagonal shape, whereas when $p_0$ is 4.8, the cell is hexagonal (dark to light). (c) Angle $\theta$ values for Fig. \ref{fig:two-peaks} for the two different relaxed cases.}
			\label{fig:octagon-transition}
		\end{figure}
\subsection{3D Vertex Model}
\subsubsection{Regular truncated octahedron model}
	
  Now that we better understand some of the more subtle points behind how the shape landscape can depend on the parameterization and, therefore, influence a system's mechanical response, we turn to the three-dimensional case. Given the results of the prior subsection, we first work with our initial parameterization of the truncated octahedron with $l_1,l_2,\theta$,$z$ and now including $l_3$ such that $l_1=l_3$ when we first introduced the three-dimensional model. Please note that our evaluation is based on a regular truncated octahedron with unit volume. If we use an irregular one with unit volume, the result would be different due to a higher $s_0$. 
  
To better understand these curves, let us first do the same analysis for the constrained case as it serves an as upper bound for the curves using the one- or two-parameter minimization. If we set $l_1=l_3= \frac{1}{2\cdot 2^{1/6}}$ and $l_2 =\frac{\sqrt{2}}{2 \cdot2^{1/6}}$, $K_V=10$, $K_S=1$, and $\theta=\frac{3\pi}{4}$, for the constrained case we obtain 
\begin{align}
\lim_{\delta\to 0}K(E(\delta))_{\text{G}}&\simeq	\frac{|((3 + 4 \sqrt{3}) (3 + 6 \sqrt{3} - 2 \cdot2^{1/3} s_0))|}{3 \cdot2^{2/3}},\\
\lim_{\delta\to 0}K(E(\delta))_{\text{B}}&\simeq|(576 (20 + (8138 + 4440 \sqrt{3})^{1/3}))| \notag\\&/(64 + 17577 \cdot2^{2/3} + 
8424 \cdot2^{2/3} \sqrt{3}  \notag\\&+ 
72 s_0 (-3 (37 + 30 \sqrt{3}) \notag\\& + 2^{1/3} (13 + 4 \sqrt{3}) s_0))^{3/2}\\& \text{(note that $z=\frac{\sqrt{2}}{2 \cdot2^{1/6}}$ is scaled by $\sqrt{1+\delta}$)}, \notag\\
\lim_{\delta\to 0}K(E(\delta))_{\text{Y}}&\simeq| (12 (160 + 131  \cdot2^{1/3} + 42 \cdot 2^{1/3} \sqrt{3})  \notag\\&- 
8  \cdot2^{2/3} (3 + 4 \sqrt{3}) s_0)| \notag\\
&/(3 (16 + 1953 \cdot 2^{2/3} + 
936  \cdot2^{2/3} \sqrt{3}  \notag\\&+ 
8 s_0 (-3 (37 + 30 \sqrt{3})  \notag\\&+ 2^{1/3} (13 + 4 \sqrt{3}) s_0))^{3/2}).
\end{align}
As in the two-dimensional case, the curvature as a function of $s_0$ for shear modulus is a linear away from the rigidity transition location of the regular truncated octahedron with unit volume. Moreover, the other two moduli have the form of $\frac{|a+bx|}{(c+d(e-x)^2)^{3/2}}$. The peaks observed in the constrained case for the Young's modulus and the bulk modulus are present here given the choice of $K_V$ and $K_S$, which was motivated by the choice in parameters in Ref.~\cite{Zhang2023} and is different form the equivalent relative stiffness in two dimensions, which is why such peaks were not prominent in the earlier case. The mechanical response of the constrained bulk modulus exhibits a higher peak compared to the response of Young's modulus due to the selection of $C_xC_yC_z=(\sqrt{1+\delta})^3$ instead of $C_xC_yC_z=({(1+\delta)}^\frac{1}{3})^3$.

For the relaxed cases where there is at least a two-parameter minimization, we also observe the same phenomena of a rigidity transition. See Fig. 18. Again, we observe a drop/cusp in the Poisson's ratio at the transition point.  
		\begin{figure}[h!]
			\centering
			\includegraphics[width=0.5\textwidth]{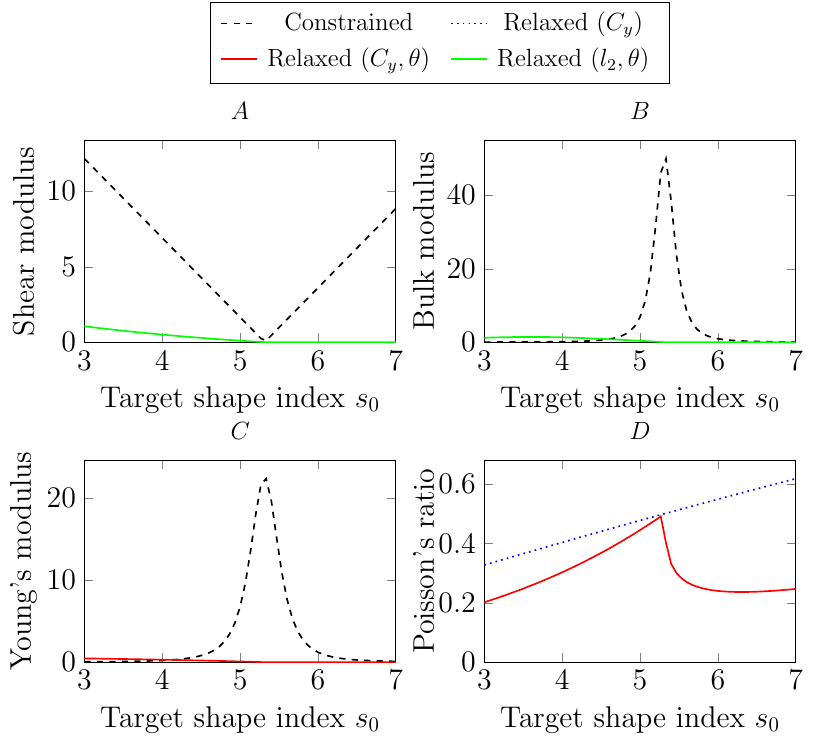}
			\caption{Mechanical response for truncated octagon.}
		\end{figure}

		\begin{center}
			\begin{tabular}{cc}
				%		\hline
				\addheight{	\includegraphics[width=2.2cm]{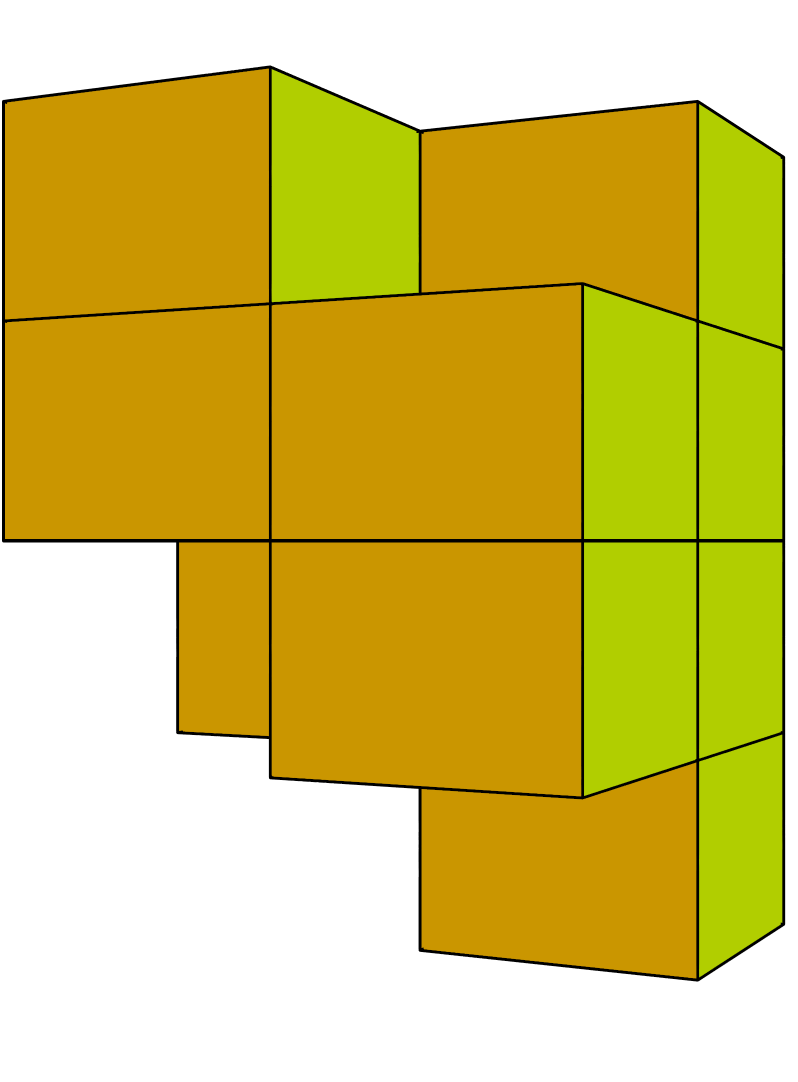}
				} &
				\addheight{\includegraphics[width=2.8cm]{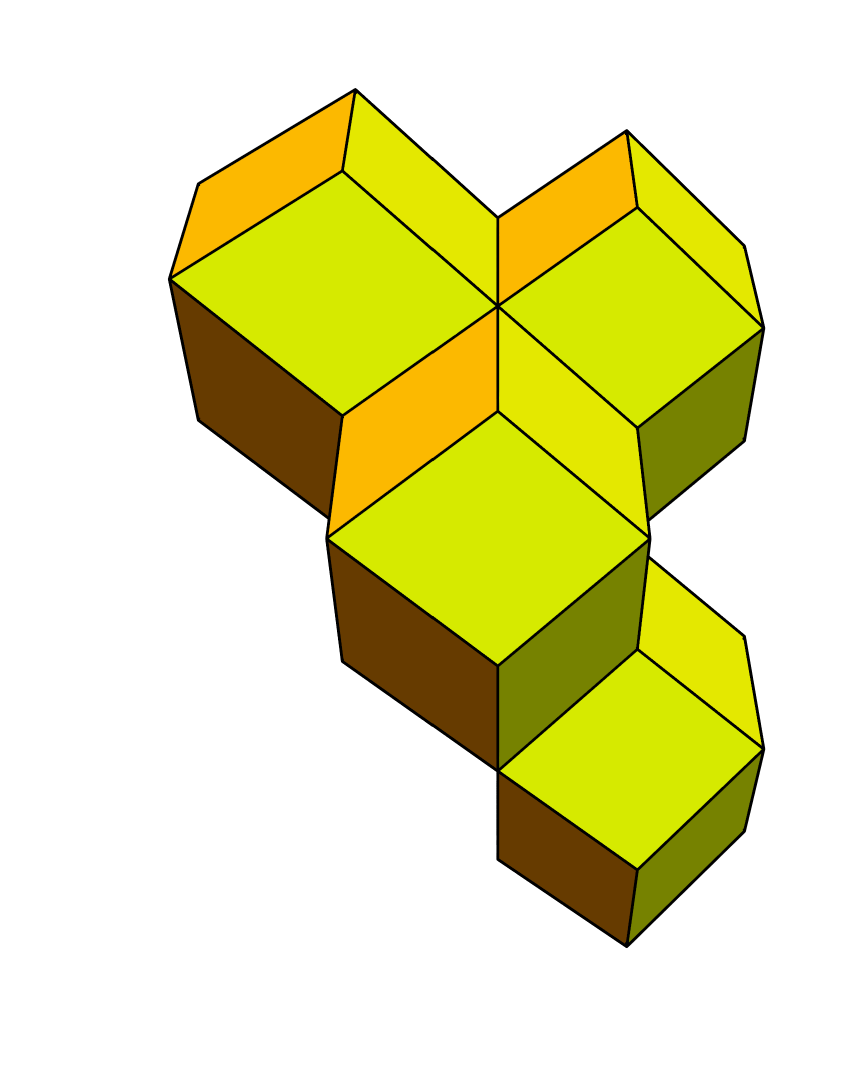}
				} \\
				\small (a) Rectangular prism &  \small (b) Rhombic dodecahedron \\
				\addheight{	\includegraphics[width=2.8cm]{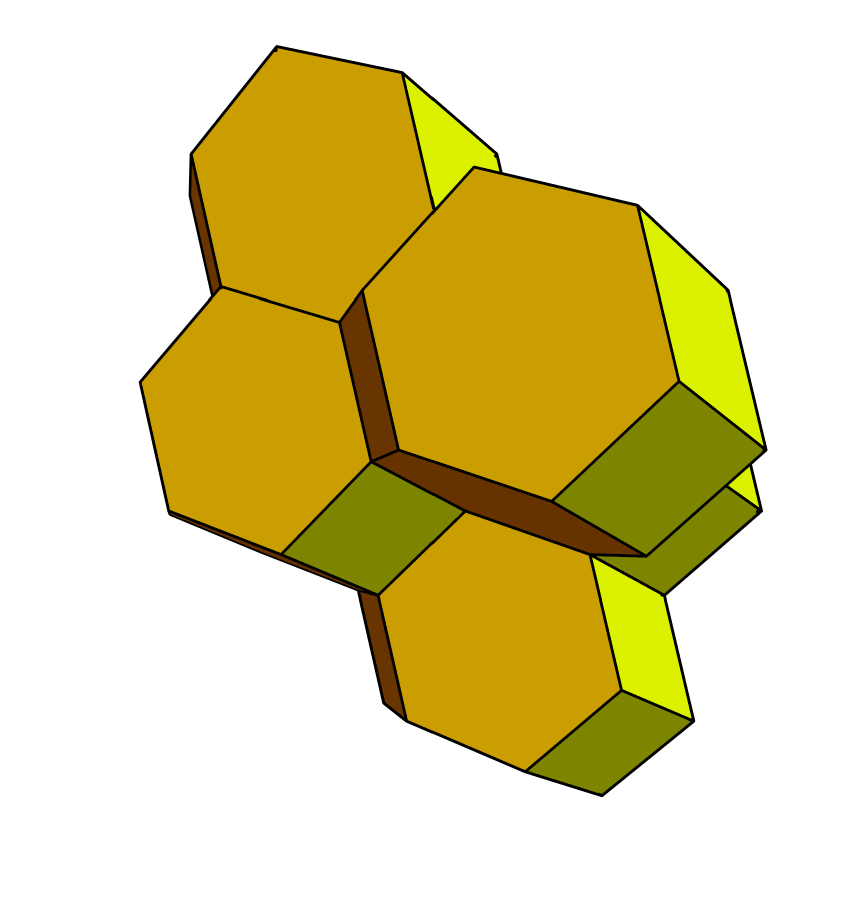}
				} &
				\addheight{\includegraphics[width=2.2cm]{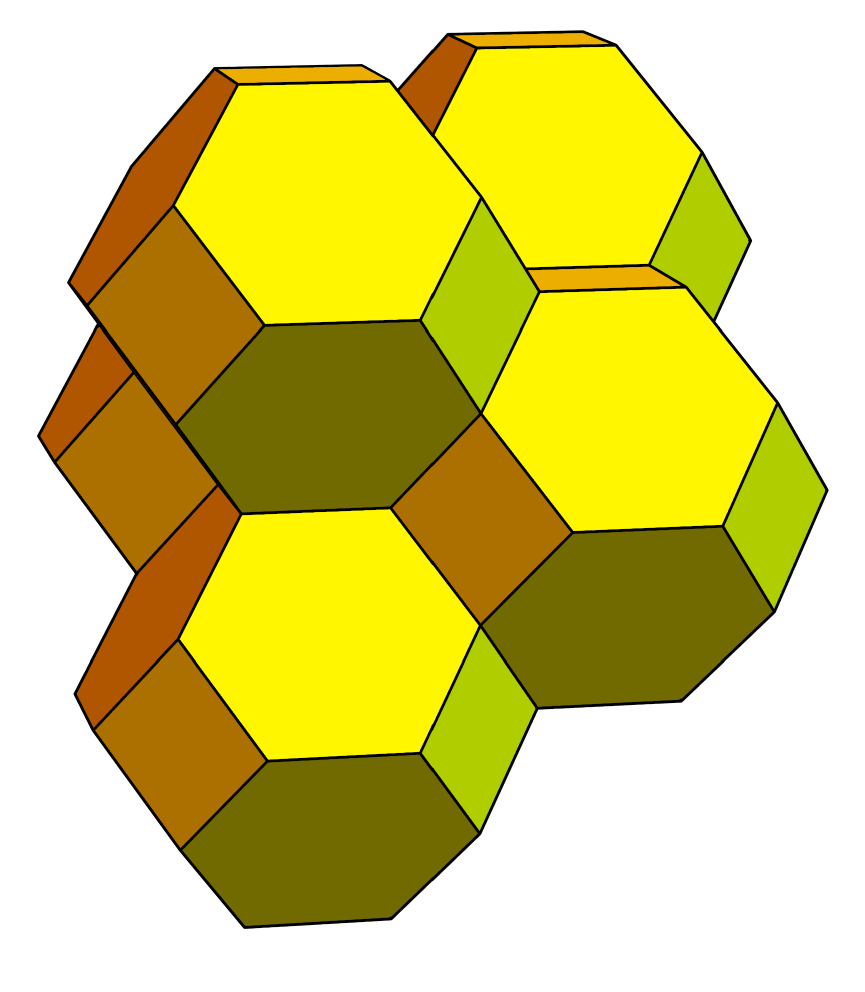}
				} \\
				\small (c) Elongated dodecahedron &  \small (d) Truncated octahedron \\
				\addheight{	\includegraphics[width=2.5cm]{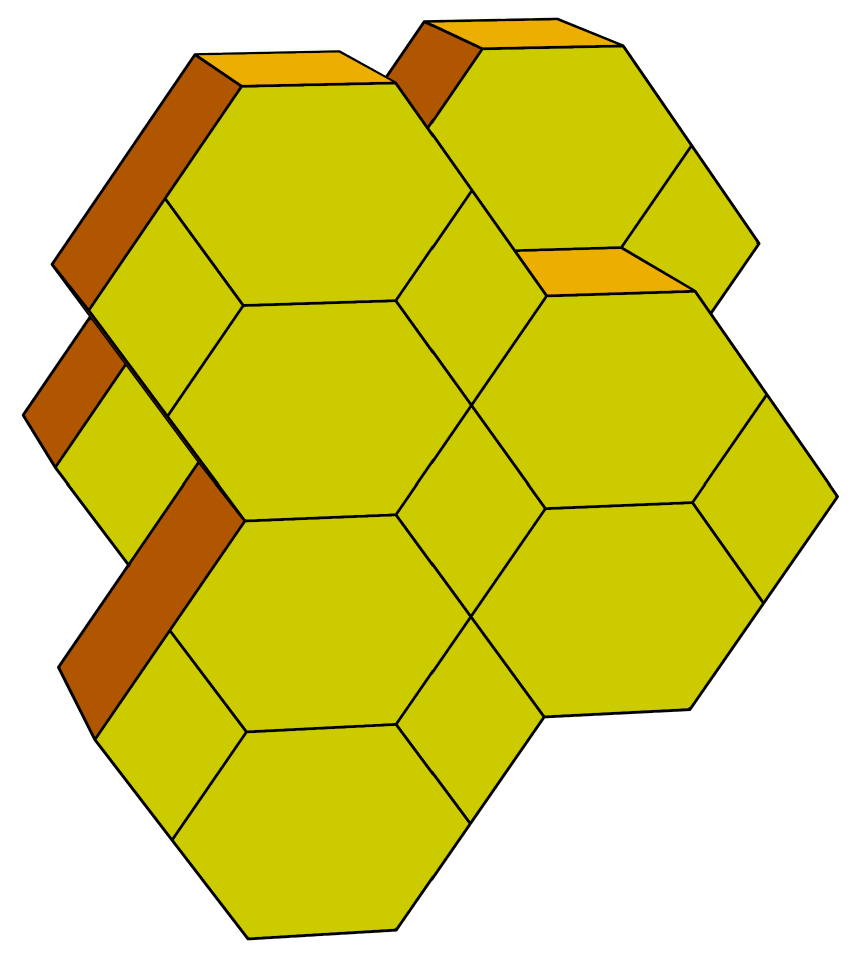}
				} &
				\addheight{	\includegraphics[width=2.5cm]{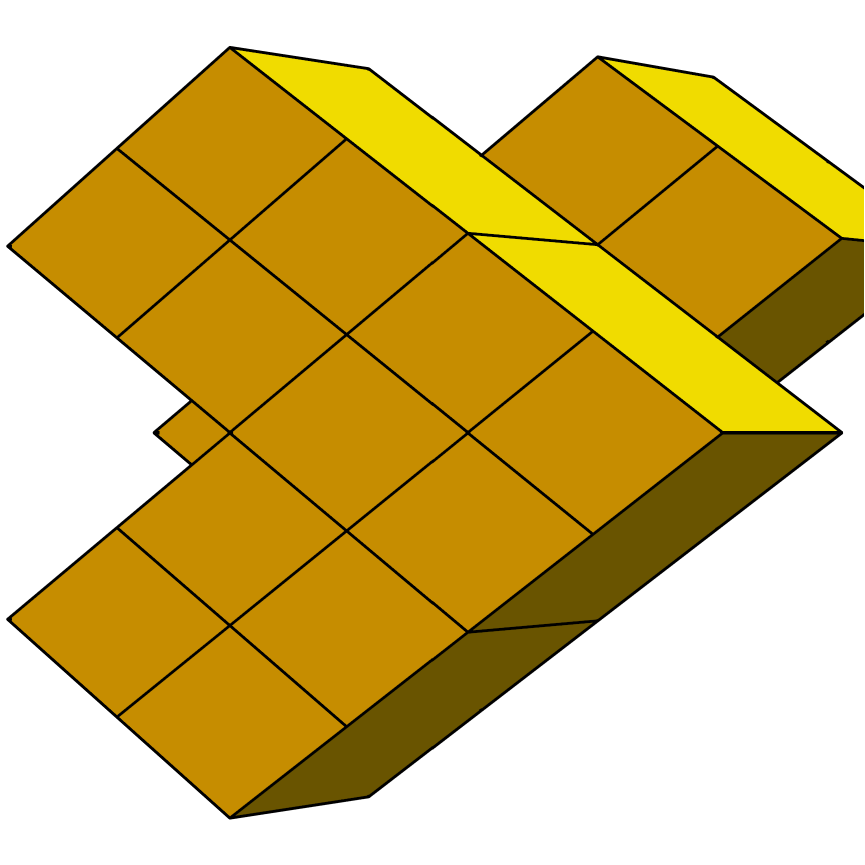}
				}\\
				\small (e) Staggered hexagonal cell & \small (f) Trigonal trapezohedron \\
			\end{tabular}
			\captionsetup{type=figure}
			\captionof{figure}{Transformable structures of the reparamterized truncated octahedron cell model.}
			\label{fig:3d-transformable-shapes}
		\end{center}

		\subsubsection{3d Model with different parameterization}
		Similar to Sec. \ref{nOM}, the 3D model utilizes the same parametrization methods by modifying the 3D coordinate vectors from the variables $l_1,l_2,l_3$ to $w,h_1,h_2$ while $\theta,z$ stay the same. The new parameterization for the truncated octahedron matches the octagonal model in the Figure \ref{fig:reparam-octagon}. With this new parameterization, additional shape transformations can be explored such that shape landscape is slightly more complex as compared to the prior parameterization. Fig. \ref{fig:3d-transformable-shapes} (a), (b), and (f) relate to the lowest values ($h_1=0$ or $h_2=0$). The value of $s_0$ for the trigonal trapezohedron in Table \ref{table:3} was determined by using the regular trapezohedron with $n=3$, and the $s_0$ for the staggered hexagonal cell was calculated from the regular truncated octahedron with an angle of $\pi/2$.
		\captionsetup{singlelinecheck = false, justification=raggedright}
		\begin{figure}[h!]
			\centering
           \begin{tabular}{c@{\qquad} c}
			\centering
			\includegraphics[width=0.5\textwidth]{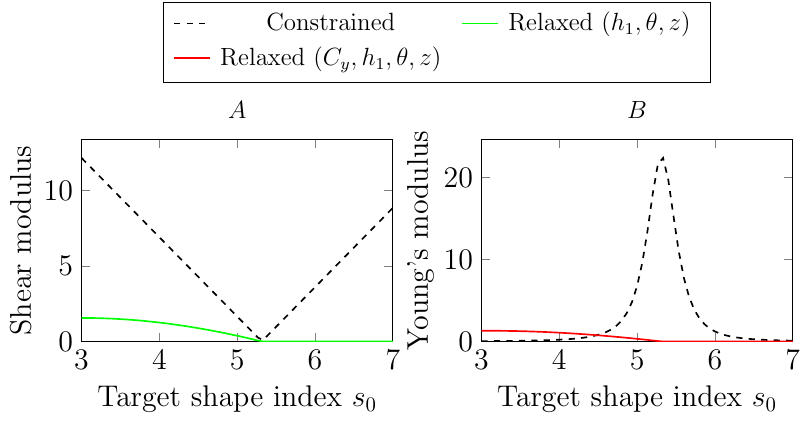}
			  &  \\
			\begin{minipage}[t]{0.95\columnwidth}
               (a) Mechanical response for the reparameterized truncated octahedron model.
               \vspace{5mm}
           \end{minipage}&\\ 
   			\centering
			\begin{tikzpicture}
					\begin{axis}[		
						width=0.4\textwidth,
						height=0.25\textwidth,
						xlabel = {$s_0$},
						ylabel = {Young's modulus},
						xlabel style={anchor=north east, inner xsep=5pt},
						ylabel style={anchor=base, inner ysep=5pt,rotate=0},
						xmin=3, xmax=7 ,
						ymin = 0, ymax = 25,legend style={at={(0.05,0.9)},anchor=north west}
						]
						\addplot[thick,blue,dashed] table {3dhw_p0table02.txt};\addlegendentry{$(\theta)$}
						\addplot[thick,orange] table {3dhw_p0table01.txt};\addlegendentry{$(C_y,\theta)$}
						\addplot[mark=none, red,thick] coordinates {(5.28,0) (5.28,25)};
						\addplot[mark=none, red,thick] coordinates {(5.8,0) (5.8,25)};
					\end{axis}
				\tikzset{SubCaption/.style={
						text width=0.45\textwidth,xshift=7mm,yshift=0mm, align=center,anchor=north}}
                \node[SubCaption] at (3,-0.5) 
                {\subcaption{Mechanical response of 3d model.}};
				\end{tikzpicture}
    & 
               \end{tabular}
			\caption{(a) Mechanical response for the reparameterized truncated octahedron model as a function of the dimensionless shape index. (b) Mechanical response of 3d model for relaxed ($\theta$) and ($C_y,\theta$). The first red line is for a value of $s_0=5.28$, and the second one is for $s_0=5.8$.}
		\end{figure}
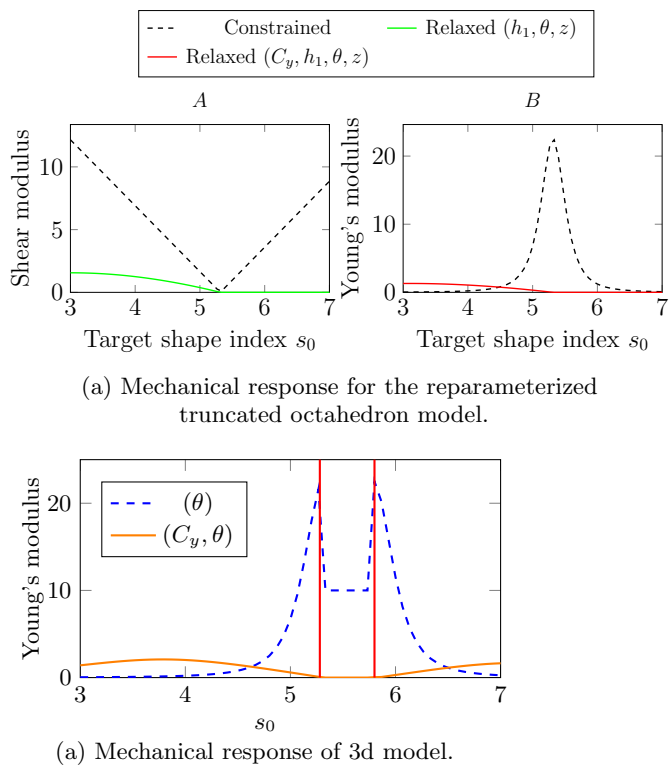
		For the reparameterized version of the truncated octahedron, we find similar behavior for the shear and bulk modulus as for the first parameterization. However, for the reparameterized version, even when energy is optimized using $\theta$, Young's modulus remains high, since decreasing the energy requires more than just changing $\theta$. If we measure Young’s modulus for $h_2=0$ using ($C_y,\theta$) and minimize energy $e_S$, we can observe a shape transition in the three-dimensional model from elongated dodecahedron to a near rhombic dodecahedron and then to trigonal trapezohedron. 
		\begin{figure}[H]
			\captionsetup{singlelinecheck = false, justification=raggedright}	
			\begin{center}
				\begin{tabular}{cc}
         				\begin{tikzpicture}
					\begin{axis}[		
						width=0.4\textwidth,
						height=0.25\textwidth,
						xlabel = {$p_0$},
						ylabel = {Angle $\theta$},
						xlabel style={anchor=north east, inner xsep=5pt},
						ylabel style={anchor=base, inner ysep=5pt,rotate=0},
						xmin=3, xmax=7 ,
						ymin = 1, ymax = 3, legend style={at={(0.05,0.45)},anchor=north west}
						]
						\addplot[thick,blue,dashed] table {3dhw_p0angtable02.txt};\addlegendentry{($\theta$)}
						\addplot[thick,orange] table {3dhw_p0angtable01.txt};\addlegendentry{($C_y,\theta$)}
						\node [label={-90:{\large$S_1$}},inner sep=2pt] at (axis cs:5.0,2){};
						\draw[->](axis cs:5.0,2.05)--(axis cs:5.2666,2.3);
						\node [label={0:{\large$S_2$}},inner sep=2pt] at (axis cs:6.0,2.5){};
						\draw[->](axis cs:6.0,2.5)--(axis cs:5.55,2.7);
						\node [label={0:{\large$S_3$}},inner sep=2pt] at (axis cs:6.2,2){};
						\draw[->](axis cs:6.2,2)--(axis cs:5.8,1.7);
					\end{axis}
				\tikzset{SubCaption/.style={
						text width=0.45\textwidth,xshift=7mm,yshift=0mm, align=center,anchor=north}}
                \node[SubCaption] at (3,-0.5) 
                {\subcaption{Angle values for relaxed ($\theta$) and ($C_y,\theta$).}};
				\end{tikzpicture}
        &\\
				\begin{tabular}{cc}
					\addheight{	\includegraphics[width=2.8cm]{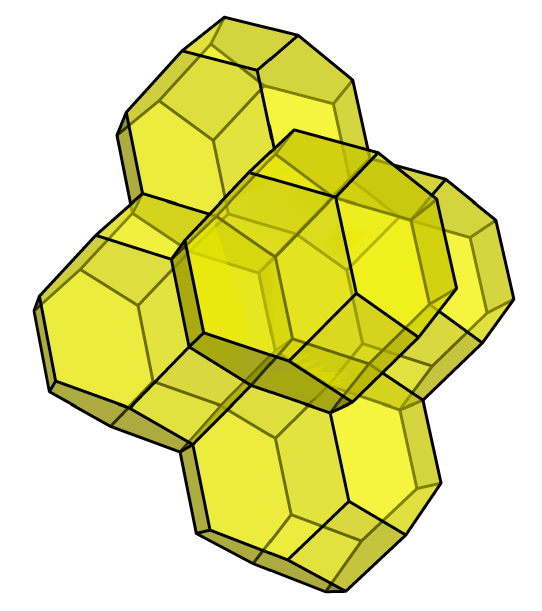}
					} &
					\addheight{\includegraphics[width=2.8cm]{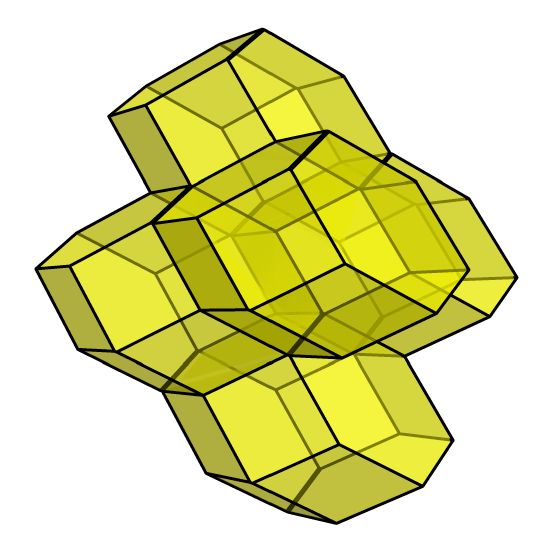}
					} \\
					(b) $S_1$ ($s_0\simeq 5.27$) & (c) $S_2$ ($s_0\simeq 5.53$)\\
					\addheight{\includegraphics[width=2.8cm]{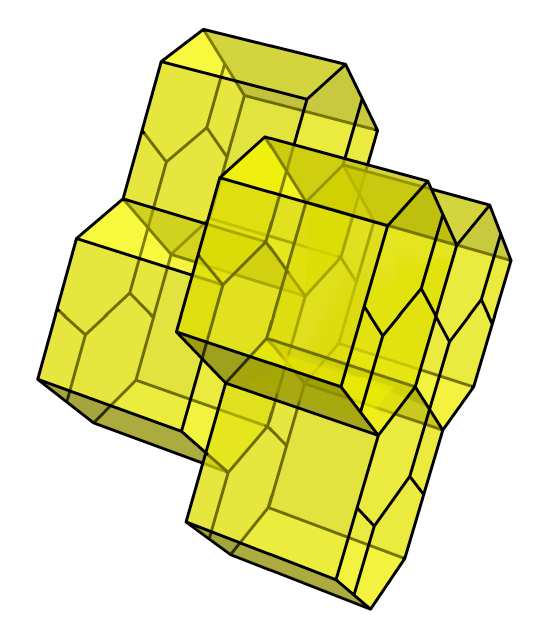}
					} & \\
					(d) $S_3$ ($s_0\simeq 5.8$) & \\
				\end{tabular}
                \end{tabular}
			\end{center}	
				\caption{(a) Mechanical response of 3d model for relaxed ($\theta$) and ($C_y,\theta$). The first red line is for a value of $s_0=5.28$, and the second one is for $s_0=5.8$. (b), (c), and (d) $S_1,S_2,S_3$ of (a) for each target shape index $s_0$.}
    \label{fig:angles2}
		\end{figure}
		\begin{table}[ht]\caption{Dimensionless shape index $s_0$ for various shapes.}
			\centering 
			\begin{tabular}{c c }
				\hline\hline                        
				& $s_0$ for regular polyhedron \\ [0.5ex]
				\hline                  
				Rectangular prism (cube) & 6\\
				Rhombic dodecahedron & 5.34539 \\
				Elongated dodecahedron & 5.49324 \\
				Truncated octahedron &  5.31474\\
				Staggered hexagonal cell &  5.78808\\
				Trigonal trapezohedron & 6 \\
				[1ex]      
				\hline
			\end{tabular}\label{table:3}
		\end{table}
				
		\section{Discussion}
		We have examined the mechanical response to infinitesimal strains of an ordered, cellular-based vertex model in both two and three dimensions by calculating the shear modulus, bulk modulus, Young’s modulus and Poisson’s ratio within mean field. We find a compatible-incompatible transition in three dimensions with the target shape index as the tuning parameter, similar to the transition found in two dimensions~\cite{Staddon2023}. The transition point location occurs at the target shape index of the truncated octahedron (with $s_0\simeq5.315$), which is a bit a lower than the rigidity transition found earlier~\cite{Zhang2023}. In this earlier three-dimensional work, cellular reconnection events are allowed and the system is disordered. The trend in the shift of the location of the transition resembles the two-dimensional case where the transition from fluid to rigid state occurs at $p_0=3.813$ when T1 events allowed~\cite{Bi_2015}, while the hexagonal model with no T1 events shows a transition at $p_0=3.72$ \cite{Staple2010,Staddon2023}. 
  
\begin{figure}[h]
			\centering{
				\resizebox{75mm}{!}{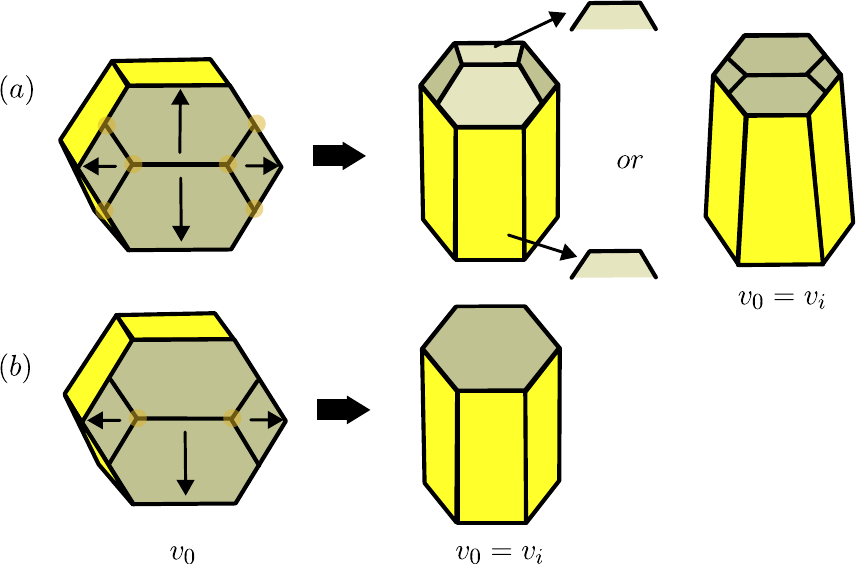}
				\caption{Staggered hexagonal cell can be morphed into various structures with the same volume. (a) Circled vertices can move up and down under the same edge length. Moreover, the top and bottom can change its size while the surface remains flat (scutoids). (b) Vertices can merge together and construct a hexagonal cylinder.}
		\label{fig:staggered-hexagonal-cell}
			}
		\end{figure}

  Shape indeed has an effect on three-dimensional rigidity transitions as it can either more readily or less readily allow a cell to achieve both the target surface area and volume. The target shape index $s_0$ explores this readiness. When the shape index is $s_0\geq5.8$, we find the cell is more likely to be a staggered hexagonal cell (for $h_2\neq0$). See Fig. \ref{fig:staggered-hexagonal-cell}. As the staggered hexagonal cell has co-planar faces, it can be thought of as a hexagonal cylinder.  We observe that two inner vertices located on the top and bottom faces can move in either the $x$ or $y$ directions as long as they remain inside of the hexagon. This can be considered as decreasing the number of constraints as four independent planes for the top of the structure becomes one. Moreover, this design makes it easier to maintain the same volume as the cell area changes. The structure becomes floppy, provided there are a sufficient number of relaxation parameters. 
  
  We have also found that for nontrivial shape transition pathways, particularly in three dimensions, there can be interesting mechanical properties even without changes in topology. For instance, the staggered hexagonal cell shape is achieved via a particular parameterization of the truncated octahedron such that there are no singularities. This change in the shape landscape, as described by its shape transition pathways, influences the Young's modulus, again, depending on the dimensionality of the relaxation parameter space. We also observe this effect in two dimensions with a particular parameterization of the hexagon in which there is a re-emergence of a nonzero Young's modulus at higher shape indices. 

  As for a very simple argument for the rigidity transition, when we look at the shape index of the tetrahedron, cube, and icosahedron, which are $s_0=7.21$, $s_0=6$, and $s_0=5.148$ respectively, the larger the value of $s_0$, the less faces the regular polyhedron has. From a mechanical standpoint, it is easier to adjust and locate the simpler polyhedron in the space as it has fewer faces. For example, the hexagonal cylinder is more advantageous than the truncated octahedron since it requires less symmetries to fill space.  Moreover, hexagonal cylinders can readily be transformed into scutoids, which are found in many biological cellular structures \cite{Silvanus2017,Gomez-Galvez2018,LEMKE2021R1098}. In comparison, the truncated octahedron requires symmetry in the diagonal direction to fill space, and it is more commonly seen in solid structures like \cite{Elechiguerra2016} or regularly packed droplets \cite{Alcinesio2020}. The previous section shows that, if there is an intermediate shape able to reduce stress, the cell will transform its shape to relax as can be seen in Fig. \ref{fig:angles2}. To go beyond this simple argument in terms of the interplay between geometry and rigidity, as is done in Refs.~\cite{Cauchy1813,Alexandrov2005,Gandikota2022}, as well as higher-order rigidity~\cite{Yan2019}, more work is being done. 
		
\begin{figure}[h]
			\centering{				
				\includegraphics[width=5cm]{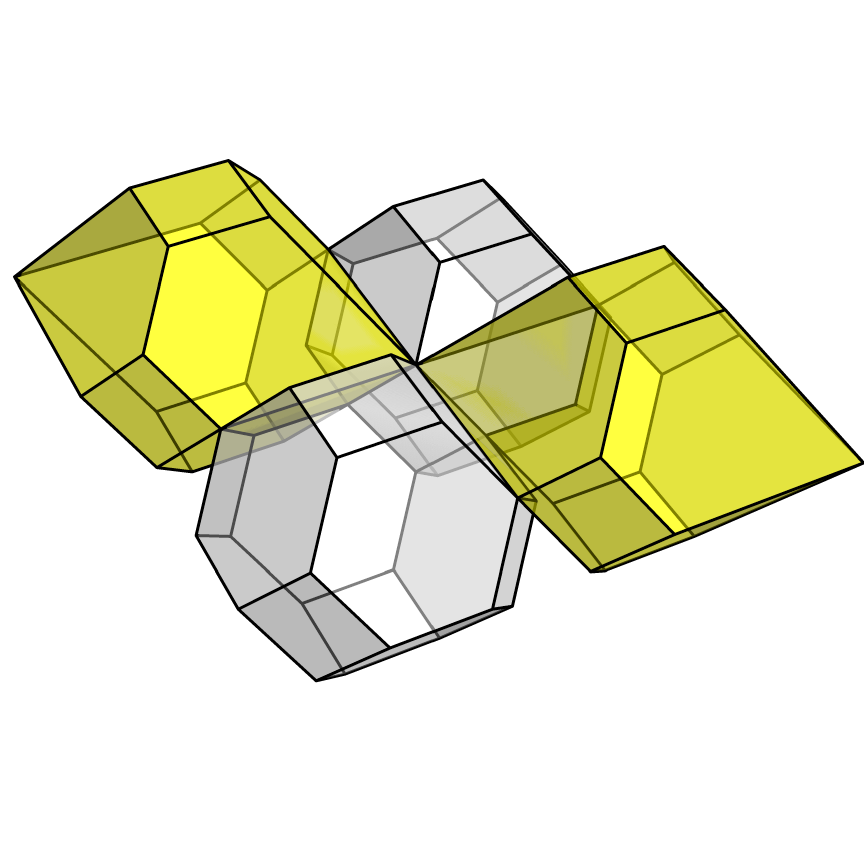}
				\caption{The 3d reconnection model (Type O) developed by Honda et al. \cite{HONDA2004439} was implemented on the truncated octahedrons, represented by the yellow cells ($s\simeq5.45477$). In order to fill space, it is necessary to replace the gray-colored cells with other shapes.}
				\label{fig:reconnection-model2}
			}
		\end{figure}

   What have we learned here that will help us understand better the rigidity transition when cellular reconnection events are allowed? Such rules have been studied using metallic solids, as discussed in \cite{SMITH1953295} and \cite{Ranganathan2008}, and have also been applied to biology~\cite{HONDA2004439,Okuda2012}. Interestingly, Williams in 1968 \cite{Williams1968} demonstrated the reconnection event sequence of a truncated octahedron for vegetable cells (or soap bubbles, metal grains) based on the cell’s own edges and vertices. Honda {\it et al.} \cite{HONDA2004439} chose a square situated on the side of the truncated octahedron. In the ordered vertex model, the dimensionless shape index $s$ $\simeq5.45477$ structures with two sides, as demonstrated in Fig. \ref{fig:reconnection-model2}, and $\simeq5.58941$ for those with four sides. Assuming a reconnection event from one side in Fig. \ref{fig:reconnection-model2}, the value of $s$ is $\simeq5.38742$ for a transition point. Note that $s\simeq5.38742\sim5.4$ is supported by the 3D Voronoi model analysis~\cite{Merkel_2018,Nogucci2019} and the 3D vertex model analysis~\cite{Zhang2023}. However, the task of achieving complete space filling of a single shape cell is hampered by the formation of gaps. The use of top and bottom squares can help achieve space filling in a single vertex model. 
   
\begin{figure}[h]
			\centering{				
				\includegraphics[width=7.5cm]{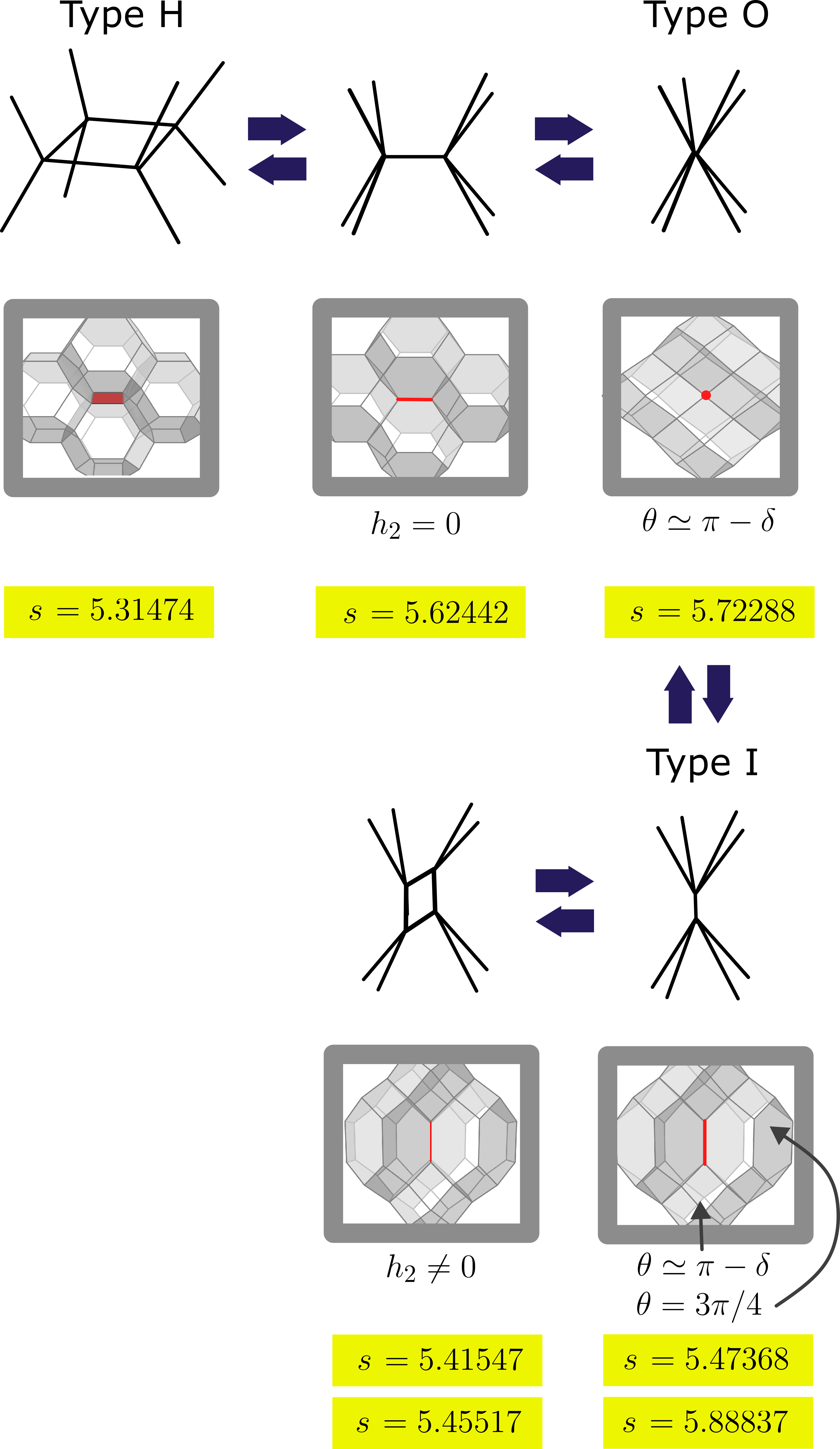}
				\caption{Modified reconnection model in an ordered truncated octahedron model.}
				\label{fig:reconnection-model}
			}
		\end{figure}

   As for another reconnection event configuration where gaps are not an issue, consider Fig. \ref{fig:T1}, a transition is shown from horizontal face (red line) to a vertical face instead of a horizontal to vertical line transition, which is a T1-like event in three dimensions. Instead of the hexagon to pentagon transition in two dimensions, we chose a hexagon to rectangular transition in three dimensions (Refer to Fig.  \ref{fig:T1}, hexagonal and rectangular faces are connected through a red line.). We can augment the T1-like transition with more steps, as depicted in Fig. \ref{fig:reconnection-model} in accordance with the reconnection model articulated in Ref.~\cite{Okuda2012}. We envision cells contracting from face, line, to a point to alter its direction. In addition, by changing the lengths of the edges (when $h_2\neq0$) in the Type I state, horizontal cells are converted into a truncated octahedrons and vertical cells become elongated dodecahedrons. Moreover, the same process can be done with hexagonal cylinder (staggered hexagonal cells), implying there are numerous reconnection choices in three dimensions~\cite{Okuda2012}. Indeed, we are only beginning to understand the nuances of rigidity transitions in three dimensions that occur in cellular-based vertex models with and without reconnection events.

	\begin{center}
	\begin{tabular}{cc}
		\addheight{	\includegraphics[width=2.95cm]{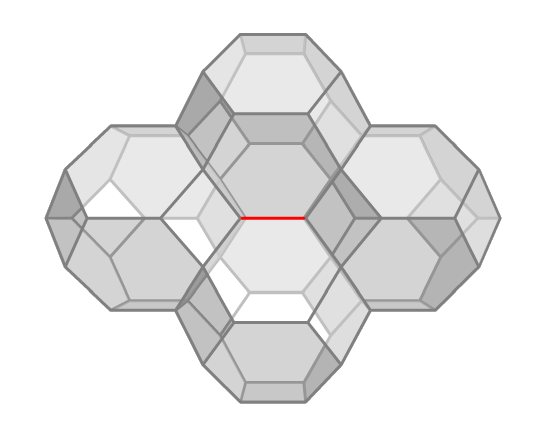}
		} &
		\addheight{\includegraphics[width=2.95cm]{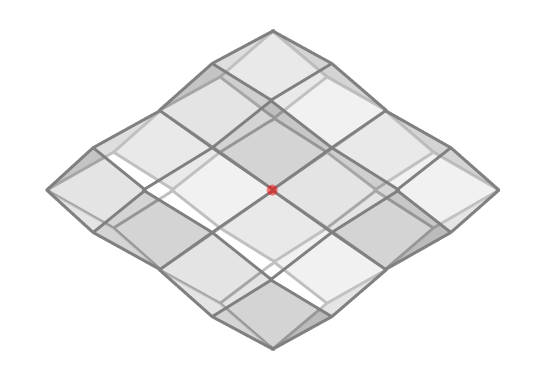}
		} \\
		\small (a) Cells before T1 swap &  \small (b) interface contraction\\
		\addheight{	\includegraphics[width=2.75cm, angle=-90]{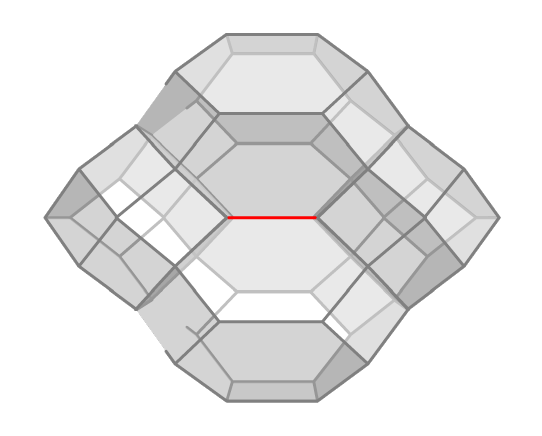}
		} & \\
		\small (c) Cells after T1 swap &  \\
	\end{tabular}
	\captionsetup{type=figure}
	\captionof{figure}{Constructing truncated octahedrons that have a corresponding T1-like transition. (a) Cell-cell interface indicated by a red line that is contracted to a point (b), and subsequently stretched in the perpendicular direction (c).}
	\label{fig:T1}
\end{center}		

		\begin{acknowledgments}
			JMS acknowledges financial support from NSF-PHY-PoLS-2014192. The authors also acknowledge insightful discussion with Shabeeb Ameen, Danilo Liarte, James Sethna, and Stephen Thornton. 
		\end{acknowledgments}
		\newpage
		\section{Appendix A: Finding minima of the energy functional $E$}
  \subsection{Two dimensions}
	Suppose energy $E$ is a function of $x$ with $a(x)$ and $p(x)$, then we have
\renewcommand{\theequation}{A.\arabic{equation}}
\setcounter{equation}{0}
\begin{align}
E(x)&=\frac{1}{2}(a(x)-1)^2+\frac{1}{2}k_r(p(x)-p_0)^2,\\
\frac{\partial E
}{\partial x}&=\frac{\partial a(x)
}{\partial x}(a(x)-1)+k_r\frac{\partial p(x)
}{\partial x}(p(x)-p_0).
\end{align}
If $\frac{\partial a(x)
}{\partial x}\neq0$, then the function has to satisfy $a(x)
=1$ to make a $E(x)=0$. We can apply same strategy for $\frac{\partial p(x)
}{\partial x}\neq0$ and find necessary conditions to achieve $E(x)=0$.
For hexagon, we have
\begin{align}
p&=2l_1+4l_2,\\
a&=2l_2(l_1-l_2\cos\theta)\sin\theta.
\end{align}
Notice that $\frac{\partial p}{\partial\theta}=0$ and nonzero for other cases ($\frac{\partial p}{\partial l_1},\frac{\partial p}{\partial l_2},\frac{\partial a}{\partial \theta},\frac{\partial a}{\partial l_1},\frac{\partial a}{\partial l_2}$).
We can write more explicitly
\begingroup
\allowdisplaybreaks
\begin{align}
E&=\frac{1}{2}(2l_2(l_1-l_2\cos\theta)\sin\theta-1)^2+\frac{1}{2}k_r(2l_1+4l_2-p_0)^2\notag\\
&=\frac{1}{2}(a-1)^2+\frac{1}{2}k_r(p-p_0)^2,\\
\frac{\partial E
}{\partial \theta}&=2l_2(l_2\cos2\theta-l_1\cos\theta)(1+2l_2(l_2\cos\theta-l_1)\sin\theta)\notag\\
&=2l_2(l_2\cos2\theta-l_1\cos\theta)(1-a),\\
\frac{\partial E
}{\partial l_1}&=2k_r(2l_1+4l_2-p_0)\notag\\
&+2l_2\sin\theta(-1+2l_2(l_1-l_2\cos\theta)\sin\theta)\notag\\
&=2k_r(p-p_0)+2l_2\sin\theta(a-1),\\
\frac{\partial E
}{\partial l_2}&=4k_r(2l_1+4l_2-p_0)\notag\\
&+(-2l_2\cos\theta\sin\theta+2(l_1-l_2\cos\theta)\sin\theta)\notag\\
&\cdot(-1+2l_2(l_1-l_2\cos\theta)\sin\theta)\notag\\
&=4k_r(p-p_0)\notag\\
&+(-2l_2\cos\theta\sin\theta+2(l_1-l_2\cos\theta)\sin\theta)(a-1).
\end{align}
\endgroup
Therefore, we need $a=1$ for $\frac{\partial E
}{\partial \theta}=0$ and $a=1,p=p_0$ for $\frac{\partial E
}{\partial l_1}=0,\frac{\partial E
}{\partial l_2}=0$. 

\subsection{Three dimensions}
For the truncated octahedron, we have $E(x)=\frac{1}{2}K_V(v(x)-1)^2+\frac{1}{2}K_S(s(x)-s_0)^2$ and
\begin{align}
	s&=2 \bigg(l_1 l_3 + l_2 z + 2 l_3 \sqrt{z^2 + l_2^2 \cos\theta^2}\notag\\
 &+ 
	l_2 \sin\theta \sqrt{z^2 + l_2^2 \cos\theta^2} \notag\\
 &+ 
	2 l_1 \sqrt{z^2 + l_2^2 \sin\theta^2} - 
	l_2 \cos\theta \sqrt{z^2 + l_2^2 \sin\theta^2}\bigg),\\
	v&=2z (l_1 - l_2 \cos\theta) (l_3 + l_2 \sin\theta).
\end{align}
Both $s$ and $v$ satisfy $\frac{\partial s,v}{\partial l_1},\frac{\partial s,v}{\partial l_2},\frac{\partial s,v}{\partial l_3},\frac{\partial s,v}{\partial \theta},\frac{\partial s,v}{\partial z}\neq0$, we need $v=1$ and $s=s_0$ for $E=0$. For $K_S=1$, we can also write explicitly as
\begingroup
\allowdisplaybreaks
\begin{align}
\frac{\partial E
}{\partial \theta}=&
\frac{1}{2}\bigg(2 K_V \big(-1 + 
2 z (l_1 - l_2 \cos\theta)\notag\\& (l_3 + 
l_2 \sin\theta)\big) \big(2 l_2 z \cos\theta (l_1 - 
l_2 \cos\theta) + \notag\\
&2 l_2 z \sin\theta (l_3 + l_2 \sin\theta)\big) \notag\\&+ 
4 \bigg(l_2 \cos\theta \sqrt{z^2 + l_2^2 \cos\theta^2}\notag\\& - (
2 l_2^2 l_3 \cos\theta \sin\theta)/\sqrt{z^2 + l_2^2 \cos\theta^2}\notag\\
& - (
l_2^3 \cos\theta \sin\theta^2)/\sqrt{z^2 + l_2^2 \cos\theta^2}\notag\\&+ (
2 l_1 l_2^2 \cos\theta \sin\theta)/\sqrt{z^2 + l_2^2 \sin\theta^2}\notag\\
& - (
l_2^3 \cos\theta^2 \sin\theta)/\sqrt{z^2 + l_2^2 \sin\theta^2}\notag\\&+ 
l_2 \sin\theta \sqrt{z^2 + l_2^2 \sin\theta^2}\bigg)\notag\\
& \bigg(-s_0 + 
2 (l_1 l_3 + l_2 z + 2 l_3 \sqrt{z^2 + l_2^2 \cos\theta^2}\notag\\& + 
l_2\sin\theta \sqrt{z^2 + l_2^2 \cos\theta^2}  \notag\\
&+ 2 l_1 \sqrt{z^2 + l_2^2 \sin\theta^2} \notag\\&- 
l_2 \cos\theta \sqrt{z^2 + l_2^2 \sin\theta^2})\bigg)\bigg)\\
=&\frac{1}{2}\bigg(2 K_V \big(-1 + 
v\big) \big(2 l_2 z \cos\theta (l_1 - 
l_2 \cos\theta)\notag\\& + 2 l_2 z \sin\theta (l_3 + l_2 \sin\theta)\big)\notag\\
& + 
4 \bigg(l_2 \cos\theta \sqrt{z^2 + l_2^2 \cos\theta^2} \notag\\&- (
2 l_2^2 l_3 \cos\theta \sin\theta)/\sqrt{z^2 + l_2^2 \cos\theta^2}\notag\\
& - (
l_2^3 \cos\theta \sin\theta^2)/\sqrt{z^2 + l_2^2 \cos\theta^2}\notag\\&+ (
2 l_1 l_2^2 \cos\theta \sin\theta)/\sqrt{z^2 + l_2^2 \sin\theta^2}\notag\\
& - (
l_2^3 \cos\theta^2 \sin\theta)/\sqrt{z^2 + l_2^2 \sin\theta^2}\notag\\&+ 
l_2 \sin\theta \sqrt{z^2 + l_2^2 \sin\theta^2}\bigg) \bigg(-s_0 + 
s\bigg)\bigg),
\end{align}
\begin{align}
\frac{\partial E
}{\partial z}=&\frac{1}{2}\bigg(4 K_V (l_1 - l_2 \cos\theta) (l_3 + l_2 \sin\theta)\notag\\& \big(-1 + 
2 z (l_1 - l_2 \cos\theta) (l_3 + l_2 \sin\theta)\big) \notag\\
&+ 
4 \bigg(l_2 + (2 l_3 z)/\sqrt{z^2 + l_2^2 \cos\theta^2} \notag\\&+ (
l_2 z \sin\theta)/\sqrt{z^2 + l_2^2 \cos\theta^2} \notag\\&+ (2 l_1 z)/
\sqrt{z^2 + l_2^2 \sin\theta^2} \notag\\&- (l_2 z \cos\theta)/\sqrt{z^2 + l_2^2 \sin\theta^2}\bigg)\notag\\
& \big(-s_0 + 
2 (l_1 l_3 + l_2 z + 2 l_3 \sqrt{z^2 + l_2^2 \cos\theta^2}\notag\\& + 
l_2\sin\theta \sqrt{z^2 + l_2^2 \cos\theta^2} + 
2 l_1 \sqrt{z^2 + l_2^2 \sin\theta^2} \notag\\&- 
l_2 \cos\theta \sqrt{z^2 + l_2^2 \sin\theta^2})\big)\bigg)\\
=&\frac{1}{2}\bigg(4 K_V (l_1 - l_2 \cos\theta) (l_3 + l_2 \sin\theta) \big(-1 + v\big) + \notag\\
&4 \bigg(l_2 + (2 l_3 z)/\sqrt{z^2 + l_2^2 \cos\theta^2} \notag\\&+ (
l_2 z \sin\theta)/\sqrt{z^2 + l_2^2 \cos\theta^2} + (2 l_1 z)/\sqrt{z^2 + l_2^2 \sin\theta^2} \notag\\
&- (l_2 z \cos\theta)/\sqrt{z^2 + l_2^2 \sin\theta^2}\bigg) \big(-s_0 +s\big)\bigg),
\end{align}
\begin{align}
\frac{\partial E
}{\partial l_1}=&\frac{1}{2} \bigg(4 K_V z (l_3 + l_2 \sin\theta) \big(-1 + 
2 z (l_1 - l_2 \cos\theta) \notag\\&(l_3 + l_2 \sin\theta)\big) + 
4 (l_3 + 2\sqrt{z^2 + l_2^2 \sin\theta^2}) \notag\\
&\bigg(-s_0 + 
2 (l_1 l_3 + l_2 z + 2 l_3 \sqrt{z^2 + l_2^2 \cos\theta^2} \notag\\&+ 
l_2 \sin\theta\sqrt{z^2 + l_2^2 \cos\theta^2} \notag\\& + 
2 l_1 \sqrt{z^2 + l_2^2 \sin\theta^2} \notag\\&- 
l2 \cos\theta \sqrt{z^2 + l_2^2 \sin\theta^2})\bigg)\bigg)\\
=&\frac{1}{2} \bigg(4 K_V z (l_3 + l_2 \sin\theta) \big(-1 + 
v\big) \notag\\&+ 
4 (l_3 + 2\sqrt{z^2 + l_2^2 \sin\theta^2}) \big(-s_0 + 
s\big)\bigg),
\end{align}
\begin{align}
\frac{\partial E
}{\partial l_2}=&\frac{1}{2}\bigg(2 K_V (2 z (l_1 - l_2 \cos\theta) \sin\theta - 
2 z \cos\theta \notag\\&(l_3 + l_2 \sin\theta))\big (-1 + 
2 z (l_1 - l_2 \cos\theta) (l_3 + l_2 \sin\theta)\big)\notag\\
& + 
4 \bigg(z + (2 l_2 l_3 \cos\theta^2)/\sqrt{z^2 + l_2^2 \cos\theta^2} \notag\\&+ (
l_2^2 \cos\theta^2 \sin\theta)/\sqrt{z^2 + l_2^2 \cos\theta^2}\notag\\& + 
\sqrt{z^2 + l_2^2 \cos\theta^2} \sin\theta \notag\\
&+ (
2 l_1 l_2 \sin\theta^2)/\sqrt{z^2 + l_2^2 \sin\theta^2} \notag\\&- (
l_2^2 \cos\theta \sin\theta^2)/\sqrt{z^2 + l_2^2 \sin\theta^2}\notag\\&- 
\cos\theta \sqrt{z^2 + l_2^2 \sin\theta^2}\bigg) \notag\\
&\big(-s_0 + 
2 (l_1 l_3 + l_2 z + 2 l_3 \sqrt{z^2 + l_2^2 \cos\theta^2} \notag\\&+ 
l_2 \sin\theta\sqrt{z^2 + l_2^2 \cos\theta^2}  + 
2 l_1 \sqrt{z^2 + l_2^2 \sin\theta^2} \notag\\&- 
l_2 \cos\theta\sqrt{z^2 + l_2^2 \sin\theta^2})\big)\bigg)\\
=&\frac{1}{2}\bigg(2 K_V (2 z (l_1 - l_2 \cos\theta) \sin\theta - 
2 z \cos\theta (l_3 + l_2 \sin\theta))\notag\\&
\big (-1 + v\big)+ 
4 \bigg(z + (2 l_2 l_3 \cos\theta^2)/\sqrt{z^2 + l_2^2 \cos\theta^2} \notag\\&+ (
l_2^2 \cos\theta^2 \sin\theta)/\sqrt{z^2 + l_2^2 \cos\theta^2} \notag\\&+ 
\sqrt{z^2 + l_2^2 \cos\theta^2} \sin\theta \notag\\
&+ (
2 l_1 l_2 \sin\theta^2)/\sqrt{z^2 + l_2^2 \sin\theta^2} \notag\\&- (
l_2^2 \cos\theta \sin\theta^2)/\sqrt{z^2 + l_2^2 \sin\theta^2}\notag\\&- 
\cos\theta \sqrt{z^2 + l_2^2 \sin\theta^2}\bigg) \big(-s_0 + 
s\big)\bigg),
\end{align}
\begin{align}
\frac{\partial E
}{\partial l_3}=&\frac{1}{2}\bigg(4 K_V z (l_1 - l_2 \cos\theta)\big (-1 + 
2 z (l_1 - l_2 \cos\theta)\notag \\&(l_3 + l_2 \sin\theta)\big) + 
4 (l_1 + 2 \sqrt{z^2 + l_2^2 \cos\theta^2})\notag\\
&\big (-s0 + 
2 (l_1 l_3 + l_2 z + 2 l_3 \sqrt{z^2 + l_2^2 \cos\theta^2}\notag\\& + 
l_2 \sin\theta\sqrt{z^2 + l_2^2 \cos\theta^2} + 
2 l_1 \sqrt{z^2 + l_2^2 \sin\theta^2} \notag\\&- 
l2 \cos\theta \sqrt{z^2 + l_2^2 \sin\theta^2})\big)\bigg)\\
=&\frac{1}{2}\bigg(4 K_V z (l_1 - l_2 \cos\theta)\big (-1 + 
v\big) \notag\\&+ 
4 (l_1 + 2 \sqrt{z^2 + l_2^2 \cos\theta^2})\big (-s0 + s\big)\bigg).
\end{align}
\endgroup
\section{Appendix B: Energy functional with scaling factors}
\subsection{Two dimensions}
\begingroup
\allowdisplaybreaks
\begin{align}
e_{S(\text{hex})}&=\frac{1}{2}(C_xC_y\cdot2l_2(l_1-l_2\cos\theta)\sin\theta-1)^2\notag\\&+\frac{1}{2}k_r(C_x\cdot2l_1\notag\\
&+\sqrt{{C_x}^2\cos{\theta}^2+{C_y}^2
\sin{\theta}^2}\cdot4l_2-p_0)^2
\end{align}
\begin{align}
e_{S(\text{oct})}&=\frac{1}{2}\bigg(C_xC_y\cdot\big( l_1(l_1-2l_2\cos\theta)+2l_1l_2\sin\theta\notag\\
&-2{l_2}^2\cos\theta\sin\theta\big)-1\bigg)^2+\frac{1}{2}k_r((C_x+C_y)\cdot2l_1\notag\\
&+\sqrt{{C_x}^2\cos{\theta}^2+{C_y}^2
\sin{\theta}^2}\cdot4l_2-p_0)^2    
\end{align}
\subsection{Three dimensions}
Recall we have $e_T=\frac{1}{2}(v-1)^2+\frac{1}{2}k_{\rho}(s-s_0)^2$, and where $s,v$ are
\begin{align}
	s&=2 \bigg(C_xC_y\cdot l_1 l_3 +  2 C_yl_3 \sqrt{{(C_z\cdot z)}^2 + {C_x}^2\cdot l_2^2 \cos\theta^2}\notag\\
 &+ 
	C_yl_2 \sin\theta \sqrt{{(C_z\cdot z)}^2 + {C_x}^2\cdot l_2^2 \cos\theta^2} \notag\\
&+	l_2 C_z z \sqrt{{C_x}^2\cos\theta^2 + {C_y}^2\sin\theta^2} \notag\\
 &+ 
	2  C_xl_1\sqrt{{(C_z\cdot z)}^2 + {C_y}^2\cdot l_2^2 \sin\theta^2} \notag\\
 &- 
	C_xl_2 \cos\theta \sqrt{{(C_z\cdot z)}^2 + {C_y}^2\cdot l_2^2 \sin\theta^2}\bigg),\\
	v&=C_xC_yC_z\cdot2z (l_1 - l_2 \cos\theta) (l_3 + l_2 \sin\theta).
\end{align}
\endgroup
		\nocite{*}
		
		\bibliography{3D_ordered_vertex_model_moduli_arXiv_submit_v1}

\begin{thebibliography}{53}%
\makeatletter
\providecommand \@ifxundefined [1]{%
 \@ifx{#1\undefined}
}%
\providecommand \@ifnum [1]{%
 \ifnum #1\expandafter \@firstoftwo
 \else \expandafter \@secondoftwo
 \fi
}%
\providecommand \@ifx [1]{%
 \ifx #1\expandafter \@firstoftwo
 \else \expandafter \@secondoftwo
 \fi
}%
\providecommand \natexlab [1]{#1}%
\providecommand \enquote  [1]{``#1''}%
\providecommand \bibnamefont  [1]{#1}%
\providecommand \bibfnamefont [1]{#1}%
\providecommand \citenamefont [1]{#1}%
\providecommand \href@noop [0]{\@secondoftwo}%
\providecommand \href [0]{\begingroup \@sanitize@url \@href}%
\providecommand \@href[1]{\@@startlink{#1}\@@href}%
\providecommand \@@href[1]{\endgroup#1\@@endlink}%
\providecommand \@sanitize@url [0]{\catcode `\\12\catcode `\$12\catcode
  `\&12\catcode `\#12\catcode `\^12\catcode `\_12\catcode `\%12\relax}%
\providecommand \@@startlink[1]{}%
\providecommand \@@endlink[0]{}%
\providecommand \url  [0]{\begingroup\@sanitize@url \@url }%
\providecommand \@url [1]{\endgroup\@href {#1}{\urlprefix }}%
\providecommand \urlprefix  [0]{URL }%
\providecommand \Eprint [0]{\href }%
\providecommand \doibase [0]{https://doi.org/}%
\providecommand \selectlanguage [0]{\@gobble}%
\providecommand \bibinfo  [0]{\@secondoftwo}%
\providecommand \bibfield  [0]{\@secondoftwo}%
\providecommand \translation [1]{[#1]}%
\providecommand \BibitemOpen [0]{}%
\providecommand \bibitemStop [0]{}%
\providecommand \bibitemNoStop [0]{.\EOS\space}%
\providecommand \EOS [0]{\spacefactor3000\relax}%
\providecommand \BibitemShut  [1]{\csname bibitem#1\endcsname}%
\let\auto@bib@innerbib\@empty
%</preamble>
\bibitem [{\citenamefont {Kelvin}(1887)}]{Kelvin1887}%
  \BibitemOpen
  \bibfield  {author} {\bibinfo {author} {\bibfnamefont {L.}~\bibnamefont
  {Kelvin}},\ }\href@noop {} {\bibfield  {journal} {\bibinfo  {journal} {Phil.
  Mag.}\ }\textbf {\bibinfo {volume} {24}},\ \bibinfo {pages} {503} (\bibinfo
  {year} {1887})}\BibitemShut {NoStop}%
\bibitem [{\citenamefont {Weaire}\ and\ \citenamefont
  {Phelan}(1994)}]{Weaire1994}%
  \BibitemOpen
  \bibfield  {author} {\bibinfo {author} {\bibfnamefont {D.}~\bibnamefont
  {Weaire}}\ and\ \bibinfo {author} {\bibfnamefont {R.}~\bibnamefont
  {Phelan}},\ }\href@noop {} {\bibfield  {journal} {\bibinfo  {journal}
  {Philosophical Magazine Letters}\ }\textbf {\bibinfo {volume} {69}},\
  \bibinfo {pages} {107} (\bibinfo {year} {1994})}\BibitemShut {NoStop}%
\bibitem [{\citenamefont {Bi}\ \emph {et~al.}(2015)\citenamefont {Bi},
  \citenamefont {Lopez}, \citenamefont {Schwarz},\ and\ \citenamefont
  {Manning}}]{Bi_2015}%
  \BibitemOpen
  \bibfield  {author} {\bibinfo {author} {\bibfnamefont {D.}~\bibnamefont
  {Bi}}, \bibinfo {author} {\bibfnamefont {J.~H.}\ \bibnamefont {Lopez}},
  \bibinfo {author} {\bibfnamefont {J.~M.}\ \bibnamefont {Schwarz}},\ and\
  \bibinfo {author} {\bibfnamefont {M.~L.}\ \bibnamefont {Manning}},\ }\href
  {https://doi.org/10.1038/nphys3471} {\bibfield  {journal} {\bibinfo
  {journal} {Nature Physics}\ }\textbf {\bibinfo {volume} {11}},\ \bibinfo
  {pages} {1074} (\bibinfo {year} {2015})}\BibitemShut {NoStop}%
\bibitem [{\citenamefont {Merkel}\ and\ \citenamefont
  {Manning}(2018)}]{Merkel_2018}%
  \BibitemOpen
  \bibfield  {author} {\bibinfo {author} {\bibfnamefont {M.}~\bibnamefont
  {Merkel}}\ and\ \bibinfo {author} {\bibfnamefont {M.~L.}\ \bibnamefont
  {Manning}},\ }\href {https://doi.org/10.1088/1367-2630/aaaa13} {\bibfield
  {journal} {\bibinfo  {journal} {New Journal of Physics}\ }\textbf {\bibinfo
  {volume} {20}},\ \bibinfo {pages} {022002} (\bibinfo {year}
  {2018})}\BibitemShut {NoStop}%
\bibitem [{\citenamefont {Zhang}\ and\ \citenamefont
  {Schwarz}(2022)}]{Zhang2023}%
  \BibitemOpen
  \bibfield  {author} {\bibinfo {author} {\bibfnamefont {T.}~\bibnamefont
  {Zhang}}\ and\ \bibinfo {author} {\bibfnamefont {J.~M.}\ \bibnamefont
  {Schwarz}},\ }\href {https://doi.org/10.1103/PhysRevResearch.4.043148}
  {\bibfield  {journal} {\bibinfo  {journal} {Phys. Rev. Res.}\ }\textbf
  {\bibinfo {volume} {4}},\ \bibinfo {pages} {043148} (\bibinfo {year}
  {2022})}\BibitemShut {NoStop}%
\bibitem [{\citenamefont {Cantat}\ \emph {et~al.}(2013)\citenamefont {Cantat},
  \citenamefont {Cohen-Addad}, \citenamefont {Elias}, \citenamefont {Graner},
  \citenamefont {H{\"o}hler}, \citenamefont {Pitois}, \citenamefont {Rouyer},\
  and\ \citenamefont {Saint-Jalmes}}]{Cantat2013}%
  \BibitemOpen
  \bibfield  {author} {\bibinfo {author} {\bibfnamefont {I.}~\bibnamefont
  {Cantat}}, \bibinfo {author} {\bibfnamefont {S.}~\bibnamefont {Cohen-Addad}},
  \bibinfo {author} {\bibfnamefont {F.}~\bibnamefont {Elias}}, \bibinfo
  {author} {\bibfnamefont {F.}~\bibnamefont {Graner}}, \bibinfo {author}
  {\bibfnamefont {R.}~\bibnamefont {H{\"o}hler}}, \bibinfo {author}
  {\bibfnamefont {O.}~\bibnamefont {Pitois}}, \bibinfo {author} {\bibfnamefont
  {F.}~\bibnamefont {Rouyer}},\ and\ \bibinfo {author} {\bibfnamefont
  {A.}~\bibnamefont {Saint-Jalmes}},\ }\href@noop {} {\emph {\bibinfo {title}
  {Foams: structure and dynamics}}}\ (\bibinfo  {publisher} {OUP Oxford},\
  \bibinfo {year} {2013})\BibitemShut {NoStop}%
\bibitem [{\citenamefont {Honda}\ \emph {et~al.}(1982)\citenamefont {Honda},
  \citenamefont {Ogita}, \citenamefont {Higuchi},\ and\ \citenamefont
  {Kani}}]{Honda1982}%
  \BibitemOpen
  \bibfield  {author} {\bibinfo {author} {\bibfnamefont {H.}~\bibnamefont
  {Honda}}, \bibinfo {author} {\bibfnamefont {Y.}~\bibnamefont {Ogita}},
  \bibinfo {author} {\bibfnamefont {S.}~\bibnamefont {Higuchi}},\ and\ \bibinfo
  {author} {\bibfnamefont {K.}~\bibnamefont {Kani}},\ }\href@noop {} {\bibfield
   {journal} {\bibinfo  {journal} {Journal of Morphology}\ }\textbf {\bibinfo
  {volume} {174}},\ \bibinfo {pages} {25} (\bibinfo {year} {1982})}\BibitemShut
  {NoStop}%
\bibitem [{\citenamefont {Honda}\ and\ \citenamefont
  {Nagai}(2022)}]{Honda2022}%
  \BibitemOpen
  \bibfield  {author} {\bibinfo {author} {\bibfnamefont {H.}~\bibnamefont
  {Honda}}\ and\ \bibinfo {author} {\bibfnamefont {T.}~\bibnamefont {Nagai}},\
  }\href@noop {} {\emph {\bibinfo {title} {Mathematical Models of Cell-Based
  Morphogenesis}}}\ (\bibinfo  {publisher} {Springer},\ \bibinfo {year}
  {2022})\BibitemShut {NoStop}%
\bibitem [{\citenamefont {Farhadifar}\ \emph {et~al.}(2007)\citenamefont
  {Farhadifar}, \citenamefont {R\"oper}, \citenamefont {Aigouy}, \citenamefont
  {Eaton},\ and\ \citenamefont {J\"ulicher}}]{Farhadifar2007}%
  \BibitemOpen
  \bibfield  {author} {\bibinfo {author} {\bibfnamefont {R.}~\bibnamefont
  {Farhadifar}}, \bibinfo {author} {\bibfnamefont {J.-C.}\ \bibnamefont
  {R\"oper}}, \bibinfo {author} {\bibfnamefont {B.}~\bibnamefont {Aigouy}},
  \bibinfo {author} {\bibfnamefont {S.}~\bibnamefont {Eaton}},\ and\ \bibinfo
  {author} {\bibfnamefont {F.}~\bibnamefont {J\"ulicher}},\ }\href
  {https://doi.org/https://doi.org/10.1016/j.cub.2007.11.049} {\bibfield
  {journal} {\bibinfo  {journal} {Current Biology}\ }\textbf {\bibinfo {volume}
  {17}},\ \bibinfo {pages} {2095} (\bibinfo {year} {2007})}\BibitemShut
  {NoStop}%
\bibitem [{\citenamefont {Staple}\ \emph {et~al.}(2010)\citenamefont {Staple},
  \citenamefont {Farhadifar}, \citenamefont {Röper}, \citenamefont {Aigouy},
  \citenamefont {Eaton},\ and\ \citenamefont {Jülicher}}]{Staple2010}%
  \BibitemOpen
  \bibfield  {author} {\bibinfo {author} {\bibfnamefont {D.~B.}\ \bibnamefont
  {Staple}}, \bibinfo {author} {\bibfnamefont {R.}~\bibnamefont {Farhadifar}},
  \bibinfo {author} {\bibfnamefont {J.~C.}\ \bibnamefont {Röper}}, \bibinfo
  {author} {\bibfnamefont {B.}~\bibnamefont {Aigouy}}, \bibinfo {author}
  {\bibfnamefont {S.}~\bibnamefont {Eaton}},\ and\ \bibinfo {author}
  {\bibfnamefont {F.}~\bibnamefont {Jülicher}},\ }\href
  {https://doi.org/10.1140/epje/i2010-10677-0} {\bibfield  {journal} {\bibinfo
  {journal} {The European Physical Journal E}\ }\textbf {\bibinfo {volume}
  {33}},\ \bibinfo {pages} {117} (\bibinfo {year} {2010})}\BibitemShut
  {NoStop}%
\bibitem [{\citenamefont {Mirams}\ \emph {et~al.}(2013)\citenamefont {Mirams},
  \citenamefont {Arthurs}, \citenamefont {Bernabeu}, \citenamefont {Bordas},
  \citenamefont {Cooper}, \citenamefont {Corrias}, \citenamefont {Davit},
  \citenamefont {Dunn}, \citenamefont {Fletcher}, \citenamefont {Harvey} \emph
  {et~al.}}]{Mirams2013}%
  \BibitemOpen
  \bibfield  {author} {\bibinfo {author} {\bibfnamefont {G.~R.}\ \bibnamefont
  {Mirams}}, \bibinfo {author} {\bibfnamefont {C.~J.}\ \bibnamefont {Arthurs}},
  \bibinfo {author} {\bibfnamefont {M.~O.}\ \bibnamefont {Bernabeu}}, \bibinfo
  {author} {\bibfnamefont {R.}~\bibnamefont {Bordas}}, \bibinfo {author}
  {\bibfnamefont {J.}~\bibnamefont {Cooper}}, \bibinfo {author} {\bibfnamefont
  {A.}~\bibnamefont {Corrias}}, \bibinfo {author} {\bibfnamefont
  {Y.}~\bibnamefont {Davit}}, \bibinfo {author} {\bibfnamefont {S.-J.}\
  \bibnamefont {Dunn}}, \bibinfo {author} {\bibfnamefont {A.~G.}\ \bibnamefont
  {Fletcher}}, \bibinfo {author} {\bibfnamefont {D.~G.}\ \bibnamefont
  {Harvey}}, \emph {et~al.},\ }\href@noop {} {\bibfield  {journal} {\bibinfo
  {journal} {PLoS Computational Biology}\ }\textbf {\bibinfo {volume} {9}},\
  \bibinfo {pages} {e1002970} (\bibinfo {year} {2013})}\BibitemShut {NoStop}%
\bibitem [{\citenamefont {Sussman}(2017)}]{Sussman2017}%
  \BibitemOpen
  \bibfield  {author} {\bibinfo {author} {\bibfnamefont {D.~M.}\ \bibnamefont
  {Sussman}},\ }\href@noop {} {\bibfield  {journal} {\bibinfo  {journal}
  {Computer Physics Communications}\ }\textbf {\bibinfo {volume} {219}},\
  \bibinfo {pages} {400} (\bibinfo {year} {2017})}\BibitemShut {NoStop}%
\bibitem [{\citenamefont {Moshe}\ \emph {et~al.}(2018)\citenamefont {Moshe},
  \citenamefont {Bowick},\ and\ \citenamefont {Marchetti}}]{Moshe2018}%
  \BibitemOpen
  \bibfield  {author} {\bibinfo {author} {\bibfnamefont {M.}~\bibnamefont
  {Moshe}}, \bibinfo {author} {\bibfnamefont {M.~J.}\ \bibnamefont {Bowick}},\
  and\ \bibinfo {author} {\bibfnamefont {M.~C.}\ \bibnamefont {Marchetti}},\
  }\href@noop {} {\bibfield  {journal} {\bibinfo  {journal} {Physical Review
  Letters}\ }\textbf {\bibinfo {volume} {120}},\ \bibinfo {pages} {268105}
  (\bibinfo {year} {2018})}\BibitemShut {NoStop}%
\bibitem [{\citenamefont {Sahu}\ \emph {et~al.}(2020)\citenamefont {Sahu},
  \citenamefont {Sussman}, \citenamefont {R{\"u}bsam}, \citenamefont {Mertz},
  \citenamefont {Horsley}, \citenamefont {Dufresne}, \citenamefont {Niessen},
  \citenamefont {Marchetti}, \citenamefont {Manning},\ and\ \citenamefont
  {Schwarz}}]{Sahu2020}%
  \BibitemOpen
  \bibfield  {author} {\bibinfo {author} {\bibfnamefont {P.}~\bibnamefont
  {Sahu}}, \bibinfo {author} {\bibfnamefont {D.~M.}\ \bibnamefont {Sussman}},
  \bibinfo {author} {\bibfnamefont {M.}~\bibnamefont {R{\"u}bsam}}, \bibinfo
  {author} {\bibfnamefont {A.~F.}\ \bibnamefont {Mertz}}, \bibinfo {author}
  {\bibfnamefont {V.}~\bibnamefont {Horsley}}, \bibinfo {author} {\bibfnamefont
  {E.~R.}\ \bibnamefont {Dufresne}}, \bibinfo {author} {\bibfnamefont {C.~M.}\
  \bibnamefont {Niessen}}, \bibinfo {author} {\bibfnamefont {M.~C.}\
  \bibnamefont {Marchetti}}, \bibinfo {author} {\bibfnamefont {M.~L.}\
  \bibnamefont {Manning}},\ and\ \bibinfo {author} {\bibfnamefont {J.~M.}\
  \bibnamefont {Schwarz}},\ }\href@noop {} {\bibfield  {journal} {\bibinfo
  {journal} {Soft Matter}\ }\textbf {\bibinfo {volume} {16}},\ \bibinfo {pages}
  {3325} (\bibinfo {year} {2020})}\BibitemShut {NoStop}%
\bibitem [{\citenamefont {Huang}\ \emph {et~al.}(2022)\citenamefont {Huang},
  \citenamefont {Cochran}, \citenamefont {Fielding}, \citenamefont
  {Marchetti},\ and\ \citenamefont {Bi}}]{Huang2022}%
  \BibitemOpen
  \bibfield  {author} {\bibinfo {author} {\bibfnamefont {J.}~\bibnamefont
  {Huang}}, \bibinfo {author} {\bibfnamefont {J.~O.}\ \bibnamefont {Cochran}},
  \bibinfo {author} {\bibfnamefont {S.~M.}\ \bibnamefont {Fielding}}, \bibinfo
  {author} {\bibfnamefont {M.~C.}\ \bibnamefont {Marchetti}},\ and\ \bibinfo
  {author} {\bibfnamefont {D.}~\bibnamefont {Bi}},\ }\href@noop {} {\bibfield
  {journal} {\bibinfo  {journal} {Physical Review Letters}\ }\textbf {\bibinfo
  {volume} {128}},\ \bibinfo {pages} {178001} (\bibinfo {year}
  {2022})}\BibitemShut {NoStop}%
\bibitem [{\citenamefont {Chen}\ \emph {et~al.}(2022)\citenamefont {Chen},
  \citenamefont {Gao}, \citenamefont {Li}, \citenamefont {Mao}, \citenamefont
  {Tang},\ and\ \citenamefont {Jiang}}]{Chen2022}%
  \BibitemOpen
  \bibfield  {author} {\bibinfo {author} {\bibfnamefont {Y.}~\bibnamefont
  {Chen}}, \bibinfo {author} {\bibfnamefont {Q.}~\bibnamefont {Gao}}, \bibinfo
  {author} {\bibfnamefont {J.}~\bibnamefont {Li}}, \bibinfo {author}
  {\bibfnamefont {F.}~\bibnamefont {Mao}}, \bibinfo {author} {\bibfnamefont
  {R.}~\bibnamefont {Tang}},\ and\ \bibinfo {author} {\bibfnamefont
  {H.}~\bibnamefont {Jiang}},\ }\href@noop {} {\bibfield  {journal} {\bibinfo
  {journal} {Physical Review Letters}\ }\textbf {\bibinfo {volume} {128}},\
  \bibinfo {pages} {018101} (\bibinfo {year} {2022})}\BibitemShut {NoStop}%
\bibitem [{\citenamefont {Park}\ \emph {et~al.}(2015)\citenamefont {Park},
  \citenamefont {Kim}, \citenamefont {Bi}, \citenamefont {Mitchel},
  \citenamefont {Qazvini}, \citenamefont {Tantisira}, \citenamefont {Park},
  \citenamefont {McGill}, \citenamefont {Kim}, \citenamefont {Gweon},
  \citenamefont {Notbohm}, \citenamefont {Steward~Jr}, \citenamefont {Burger},
  \citenamefont {Randell}, \citenamefont {Kho}, \citenamefont {Tambe},
  \citenamefont {Hardin}, \citenamefont {Shore}, \citenamefont {Israel},
  \citenamefont {Weitz}, \citenamefont {Tschumperlin}, \citenamefont {Henske},
  \citenamefont {Weiss}, \citenamefont {Manning}, \citenamefont {Butler},
  \citenamefont {Drazen},\ and\ \citenamefont {Fredberg}}]{Park2015}%
  \BibitemOpen
  \bibfield  {author} {\bibinfo {author} {\bibfnamefont {J.-A.}\ \bibnamefont
  {Park}}, \bibinfo {author} {\bibfnamefont {J.~H.}\ \bibnamefont {Kim}},
  \bibinfo {author} {\bibfnamefont {D.}~\bibnamefont {Bi}}, \bibinfo {author}
  {\bibfnamefont {J.~A.}\ \bibnamefont {Mitchel}}, \bibinfo {author}
  {\bibfnamefont {N.~T.}\ \bibnamefont {Qazvini}}, \bibinfo {author}
  {\bibfnamefont {K.}~\bibnamefont {Tantisira}}, \bibinfo {author}
  {\bibfnamefont {C.~Y.}\ \bibnamefont {Park}}, \bibinfo {author}
  {\bibfnamefont {M.}~\bibnamefont {McGill}}, \bibinfo {author} {\bibfnamefont
  {S.-H.}\ \bibnamefont {Kim}}, \bibinfo {author} {\bibfnamefont
  {B.}~\bibnamefont {Gweon}}, \bibinfo {author} {\bibfnamefont
  {J.}~\bibnamefont {Notbohm}}, \bibinfo {author} {\bibfnamefont
  {R.}~\bibnamefont {Steward~Jr}}, \bibinfo {author} {\bibfnamefont
  {S.}~\bibnamefont {Burger}}, \bibinfo {author} {\bibfnamefont {S.~H.}\
  \bibnamefont {Randell}}, \bibinfo {author} {\bibfnamefont {A.~T.}\
  \bibnamefont {Kho}}, \bibinfo {author} {\bibfnamefont {D.~T.}\ \bibnamefont
  {Tambe}}, \bibinfo {author} {\bibfnamefont {C.}~\bibnamefont {Hardin}},
  \bibinfo {author} {\bibfnamefont {S.~A.}\ \bibnamefont {Shore}}, \bibinfo
  {author} {\bibfnamefont {E.}~\bibnamefont {Israel}}, \bibinfo {author}
  {\bibfnamefont {D.~A.}\ \bibnamefont {Weitz}}, \bibinfo {author}
  {\bibfnamefont {D.~J.}\ \bibnamefont {Tschumperlin}}, \bibinfo {author}
  {\bibfnamefont {E.}~\bibnamefont {Henske}}, \bibinfo {author} {\bibfnamefont
  {S.~T.}\ \bibnamefont {Weiss}}, \bibinfo {author} {\bibfnamefont {M.~L.}\
  \bibnamefont {Manning}}, \bibinfo {author} {\bibfnamefont {J.~P.}\
  \bibnamefont {Butler}}, \bibinfo {author} {\bibfnamefont {J.~M.}\
  \bibnamefont {Drazen}},\ and\ \bibinfo {author} {\bibfnamefont {J.~J.}\
  \bibnamefont {Fredberg}},\ }\href
  {https://doi.org/https://doi.org/10.1038/nmat4357} {\bibfield  {journal}
  {\bibinfo  {journal} {Nature Materials}\ }\textbf {\bibinfo {volume} {14}},\
  \bibinfo {pages} {1040} (\bibinfo {year} {2015})}\BibitemShut {NoStop}%
\bibitem [{\citenamefont {Bi}\ \emph {et~al.}(2016)\citenamefont {Bi},
  \citenamefont {Yang}, \citenamefont {Marchetti},\ and\ \citenamefont
  {Manning}}]{Bi2016}%
  \BibitemOpen
  \bibfield  {author} {\bibinfo {author} {\bibfnamefont {D.}~\bibnamefont
  {Bi}}, \bibinfo {author} {\bibfnamefont {X.}~\bibnamefont {Yang}}, \bibinfo
  {author} {\bibfnamefont {M.~C.}\ \bibnamefont {Marchetti}},\ and\ \bibinfo
  {author} {\bibfnamefont {M.~L.}\ \bibnamefont {Manning}},\ }\href
  {https://doi.org/10.1103/PhysRevX.6.021011} {\bibfield  {journal} {\bibinfo
  {journal} {Phys. Rev. X}\ }\textbf {\bibinfo {volume} {6}},\ \bibinfo {pages}
  {021011} (\bibinfo {year} {2016})}\BibitemShut {NoStop}%
\bibitem [{\citenamefont {Barton}\ \emph {et~al.}(2017)\citenamefont {Barton},
  \citenamefont {Henkes}, \citenamefont {Weijer},\ and\ \citenamefont
  {Sknepnek}}]{Barton2017}%
  \BibitemOpen
  \bibfield  {author} {\bibinfo {author} {\bibfnamefont {D.~L.}\ \bibnamefont
  {Barton}}, \bibinfo {author} {\bibfnamefont {S.}~\bibnamefont {Henkes}},
  \bibinfo {author} {\bibfnamefont {C.~J.}\ \bibnamefont {Weijer}},\ and\
  \bibinfo {author} {\bibfnamefont {R.}~\bibnamefont {Sknepnek}},\ }\href@noop
  {} {\bibfield  {journal} {\bibinfo  {journal} {PLoS Computational Biology}\
  }\textbf {\bibinfo {volume} {13}},\ \bibinfo {pages} {e1005569} (\bibinfo
  {year} {2017})}\BibitemShut {NoStop}%
\bibitem [{\citenamefont {Okuda}\ \emph {et~al.}(2012)\citenamefont {Okuda},
  \citenamefont {Inoue}, \citenamefont {Eiraku}, \citenamefont {Sasai},\ and\
  \citenamefont {Adachi}}]{Okuda2012}%
  \BibitemOpen
  \bibfield  {author} {\bibinfo {author} {\bibfnamefont {S.}~\bibnamefont
  {Okuda}}, \bibinfo {author} {\bibfnamefont {Y.}~\bibnamefont {Inoue}},
  \bibinfo {author} {\bibfnamefont {M.}~\bibnamefont {Eiraku}}, \bibinfo
  {author} {\bibfnamefont {Y.}~\bibnamefont {Sasai}},\ and\ \bibinfo {author}
  {\bibfnamefont {T.}~\bibnamefont {Adachi}},\ }\href
  {https://doi.org/10.1007/s10237-012-0430-7} {\bibfield  {journal} {\bibinfo
  {journal} {Biomechanics and Modeling in Mechanobiology}\ }\textbf {\bibinfo
  {volume} {12}} (\bibinfo {year} {2012})}\BibitemShut {NoStop}%
\bibitem [{\citenamefont {Okuda}\ and\ \citenamefont {Sato}(2022)}]{Okuda2022}%
  \BibitemOpen
  \bibfield  {author} {\bibinfo {author} {\bibfnamefont {S.}~\bibnamefont
  {Okuda}}\ and\ \bibinfo {author} {\bibfnamefont {K.}~\bibnamefont {Sato}},\
  }\href@noop {} {\bibfield  {journal} {\bibinfo  {journal} {Biophysical
  Journal}\ }\textbf {\bibinfo {volume} {121}},\ \bibinfo {pages} {1856}
  (\bibinfo {year} {2022})}\BibitemShut {NoStop}%
\bibitem [{\citenamefont {Okuda}\ \emph {et~al.}(2018)\citenamefont {Okuda},
  \citenamefont {Miura}, \citenamefont {Inoue}, \citenamefont {Adachi},\ and\
  \citenamefont {Eiraku}}]{Okuda2018}%
  \BibitemOpen
  \bibfield  {author} {\bibinfo {author} {\bibfnamefont {S.}~\bibnamefont
  {Okuda}}, \bibinfo {author} {\bibfnamefont {T.}~\bibnamefont {Miura}},
  \bibinfo {author} {\bibfnamefont {Y.}~\bibnamefont {Inoue}}, \bibinfo
  {author} {\bibfnamefont {T.}~\bibnamefont {Adachi}},\ and\ \bibinfo {author}
  {\bibfnamefont {M.}~\bibnamefont {Eiraku}},\ }\href@noop {} {\bibfield
  {journal} {\bibinfo  {journal} {Scientific Reports}\ }\textbf {\bibinfo
  {volume} {8}},\ \bibinfo {pages} {2386} (\bibinfo {year} {2018})}\BibitemShut
  {NoStop}%
\bibitem [{\citenamefont {Rozman}\ \emph {et~al.}(2020)\citenamefont {Rozman},
  \citenamefont {Krajnc},\ and\ \citenamefont {Ziherl}}]{Rozman2020}%
  \BibitemOpen
  \bibfield  {author} {\bibinfo {author} {\bibfnamefont {J.}~\bibnamefont
  {Rozman}}, \bibinfo {author} {\bibfnamefont {M.}~\bibnamefont {Krajnc}},\
  and\ \bibinfo {author} {\bibfnamefont {P.}~\bibnamefont {Ziherl}},\
  }\href@noop {} {\bibfield  {journal} {\bibinfo  {journal} {Nature
  Communications}\ }\textbf {\bibinfo {volume} {11}},\ \bibinfo {pages} {3805}
  (\bibinfo {year} {2020})}\BibitemShut {NoStop}%
\bibitem [{\citenamefont {Zhang}\ \emph {et~al.}(2023)\citenamefont {Zhang},
  \citenamefont {Gupta}, \citenamefont {Lancaster},\ and\ \citenamefont
  {Schwarz}}]{ZhangSarthak2023}%
  \BibitemOpen
  \bibfield  {author} {\bibinfo {author} {\bibfnamefont {T.}~\bibnamefont
  {Zhang}}, \bibinfo {author} {\bibfnamefont {S.}~\bibnamefont {Gupta}},
  \bibinfo {author} {\bibfnamefont {M.~A.}\ \bibnamefont {Lancaster}},\ and\
  \bibinfo {author} {\bibfnamefont {J.~M.}\ \bibnamefont {Schwarz}},\
  }\href@noop {} {\bibfield  {journal} {\bibinfo  {journal} {bioRxiv}\ ,\
  \bibinfo {pages} {2023}} (\bibinfo {year} {2023})}\BibitemShut {NoStop}%
\bibitem [{\citenamefont {Staddon}\ \emph {et~al.}(2023)\citenamefont
  {Staddon}, \citenamefont {Hernandez}, \citenamefont {Bowick}, \citenamefont
  {Moshe},\ and\ \citenamefont {Marchetti}}]{Staddon2023}%
  \BibitemOpen
  \bibfield  {author} {\bibinfo {author} {\bibfnamefont {M.~F.}\ \bibnamefont
  {Staddon}}, \bibinfo {author} {\bibfnamefont {A.}~\bibnamefont {Hernandez}},
  \bibinfo {author} {\bibfnamefont {M.~J.}\ \bibnamefont {Bowick}}, \bibinfo
  {author} {\bibfnamefont {M.}~\bibnamefont {Moshe}},\ and\ \bibinfo {author}
  {\bibfnamefont {M.~C.}\ \bibnamefont {Marchetti}},\ }\href@noop {} {\bibfield
   {journal} {\bibinfo  {journal} {Soft Matter}\ }\textbf {\bibinfo {volume}
  {19}},\ \bibinfo {pages} {3080} (\bibinfo {year} {2023})}\BibitemShut
  {NoStop}%
\bibitem [{\citenamefont {Damavandi}\ \emph {et~al.}(2022)\citenamefont
  {Damavandi}, \citenamefont {Manning},\ and\ \citenamefont
  {Schwarz}}]{Damavandi2022}%
  \BibitemOpen
  \bibfield  {author} {\bibinfo {author} {\bibfnamefont {O.~K.}\ \bibnamefont
  {Damavandi}}, \bibinfo {author} {\bibfnamefont {M.~L.}\ \bibnamefont
  {Manning}},\ and\ \bibinfo {author} {\bibfnamefont {J.~M.}\ \bibnamefont
  {Schwarz}},\ }\href@noop {} {\bibfield  {journal} {\bibinfo  {journal}
  {Europhysics Letters}\ }\textbf {\bibinfo {volume} {138}},\ \bibinfo {pages}
  {27001} (\bibinfo {year} {2022})}\BibitemShut {NoStop}%
\bibitem [{\citenamefont {Inc.}(2022)}]{Mathematica}%
  \BibitemOpen
  \bibfield  {author} {\bibinfo {author} {\bibfnamefont {W.~R.}\ \bibnamefont
  {Inc.}},\ }\href {https://www.wolfram.com/mathematica} {\bibinfo {title}
  {Mathematica, {V}ersion 13.2}} (\bibinfo {year} {2022}),\ \bibinfo {note}
  {champaign, IL, 2022}\BibitemShut {NoStop}%
\bibitem [{\citenamefont {Alt}\ \emph {et~al.}(2017)\citenamefont {Alt},
  \citenamefont {Ganguly},\ and\ \citenamefont {Salbreux}}]{Silvanus2017}%
  \BibitemOpen
  \bibfield  {author} {\bibinfo {author} {\bibfnamefont {S.}~\bibnamefont
  {Alt}}, \bibinfo {author} {\bibfnamefont {P.}~\bibnamefont {Ganguly}},\ and\
  \bibinfo {author} {\bibfnamefont {G.}~\bibnamefont {Salbreux}},\ }\href@noop
  {} {\bibfield  {journal} {\bibinfo  {journal} {Philosophical Transactions of
  the Royal Society B: Biological Sciences}\ }\textbf {\bibinfo {volume}
  {372}},\ \bibinfo {pages} {20150520} (\bibinfo {year} {2017})}\BibitemShut
  {NoStop}%
\bibitem [{\citenamefont {G\'omez-G\'alvez}\ \emph {et~al.}(2018)\citenamefont
  {G\'omez-G\'alvez}, \citenamefont {Vicente-Munuera}, \citenamefont {Tagua},
  \citenamefont {Forja}, \citenamefont {Castro}, \citenamefont {Letr\'an},
  \citenamefont {Valencia-Exp\'osito}, \citenamefont {Grima}, \citenamefont
  {Berm\'udez-Gallardo}, \citenamefont {Serrano-Pérez-Higueras}, \citenamefont
  {Cavodeassi}, \citenamefont {Sotillos}, \citenamefont {Mart\'in-Bermudo},
  \citenamefont {M\'arquez}, \citenamefont {Buceta},\ and\ \citenamefont
  {Escudero}}]{Gomez-Galvez2018}%
  \BibitemOpen
  \bibfield  {author} {\bibinfo {author} {\bibfnamefont {P.}~\bibnamefont
  {G\'omez-G\'alvez}}, \bibinfo {author} {\bibfnamefont {P.}~\bibnamefont
  {Vicente-Munuera}}, \bibinfo {author} {\bibfnamefont {A.}~\bibnamefont
  {Tagua}}, \bibinfo {author} {\bibfnamefont {C.}~\bibnamefont {Forja}},
  \bibinfo {author} {\bibfnamefont {A.~M.}\ \bibnamefont {Castro}}, \bibinfo
  {author} {\bibfnamefont {M.}~\bibnamefont {Letr\'an}}, \bibinfo {author}
  {\bibfnamefont {A.}~\bibnamefont {Valencia-Exp\'osito}}, \bibinfo {author}
  {\bibfnamefont {C.}~\bibnamefont {Grima}}, \bibinfo {author} {\bibfnamefont
  {M.}~\bibnamefont {Berm\'udez-Gallardo}}, \bibinfo {author} {\bibfnamefont
  {O.}~\bibnamefont {Serrano-Pérez-Higueras}}, \bibinfo {author}
  {\bibfnamefont {F.}~\bibnamefont {Cavodeassi}}, \bibinfo {author}
  {\bibfnamefont {S.}~\bibnamefont {Sotillos}}, \bibinfo {author}
  {\bibfnamefont {M.~D.}\ \bibnamefont {Mart\'in-Bermudo}}, \bibinfo {author}
  {\bibfnamefont {A.}~\bibnamefont {M\'arquez}}, \bibinfo {author}
  {\bibfnamefont {J.}~\bibnamefont {Buceta}},\ and\ \bibinfo {author}
  {\bibfnamefont {L.~M.}\ \bibnamefont {Escudero}},\ }\href
  {https://doi.org/10.1038/s41467-018-05376-1} {\bibfield  {journal} {\bibinfo
  {journal} {Nature Communications}\ }\textbf {\bibinfo {volume} {9}},\
  \bibinfo {pages} {2960} (\bibinfo {year} {2018})}\BibitemShut {NoStop}%
\bibitem [{\citenamefont {Lemke}\ and\ \citenamefont
  {Nelson}(2021)}]{LEMKE2021R1098}%
  \BibitemOpen
  \bibfield  {author} {\bibinfo {author} {\bibfnamefont {S.~B.}\ \bibnamefont
  {Lemke}}\ and\ \bibinfo {author} {\bibfnamefont {C.~M.}\ \bibnamefont
  {Nelson}},\ }\href
  {https://doi.org/https://doi.org/10.1016/j.cub.2021.07.078} {\bibfield
  {journal} {\bibinfo  {journal} {Current Biology}\ }\textbf {\bibinfo {volume}
  {31}},\ \bibinfo {pages} {R1098} (\bibinfo {year} {2021})}\BibitemShut
  {NoStop}%
\bibitem [{\citenamefont {Elechiguerra}\ \emph {et~al.}(2006)\citenamefont
  {Elechiguerra}, \citenamefont {Reyes-Gasga},\ and\ \citenamefont
  {Yacaman}}]{Elechiguerra2016}%
  \BibitemOpen
  \bibfield  {author} {\bibinfo {author} {\bibfnamefont {J.}~\bibnamefont
  {Elechiguerra}}, \bibinfo {author} {\bibfnamefont {J.}~\bibnamefont
  {Reyes-Gasga}},\ and\ \bibinfo {author} {\bibfnamefont {M.}~\bibnamefont
  {Yacaman}},\ }\href {https://doi.org/10.1039/B607128G} {\bibfield  {journal}
  {\bibinfo  {journal} {J. Mater. Chem.}\ }\textbf {\bibinfo {volume} {16}},\
  \bibinfo {pages} {3906} (\bibinfo {year} {2006})}\BibitemShut {NoStop}%
\bibitem [{\citenamefont {Alcinesio}\ \emph {et~al.}(2020)\citenamefont
  {Alcinesio}, \citenamefont {Meacock}, \citenamefont {Allan}, \citenamefont
  {Monico}, \citenamefont {Restrepo~Schild}, \citenamefont {Cazimoglu},
  \citenamefont {Cornall}, \citenamefont {Krishna~Kumar},\ and\ \citenamefont
  {Bayley}}]{Alcinesio2020}%
  \BibitemOpen
  \bibfield  {author} {\bibinfo {author} {\bibfnamefont {A.}~\bibnamefont
  {Alcinesio}}, \bibinfo {author} {\bibfnamefont {O.~J.}\ \bibnamefont
  {Meacock}}, \bibinfo {author} {\bibfnamefont {R.~G.}\ \bibnamefont {Allan}},
  \bibinfo {author} {\bibfnamefont {C.}~\bibnamefont {Monico}}, \bibinfo
  {author} {\bibfnamefont {V.}~\bibnamefont {Restrepo~Schild}}, \bibinfo
  {author} {\bibfnamefont {I.}~\bibnamefont {Cazimoglu}}, \bibinfo {author}
  {\bibfnamefont {M.~T.}\ \bibnamefont {Cornall}}, \bibinfo {author}
  {\bibfnamefont {R.}~\bibnamefont {Krishna~Kumar}},\ and\ \bibinfo {author}
  {\bibfnamefont {H.}~\bibnamefont {Bayley}},\ }\href@noop {} {\bibfield
  {journal} {\bibinfo  {journal} {Nature Communications}\ }\textbf {\bibinfo
  {volume} {11}} (\bibinfo {year} {2020})}\BibitemShut {NoStop}%
\bibitem [{\citenamefont {Cauchy}(1813)}]{Cauchy1813}%
  \BibitemOpen
  \bibfield  {author} {\bibinfo {author} {\bibfnamefont {A.~L.}\ \bibnamefont
  {Cauchy}},\ }\href@noop {} {\bibfield  {journal} {\bibinfo  {journal} {J.
  Ecole Polytechnique}\ }\textbf {\bibinfo {volume} {9}},\ \bibinfo {pages}
  {87} (\bibinfo {year} {1813})}\BibitemShut {NoStop}%
\bibitem [{\citenamefont {Alexandrov}(2005)}]{Alexandrov2005}%
  \BibitemOpen
  \bibfield  {author} {\bibinfo {author} {\bibfnamefont {A.~D.}\ \bibnamefont
  {Alexandrov}},\ }\href@noop {} {\emph {\bibinfo {title} {Convex
  polyhedra}}},\ Vol.\ \bibinfo {volume} {109}\ (\bibinfo  {publisher}
  {Springer},\ \bibinfo {year} {2005})\BibitemShut {NoStop}%
\bibitem [{\citenamefont {Gandikota}\ \emph {et~al.}(2022)\citenamefont
  {Gandikota}, \citenamefont {Parker},\ and\ \citenamefont
  {Schwarz}}]{Gandikota2022}%
  \BibitemOpen
  \bibfield  {author} {\bibinfo {author} {\bibfnamefont {M.}~\bibnamefont
  {Gandikota}}, \bibinfo {author} {\bibfnamefont {A.}~\bibnamefont {Parker}},\
  and\ \bibinfo {author} {\bibfnamefont {J.}~\bibnamefont {Schwarz}},\
  }\href@noop {} {\bibfield  {journal} {\bibinfo  {journal} {Physical Review
  E}\ }\textbf {\bibinfo {volume} {106}},\ \bibinfo {pages} {055003} (\bibinfo
  {year} {2022})}\BibitemShut {NoStop}%
\bibitem [{\citenamefont {Yan}\ and\ \citenamefont {Bi}(2019)}]{Yan2019}%
  \BibitemOpen
  \bibfield  {author} {\bibinfo {author} {\bibfnamefont {L.}~\bibnamefont
  {Yan}}\ and\ \bibinfo {author} {\bibfnamefont {D.}~\bibnamefont {Bi}},\
  }\href@noop {} {\bibfield  {journal} {\bibinfo  {journal} {Physical Review
  X}\ }\textbf {\bibinfo {volume} {9}},\ \bibinfo {pages} {011029} (\bibinfo
  {year} {2019})}\BibitemShut {NoStop}%
\bibitem [{\citenamefont {Smith}(1953)}]{SMITH1953295}%
  \BibitemOpen
  \bibfield  {author} {\bibinfo {author} {\bibfnamefont {C.}~\bibnamefont
  {Smith}},\ }\href
  {https://doi.org/https://doi.org/10.1016/0001-6160(53)90102-3} {\bibfield
  {journal} {\bibinfo  {journal} {Acta Metallurgica}\ }\textbf {\bibinfo
  {volume} {1}},\ \bibinfo {pages} {295} (\bibinfo {year} {1953})}\BibitemShut
  {NoStop}%
\bibitem [{\citenamefont {Ranganathan}\ and\ \citenamefont
  {Lord}(2008)}]{Ranganathan2008}%
  \BibitemOpen
  \bibfield  {author} {\bibinfo {author} {\bibfnamefont {S.}~\bibnamefont
  {Ranganathan}}\ and\ \bibinfo {author} {\bibfnamefont {E.}~\bibnamefont
  {Lord}},\ }\href {https://doi.org/10.1080/09500830802112173} {\bibfield
  {journal} {\bibinfo  {journal} {Philosophical Magazine Letters}\ }\textbf
  {\bibinfo {volume} {88}},\ \bibinfo {pages} {703} (\bibinfo {year}
  {2008})}\BibitemShut {NoStop}%
\bibitem [{\citenamefont {Honda}\ \emph {et~al.}(2004)\citenamefont {Honda},
  \citenamefont {Tanemura},\ and\ \citenamefont {Nagai}}]{HONDA2004439}%
  \BibitemOpen
  \bibfield  {author} {\bibinfo {author} {\bibfnamefont {H.}~\bibnamefont
  {Honda}}, \bibinfo {author} {\bibfnamefont {M.}~\bibnamefont {Tanemura}},\
  and\ \bibinfo {author} {\bibfnamefont {T.}~\bibnamefont {Nagai}},\ }\href
  {https://doi.org/https://doi.org/10.1016/j.jtbi.2003.10.001} {\bibfield
  {journal} {\bibinfo  {journal} {Journal of Theoretical Biology}\ }\textbf
  {\bibinfo {volume} {226}},\ \bibinfo {pages} {439} (\bibinfo {year}
  {2004})}\BibitemShut {NoStop}%
\bibitem [{\citenamefont {Williams}(1968)}]{Williams1968}%
  \BibitemOpen
  \bibfield  {author} {\bibinfo {author} {\bibfnamefont {R.~E.}\ \bibnamefont
  {Williams}},\ }\href {https://doi.org/10.1126/science.161.3838.276}
  {\bibfield  {journal} {\bibinfo  {journal} {Nature}\ ,\ \bibinfo {pages}
  {276}} (\bibinfo {year} {1968})}\BibitemShut {NoStop}%
\bibitem [{\citenamefont {Nogucci}(2019)}]{Nogucci2019}%
  \BibitemOpen
  \bibfield  {author} {\bibinfo {author} {\bibfnamefont {H.}~\bibnamefont
  {Nogucci}},\ }\href {https://doi.org/10.2142/biophysico.16.0_9} {\bibfield
  {journal} {\bibinfo  {journal} {Biophysics and Physicobiology}\ }\textbf
  {\bibinfo {volume} {16}},\ \bibinfo {pages} {9} (\bibinfo {year}
  {2019})}\BibitemShut {NoStop}%
\bibitem [{\citenamefont {Lenne}\ and\ \citenamefont
  {Trivedi}(2022)}]{Lenne2022}%
  \BibitemOpen
  \bibfield  {author} {\bibinfo {author} {\bibfnamefont {P.-F.}\ \bibnamefont
  {Lenne}}\ and\ \bibinfo {author} {\bibfnamefont {V.}~\bibnamefont
  {Trivedi}},\ }\href {https://doi.org/10.1038/s41467-022-28151-9} {\bibfield
  {journal} {\bibinfo  {journal} {Nature Communications}\ }\textbf {\bibinfo
  {volume} {13}},\ \bibinfo {pages} {2041} (\bibinfo {year}
  {2022})}\BibitemShut {NoStop}%
\bibitem [{\citenamefont {Angelini}\ \emph {et~al.}(2011)\citenamefont
  {Angelini}, \citenamefont {Hannezo}, \citenamefont {Trepat}, \citenamefont
  {Marquez}, \citenamefont {Fredberg},\ and\ \citenamefont
  {Weitz}}]{Angelini2011}%
  \BibitemOpen
  \bibfield  {author} {\bibinfo {author} {\bibfnamefont {T.~E.}\ \bibnamefont
  {Angelini}}, \bibinfo {author} {\bibfnamefont {E.}~\bibnamefont {Hannezo}},
  \bibinfo {author} {\bibfnamefont {X.}~\bibnamefont {Trepat}}, \bibinfo
  {author} {\bibfnamefont {M.}~\bibnamefont {Marquez}}, \bibinfo {author}
  {\bibfnamefont {J.~J.}\ \bibnamefont {Fredberg}},\ and\ \bibinfo {author}
  {\bibfnamefont {D.~A.}\ \bibnamefont {Weitz}},\ }\href
  {https://doi.org/10.1073/pnas.1010059108} {\bibfield  {journal} {\bibinfo
  {journal} {Proceedings of the National Academy of Sciences}\ }\textbf
  {\bibinfo {volume} {108}},\ \bibinfo {pages} {4714} (\bibinfo {year}
  {2011})}\BibitemShut {NoStop}%
\bibitem [{\citenamefont {Sch\"otz}\ \emph {et~al.}(2013)\citenamefont
  {Sch\"otz}, \citenamefont {Lanio}, \citenamefont {Talbot},\ and\
  \citenamefont {Manning}}]{Schotz2013}%
  \BibitemOpen
  \bibfield  {author} {\bibinfo {author} {\bibfnamefont {E.-M.}\ \bibnamefont
  {Sch\"otz}}, \bibinfo {author} {\bibfnamefont {M.}~\bibnamefont {Lanio}},
  \bibinfo {author} {\bibfnamefont {J.~A.}\ \bibnamefont {Talbot}},\ and\
  \bibinfo {author} {\bibfnamefont {M.~L.}\ \bibnamefont {Manning}},\
  }\href@noop {} {\bibfield  {journal} {\bibinfo  {journal} {Journal of The
  Royal Society Interface}\ }\textbf {\bibinfo {volume} {10}},\ \bibinfo
  {pages} {20130726} (\bibinfo {year} {2013})}\BibitemShut {NoStop}%
\bibitem [{\citenamefont {Friedl}\ and\ \citenamefont
  {Gilmour}(2009)}]{Friedl2009}%
  \BibitemOpen
  \bibfield  {author} {\bibinfo {author} {\bibfnamefont {P.}~\bibnamefont
  {Friedl}}\ and\ \bibinfo {author} {\bibfnamefont {D.}~\bibnamefont
  {Gilmour}},\ }\href {https://doi.org/https://doi.org/10.1038/nrm2720}
  {\bibfield  {journal} {\bibinfo  {journal} {Nature Reviews Molecular Cell
  Biology}\ }\textbf {\bibinfo {volume} {10}},\ \bibinfo {pages} {445}
  (\bibinfo {year} {2009})}\BibitemShut {NoStop}%
\bibitem [{\citenamefont {Sadati}\ \emph {et~al.}(2013)\citenamefont {Sadati},
  \citenamefont {{Taheri Qazvini}}, \citenamefont {Krishnan}, \citenamefont
  {Park},\ and\ \citenamefont {Fredberg}}]{SADATI2013121}%
  \BibitemOpen
  \bibfield  {author} {\bibinfo {author} {\bibfnamefont {M.}~\bibnamefont
  {Sadati}}, \bibinfo {author} {\bibfnamefont {N.}~\bibnamefont {{Taheri
  Qazvini}}}, \bibinfo {author} {\bibfnamefont {R.}~\bibnamefont {Krishnan}},
  \bibinfo {author} {\bibfnamefont {C.~Y.}\ \bibnamefont {Park}},\ and\
  \bibinfo {author} {\bibfnamefont {J.~J.}\ \bibnamefont {Fredberg}},\ }\href
  {https://doi.org/https://doi.org/10.1016/j.diff.2013.02.005} {\bibfield
  {journal} {\bibinfo  {journal} {Differentiation}\ }\textbf {\bibinfo {volume}
  {86}},\ \bibinfo {pages} {121} (\bibinfo {year} {2013})}\BibitemShut
  {NoStop}%
\bibitem [{\citenamefont {Kim}\ \emph {et~al.}(2021)\citenamefont {Kim},
  \citenamefont {Pochitaloff}, \citenamefont {Stooke-Vaughan},\ and\
  \citenamefont {Camp\'as}}]{Kim2021}%
  \BibitemOpen
  \bibfield  {author} {\bibinfo {author} {\bibfnamefont {S.}~\bibnamefont
  {Kim}}, \bibinfo {author} {\bibfnamefont {M.}~\bibnamefont {Pochitaloff}},
  \bibinfo {author} {\bibfnamefont {G.~A.}\ \bibnamefont {Stooke-Vaughan}},\
  and\ \bibinfo {author} {\bibfnamefont {O.}~\bibnamefont {Camp\'as}},\ }\href
  {https://doi.org/https://doi.org/10.1038/s41567-021-01215-1} {\bibfield
  {journal} {\bibinfo  {journal} {Nature Physics}\ }\textbf {\bibinfo {volume}
  {17}},\ \bibinfo {pages} {859} (\bibinfo {year} {2021})}\BibitemShut
  {NoStop}%
\bibitem [{\citenamefont {Ingber}(1997)}]{Ingber1997}%
  \BibitemOpen
  \bibfield  {author} {\bibinfo {author} {\bibfnamefont {D.~E.}\ \bibnamefont
  {Ingber}},\ }\href@noop {} {\bibfield  {journal} {\bibinfo  {journal} {Annual
  Review of Physiology}\ }\textbf {\bibinfo {volume} {59}},\ \bibinfo {pages}
  {575} (\bibinfo {year} {1997})}\BibitemShut {NoStop}%
\bibitem [{\citenamefont {Viens}\ and\ \citenamefont
  {Brodland}(2007)}]{Viens2007}%
  \BibitemOpen
  \bibfield  {author} {\bibinfo {author} {\bibfnamefont {D.}~\bibnamefont
  {Viens}}\ and\ \bibinfo {author} {\bibfnamefont {G.~W.}\ \bibnamefont
  {Brodland}},\ }\href@noop {} {\bibfield  {journal} {\bibinfo  {journal}
  {Journal of Biomechanical Engineering}\ }\textbf {\bibinfo {volume} {129}},\
  \bibinfo {pages} {651} (\bibinfo {year} {2007})}\BibitemShut {NoStop}%
\bibitem [{\citenamefont {Nnetu}\ \emph {et~al.}(2012)\citenamefont {Nnetu},
  \citenamefont {Knorr}, \citenamefont {K\"as},\ and\ \citenamefont
  {Zink}}]{Nnetu_2012}%
  \BibitemOpen
  \bibfield  {author} {\bibinfo {author} {\bibfnamefont {K.~D.}\ \bibnamefont
  {Nnetu}}, \bibinfo {author} {\bibfnamefont {M.}~\bibnamefont {Knorr}},
  \bibinfo {author} {\bibfnamefont {J.}~\bibnamefont {K\"as}},\ and\ \bibinfo
  {author} {\bibfnamefont {M.}~\bibnamefont {Zink}},\ }\href
  {https://doi.org/10.1088/1367-2630/14/11/115012} {\bibfield  {journal}
  {\bibinfo  {journal} {New Journal of Physics}\ }\textbf {\bibinfo {volume}
  {14}},\ \bibinfo {pages} {115012} (\bibinfo {year} {2012})}\BibitemShut
  {NoStop}%
\bibitem [{\citenamefont {Mongera}\ \emph {et~al.}(2018)\citenamefont
  {Mongera}, \citenamefont {Rowghanian}, \citenamefont {Gustafson},
  \citenamefont {Shelton}, \citenamefont {Kealhofer}, \citenamefont {Carn},
  \citenamefont {Serwane}, \citenamefont {Lucio}, \citenamefont {Giammona},\
  and\ \citenamefont {Camp\'as}}]{Mongera2018}%
  \BibitemOpen
  \bibfield  {author} {\bibinfo {author} {\bibfnamefont {A.}~\bibnamefont
  {Mongera}}, \bibinfo {author} {\bibfnamefont {P.}~\bibnamefont {Rowghanian}},
  \bibinfo {author} {\bibfnamefont {H.~J.}\ \bibnamefont {Gustafson}}, \bibinfo
  {author} {\bibfnamefont {E.}~\bibnamefont {Shelton}}, \bibinfo {author}
  {\bibfnamefont {D.~A.}\ \bibnamefont {Kealhofer}}, \bibinfo {author}
  {\bibfnamefont {E.~K.}\ \bibnamefont {Carn}}, \bibinfo {author}
  {\bibfnamefont {F.}~\bibnamefont {Serwane}}, \bibinfo {author} {\bibfnamefont
  {A.~A.}\ \bibnamefont {Lucio}}, \bibinfo {author} {\bibfnamefont
  {J.}~\bibnamefont {Giammona}},\ and\ \bibinfo {author} {\bibfnamefont
  {O.}~\bibnamefont {Camp\'as}},\ }\href
  {https://doi.org/10.1038/s41586-018-0479-2} {\bibfield  {journal} {\bibinfo
  {journal} {Nature}\ }\textbf {\bibinfo {volume} {561}},\ \bibinfo {pages}
  {401} (\bibinfo {year} {2018})}\BibitemShut {NoStop}%
\bibitem [{\citenamefont {Oswald}\ \emph {et~al.}(2017)\citenamefont {Oswald},
  \citenamefont {Grosser}, \citenamefont {Smith},\ and\ \citenamefont
  {Käs}}]{Oswald_2017}%
  \BibitemOpen
  \bibfield  {author} {\bibinfo {author} {\bibfnamefont {L.}~\bibnamefont
  {Oswald}}, \bibinfo {author} {\bibfnamefont {S.}~\bibnamefont {Grosser}},
  \bibinfo {author} {\bibfnamefont {D.~M.}\ \bibnamefont {Smith}},\ and\
  \bibinfo {author} {\bibfnamefont {J.~A.}\ \bibnamefont {Käs}},\ }\href
  {https://doi.org/10.1088/1361-6463/aa8e83} {\bibfield  {journal} {\bibinfo
  {journal} {Journal of Physics D: Applied Physics}\ }\textbf {\bibinfo
  {volume} {50}},\ \bibinfo {pages} {483001} (\bibinfo {year}
  {2017})}\BibitemShut {NoStop}%
\bibitem [{\citenamefont {Das}\ \emph {et~al.}(2021)\citenamefont {Das},
  \citenamefont {Sastry},\ and\ \citenamefont {Bi}}]{Das2020}%
  \BibitemOpen
  \bibfield  {author} {\bibinfo {author} {\bibfnamefont {A.}~\bibnamefont
  {Das}}, \bibinfo {author} {\bibfnamefont {S.}~\bibnamefont {Sastry}},\ and\
  \bibinfo {author} {\bibfnamefont {D.}~\bibnamefont {Bi}},\ }\href
  {https://doi.org/10.1103/PhysRevX.11.041037} {\bibfield  {journal} {\bibinfo
  {journal} {Phys. Rev. X}\ }\textbf {\bibinfo {volume} {11}},\ \bibinfo
  {pages} {041037} (\bibinfo {year} {2021})}\BibitemShut {NoStop}%
\end{thebibliography}%

	\end{document}